%% file: main.tex
\documentclass[11pt,a4paper]{article}
\usepackage{jheppub}
\usepackage{epsfig,subfigure}
\usepackage{epstopdf}
\usepackage[utf8]{inputenc}
\usepackage{graphicx}
\usepackage{amssymb}
\usepackage{amsxtra}
\usepackage{amsmath}
\usepackage[export]{adjustbox}
\usepackage{booktabs,multirow,tabularx}
\usepackage{slashed}
\usepackage{float}
\usepackage{physics}
\usepackage{placeins}
\usepackage{xcolor}
\usepackage{hyperref}
\usepackage{soul}
\usepackage{bm}

\newcommand{\bea}{\begin{eqnarray}}
\newcommand{\eea}{\end{eqnarray}}

\def\beq{\begin{equation}}
\def\eeq{\end{equation}}

\newcommand{\M}{\mathcal{M}}
\usepackage{siunitx}
\usepackage{stackengine}

\preprint{PITT-PACC-2622}

\title{Beyond the Standard Model EFT: Stress Testing the Electroweak Phase Transition in the Real Singlet Model}

\author[a]{Morgan E. Cassidy,}
\author[b]{Zhongtian Dong,}
\author[c]{Peisi Huang}
\author[b]{Ian M. Lewis}
\affiliation[a]{Pittsburgh Particle Physics, Astrophysics, and Cosmology Center, Department of Physics and Astronomy, University of Pittsburgh, Pittsburgh, PA 15206, USA}
\affiliation[b]{Department of Physics and Astronomy, University of Kansas, Lawrence, Kansas, 66045~ U.S.A.}
\affiliation[c]{Department of Physics and Astronomy, University of Nebraska, Lincoln, NE 68588, USA}
\emailAdd{mec400@pitt.edu}
\emailAdd{cdong@ku.edu}
\emailAdd{peisi.huang@unl.edu}
\emailAdd{ian.lewis@ku.edu}
\abstract{The real singlet scalar extension of the Standard Model is a much studied benchmark model for a strong first order electroweak phase transition (EWPT) and its role in electroweak baryogenesis. However, this model is not sufficient to generate the baryon asymmetry of the Universe, since it does not introduce any additional sources of CP violation. That is, additional new physics is still expected in the TeV range. In this paper we use effective field theory (EFT) methods to study the effects of new physics on the EWPT in the real singlet model with no $Z_2$. Effective operators up to dimension six with a cutoff scale of $\Lambda=1$~TeV are included in the scalar potential. We study the correlations between a strong first order EWPT, the Higgs trilinear coupling, resonant di-Higgs production, and gravitational wave signals in the EFT. We find that in the real singlet model with a strong first order EWPT the allowed region of the Higgs trilinear coupling is sensitive to TeV scale new physics, while the predictions of the gravitational wave signal and maximum resonant di-Higgs rate (with no additional production mechanisms) are robust against additional weakly coupled physics at or above the TeV scale. 
}

\begin{document}
\pagestyle{plain}

\maketitle



\newpage
\section{Introduction}
\label{sec:intro}
\input{intro.tex}

\section{Model}
\label{sec:model}
\input{model.tex}

\section{Electroweak Phase Transition}
\label{sec:EWPT}
\input{EWPT.tex}

\section{Phenomenology}
\label{sec:pheno}
\input{pheno.tex}

\section{Conclusions}
\label{sec:conc}
\input{conc.tex}

\section*{Acknowledgements}
MC is supported by the National Science
Foundation Graduate Research Fellowship Program. PH
is supported by the National Science Foundation under
grant number PHY-2412875. IML is supported in part by the United States Department of Energy grant number DE-SC0024673. Data to reproduce the figures are available upon request.

\appendix

\section{Experimental Searches}
\label{app:Exp}
\input{App_Exp.tex}

\section{Trilinear and Quartic Scalar Couplings}
\label{app:trilinears}
\input{App_Coup.tex}

\section{Truncating $V^{(1)}$ for 1 loop trilinears}
\label{app:V1_cutoff}
\input{App_V1_cutoff.tex}

\section{Temperature Corrected Masses}
\label{app:TempMasses}
\input{App_TempMasses.tex}

\section{Gravitational Wave Signal}
\label{app:gw}
\input{App_gravitational_waves.tex}

\section{Supplementary Numerical Results}
\label{app:Details}
\input{App_Details.tex}
\FloatBarrier
\bibliographystyle{JHEP} 
\bibliography{refnew}

\end{document}

%% file: intro.tex
Extended scalar sectors are well-motivated extensions of the Standard Model of Particle Physics (SM)~\cite{Ivanov:2017dad,Robens:2025nev} that can help solve many of the puzzles of particle physics. A particularly interesting problem they can help solve is the baryon asymmetry of the Universe (BAU).  
The introduction of new scalars  can alter the scalar potential away from the SM predictions, potentially leading to a strong first order electroweak phase transition (EWPT), which is a necessary ingredient for electroweak baryogenesis (EWBG). This EWBG scenario is interesting from the collider physics perspective in that these new scalars must couple to the Higgs and have masses in the TeV range~\cite{Ramsey-Musolf:2019lsf,Goncalves:2021egx}. Indeed, measuring the properties of the scalar potential is one of the long term goals of the LHC~\cite{CMS:2026nuu,ATLAS:2025eii} and future collider program~\cite{Profumo:2014opa,Huang:2016cjm,Curtin:2014jma,Kotwal:2016tex,Chen:2017qcz,Papaefstathiou:2021glr}. Additionally, there are many searches for new scalars [see App.~\ref{app:Exp}].

Even the simplest extension of the SM, a real scalar singlet, can provide a strong first order EWPT (SFOEWPT). This is a much studied scenario, see \cite{Profumo:2007wc,Barger:2007im,Espinosa:2011ax,Curtin:2014jma,Profumo:2014opa,Kotwal:2016tex,Huang:2016cjm,Noble:2007kk,Huang:2015tdv,Kurup:2017dzf,Chen:2017qcz,Beniwal:2017eik,Carena:2018vpt,Carena:2019une,Niemi:2021qvp,Papaefstathiou:2021glr,Ellis:2022lft,Huang:2022him,Aboudonia:2024frg,Niemi:2024axp,Ramsey-Musolf:2024ykk,Oikonomou:2024jms,Braathen:2025svl,Bosse:2026bdk,Damgaard:2015con,Vaskonen:2016yiu,Cline:2021iff} and references therein. However, such a real scalar singlet is expected to be only one particle in a larger sector. Indeed, while the real singlet scalar can provide a SFOEWPT, at the renormalizable level it does not introduce any additional CP violation that is necessary for EWBG to satisfy the Sakharov conditions~\cite{Sakharov:1967dj}. Hence, 
even within the context of generating the BAU, it is expected that there would be additional new physics beyond the real singlet scalar~\cite{Xiao:2015tja,Cline:2017jvp,Chao:2017oux,Bell:2019mbn,Espinosa:2011eu,Cline:2012hg,Huang:2018aja,Vaskonen:2016yiu,Cline:2021iff,Carena:2022qpf}. Assuming the scalar singlet is the lowest lying state and the only state accessible at the LHC, effective field theory (EFT) methods can be used to study the stability of the predictions of the simplified model against the addition of heavy new physics.  
Often beyond the Standard Model (BSM) EFTs can introduce new search channels or change the interpretation of LHC measurements and searches within a simplified model~\cite{Kim:2018mks,Alhazmi:2018whk,Criado:2019mvu,Li:2026kqk,Roy:2026kud,Crivellin:2016ihg,Carmona:2016qgo,Anisha:2019nzx,Karmakar:2019vnq,Diaz-Cruz:2001thx,Bar-Shalom:2018ure,Chala:2017sjk,Kamenik:2016tuv,Banerjee:2020jun,Anisha:2021jlz,Mitra:2022nri,Mitra:2026kxt,Dawson:2016ugw,Bauer:2016hcu,Adhikari:2020vqo}.  For example, in the real singlet model, it has been shown that effective operators can significantly change the interpretation of LHC Higgs measurements and scalar searches~\cite{Dawson:2016ugw,Bauer:2016hcu,Adhikari:2020vqo}.

In this paper we will study the EWPT in the real scalar singlet model with the scalar potential extended up to dimension-6 operators\footnote{There are additional non-renormalizable operators that couple the scalar singlet to SM fermions and gauge bosons~\cite{Gripaios:2016xuo,Adhikari:2020vqo} that can change the interpretation of collider data. Since we are focused on studying the EWPT, in this paper we limit ourselves to non-renormalizable additions to the scalar potential.}. We will consider the real singlet model with no additional symmetries\footnote{The scalar potential extended to dimension-6 operators in the $Z_2$ symmetric real singlet model extension was previously studied in~\cite{Oikonomou:2024jms} and in a gauged $U(1)_D$ model in~\cite{Bhattacharyya:2026esv}.}. The goal is to determine if the parameter space that can produce a strong first order EWPT is expanded when the higher dimensional operators are considered.  Once the allowed parameter space is determined, we will study the phenomenology of this model. The singlet model will alter the Higgs trilinear coupling and, if kinematically allowed, can give rise to resonant di-Higgs signals at the LHC, see Refs.~\cite{DiMicco:2019ngk,Dawson:2015haa,Lewis:2024yvj,Lane:2024vur,Robens:2015gla,Robens:2016xkb,Chen:2014ask,Lewis:2017dme} and references therein. Additionally, strong first order phase transitions in the early Universe will produce gravitational waves (GWs). We will study how large the resonant di-Higgs rates and variations in the trilinear couplings can be in the parameter space that successfully produces a SFOEWPT. We will also study the effects of including these operators and the prospects for detection at future GW detectors. As stated previously, these signals have been studied extensively in the renormalizable model~\cite{DiMicco:2019ngk}. Our study extends this previous work to determine the impact of effective operators on these signals in the context of a SFOEWPT. This study will help determine if the conclusions based on the renormalizable model are stable against heavy new physics, and if the signals can be enhanced when heavy off-shell new physics is included.

In Sec.~\ref{sec:model} we give an overview of the model, including the constraints on the zero temperature tree level and one-loop potential. The EWPT is studied in Sec.~\ref{sec:EWPT}.  In Sec.~\ref{sec:pheno} we study the collider and GW phenomenology. We conclude in Sec.~\ref{sec:conc}. In App.~\ref{app:Exp} we list the Higgs measurements and scalar resonant searches used to find current bounds on the model. The trilinear and quartic couplings in the EFT at tree level are provided in App.~\ref{app:trilinears}, and the one loop trilinears are in App.~\ref{app:V1_cutoff}. The thermal masses used for the phase transition are discussed in App.~\ref{app:TempMasses}. In App.~\ref{app:gw} we give the details of the GW calculation. Supplementary results studying collider observables at tree level while requiring a SFOEWPT as well as in the strongly interacting EFT are given in App~\ref{app:Details}.

%% file: model.tex
We consider the SM extended by a real gauge singlet scalar. At the renormalizable level, the singlet scalar can only couple to the SM through the scalar potential~\cite{Profumo:2007wc,Chen:2014ask}:
\begin{eqnarray}
V_{\rm ren}(\Phi,S)=-\mu^2 |\Phi|^2+\lambda |\Phi|^4+\frac{a_1}{2}S\,|\Phi|^2+\frac{a_2}{2}S^2\,|\Phi|^2+b_1\,S+\frac{b_2}{2}S^2+\frac{b_3}{3}S^3+\frac{b_4}{4}S^4,\label{eq:PotRen}
\end{eqnarray}
where $\Phi$ is the SM Higgs doublet and $S$ is the gauge singlet scalar. The renormalizable potential will be extended to include up to dimension-6 operators consisting of the Higgs doublet and scalar singlet~\cite{Gripaios:2016xuo,Bauer:2016hcu}:
\begin{eqnarray}
V_{\rm non-ren}(\Phi,S)&=&\frac{a_3}{2\,\Lambda}S^3 |\Phi|^2+\frac{a_4}{2\Lambda}|\Phi|^4 S+\frac{b_5}{5\Lambda}S^5+\frac{a_5}{2\,\Lambda^2}S^4|\Phi|^2+\frac{a_6}{2\,\Lambda^2}S^2|\Phi|^4\label{eq:PotNonRen}\\
&&+\frac{C_H}{\Lambda^2} |\Phi|^6+\frac{b_6}{6\,\Lambda^2}S^6+\mathcal{O}(\Lambda^{-3})\nonumber,
\end{eqnarray}
where $\Lambda$ is the scale of new physics. The full tree level potential is then 
\begin{eqnarray}
V_{\rm tree}(\Phi,S) &=& V_{\rm ren}(\Phi,S)+V_{\rm non-ren}(\Phi,S).\label{eq:PotTree}
\end{eqnarray}
As with the renormalizable model~\cite{Chen:2014ask}, since $S$ is a true singlet, the theory is invariant under a shift $S\rightarrow v_s+S$. Hence, there is freedom to set a parameter to zero. 
For simplicity we choose to set the vacuum expectation value of $S$ to zero\footnote{This choice is valid even when effective operators involving fermions and gauge bosons are considered \cite{Gripaios:2016xuo,Bauer:2016hcu,Adhikari:2020vqo}.}: $\langle S\rangle =0$.  

To ensure that the SM mass spectrum can be reproduced, there must be a minimum at
\begin{eqnarray}
\langle \Phi^\dagger\Phi\rangle=\frac{v_{\rm EW}^2}{2},\quad\langle S\rangle = 0,
\end{eqnarray}
where $v_{\rm EW}=246$~GeV. This requirement leads to the relationships
\begin{eqnarray}
\mu^2&=&v_{\rm EW}^2\lambda +\frac{3}{4}\frac{v_{\rm EW}^4}{\Lambda^2}C_H+\mathcal{O}(\Lambda^{-3})\label{eq:min}\\
b_1&=&-\frac{1}{4}v_{EW}^2a_1-\frac{1}{8}\frac{v_{\rm EW}^4}{\Lambda}a_4+\mathcal{O}(\Lambda^{-3}).\nonumber
\end{eqnarray}

After EWSB, there are two scalar mass eigenstates $h_1$ and $h_2$. These are admixtures of the SM Higgs boson $h$ and scalar singlet $S$:
\begin{eqnarray}
\begin{pmatrix} 
h_1\\ h_2
\end{pmatrix}
=
\begin{pmatrix} 
\cos\theta & \sin\theta\\
-\sin\theta & \cos\theta
\end{pmatrix}
\begin{pmatrix}
h\\ S
\end{pmatrix},
\end{eqnarray}
where in the unitary gauge $\Phi = (0,\phi_0)^{\rm T}/\sqrt{2}$, $\phi_0=v_{\rm EW}+h$, and $h_1,\,h_2$ have masses $m_1,m_2$, respectively. The mass matrix can then be used to solve for the relationships
\begin{eqnarray}
a_1&=&\frac{m_1^2-m_2^2}{v_{\rm EW}}\sin\,2\theta-\frac{v_{EW}^2}{\Lambda}a_4+\mathcal{O}(\Lambda^{-3})\label{eq:mass}\\
b_2+\frac{a_2}{2}v_{\rm EW}^2&=&m_1^2\sin^2\theta+m_2^2\cos^2\theta-\frac{1}{4}\frac{v_{\rm EW}^4}{\Lambda^2}a_6+\mathcal{O}(\Lambda^{-3})\nonumber\\
\lambda&=&\frac{m_1^2 \cos^2\theta+m_2^2\sin^2\theta}{2\,v_{\rm EW}^2}-\frac{3}{2}\frac{v_{\rm EW}^2}{\Lambda^2}C_H+\mathcal{O}(\Lambda^{-3}). \nonumber
\end{eqnarray}
Equations~(\ref{eq:min},\,\ref{eq:mass}) reproduce the results of \cite{Chen:2014ask} with corrections from the EFT.

We will identify $h_1$ as the discovered SM-like Higgs boson with mass $m_1=125$~GeV~\cite{ParticleDataGroup:2026aaa,ATLAS:2023oaq,CMS:2026nce} and assume $m_2>m_1$. The free parameters are then
\begin{eqnarray}
m_1=125~{\rm GeV},\,m_2,\,v_{\rm EW}=246~{\rm GeV},\,\langle S\rangle=0,\,\theta,\,a_2,\,b_3,\,b_4,\,a_3,\,a_4,\,a_5,\,a_6,\,b_5,\,b_6,\,C_H.
\end{eqnarray}

For di-Higgs phenomenology, the most important trilinear couplings are $h_1-h_1-h_1$ and $h_1-h_1-h_2$:
\begin{eqnarray}
V(\Phi,S)\supset \frac{1}{3!}\lambda_{111}h_1^3+\frac{1}{2}\lambda_{112}h_1^2h_2
\end{eqnarray}
At tree level, the trilinear couplings are
\begin{eqnarray}
\lambda_{111}&=&3\frac{m_1^2}{v_{\rm EW}}\cos^3\theta+\left(3\,a_2\,v_{\rm EW}\,\cos\theta+2\,b_3\,\sin\theta\right)\sin^2\theta+3\frac{v_{\rm EW}^2}{\Lambda}\sin\theta\left(\frac{1}{2}a_3\sin^2\theta+a_4\cos^2\theta\right)\nonumber\\
&&+6\frac{v_{\rm EW}^3}{\Lambda^2}\cos\theta\left(C_H\cos^2\theta+\frac{1}{2}a_6\sin^2\theta\right)+\mathcal{O}(\Lambda^{-3})\label{eq:Trilin}\\
\lambda_{112}&=&-\frac{2\,m_1^2+m_2^2}{v_{\rm EW}}\cos^2\theta\,\sin\theta+2\left(a_2\, v_{\rm EW}\left(1-\frac{3}{2}\sin^2\theta\right)+b_3\,\cos\theta\,\sin\theta\right)\sin\theta\nonumber\\
&&+\frac{v_{\rm EW}^2}{\Lambda}\cos\theta\left(\frac{3}{2}a_3\,\sin^2\theta+a_4\left(1-3\,\sin^2\theta\right)\right)\nonumber\\
&&-6\,\frac{v_{\rm EW}^3}{\Lambda^2}\sin\theta\left(C_H\cos^2\theta-\frac{1}{3}a_6\left(1-\frac{3}{2}\sin^2\theta\right)\right)+\mathcal{O}(\Lambda^{-3})\label{eq:lam112}
\end{eqnarray}
For completeness, other trilinear and quartic couplings are given at tree level in App.~\ref{app:trilinears}. The partial width for $h_2\rightarrow h_1h_1$ is
\begin{eqnarray}
\Gamma(h_2\rightarrow h_1h_1)&=&\frac{\lambda_{112}^2}{32\,\pi\,m_2}\sqrt{1-4\frac{m_1^2}{m_2^2}}.
\end{eqnarray}
If the trilinear coupling $\lambda_{112}$ in Eq.~(\ref{eq:lam112}) is used in $\Gamma(h_2\rightarrow h_1h_1)$ this would introduce $\Lambda^{-3}$ and $\Lambda^{-4}$ in the partial width. To be consistent with the EFT, we would then need to add dimension-7 and dimension-8 terms in the scalar potential. For consistent EFT counting, we introduce
\begin{eqnarray}
\left(\lambda_{112}^{\rm eff}\right)^2&\equiv&\left(\lambda_{112}^{(0)}+\frac{1}{\Lambda}\lambda_{112}^{(1)}\right)^2+\frac{2}{\Lambda^2}\lambda_{112}^{(2)}\lambda_{112}^{(0)},\label{eq:lam112eff}
\end{eqnarray}
where $\lambda^{(n)}_{112}$ are the coefficients of the $\Lambda^{-n}$ terms in  Eq.~(\ref{eq:lam112}):
\begin{eqnarray}
\lambda_{112}&=&\lambda_{112}^{(0)}+\frac{1}{\Lambda}\lambda^{(1)}_{112}+\frac{1}{\Lambda^2}\lambda^{(2)}_{112}+\mathcal{O}(\Lambda^{-3}).
\end{eqnarray}
With this definition, we have 
\begin{eqnarray}
\left(\lambda_{112}\right)^2&=&(\lambda^{\rm eff}_{112})^2+\mathcal{O}(\Lambda^{-3})
\end{eqnarray}
and the calculation of the total width using $\lambda_{112}^{\rm eff}$ is self-consistently truncated at dimension-6:
\begin{eqnarray}
\Gamma^{\rm eff}(h_2\rightarrow h_1h_1)&=&\frac{(\lambda_{112}^{\rm eff})^2}{32\,\pi\,m_2}\sqrt{1-4\frac{m_1^2}{m_2^2}}
\end{eqnarray}

\subsection{Zero Temperature One-Loop Potential}
In the on-shell renormalization scheme and Landau gauge, the one-loop effective potential, i.e., Coleman-Weinberg potential~\cite{Coleman:1973jx}, is~\cite{Anderson:1991zb,Delaunay:2007wb}
\begin{eqnarray}
\Delta V^{(1)}_{T=0}(\phi_0, S)&=&\sum_{i = W,Z,t,\phi_0,S} \frac{n_i}{64 \pi^2}\left[m_i^4(\phi_0,S)\left(\log\frac{m_i^2(\phi_0,S)}{m_i^2(v_{EW},0)} -\frac{3}{2} \right) +2 m_i^2(\phi_0,S)m_i^2(v_{EW},0)\right] \nonumber\\ 
&&+ \frac{n_\chi}{64\pi^2}m_\chi^4(\phi_0,S)\left[ \log\frac{m_\chi^2(\phi_0,S)}{m_h^2(v_{EW},0)} - \frac{3}{2} \right]\label{eq:VeffOS}
\end{eqnarray}
where $\phi_0$ is the neutral scalar component of the SM Higgs doublet; the sum is over $W,Z$, top, $\phi_0$, and $S$; $\chi$ are the Goldstone bosons; $n_i$ is the number of degrees of freedom for each particle; and $m_i(\phi_0,S)$ are scalar field-dependent masses. 
The degrees of freedom are $n_W=6$ for $W$s, $n_Z=3$ for Zs, $n_i=1$ for the Higgs boson and scalar $S$, $n_\chi=3$ for the Goldstone bosons, and $n_t=-12$ for top quarks.  This form of the effective potential ensures that the one-loop corrections do not shift the tree level mass and vev relations from Eqs.~(\ref{eq:min},\ref{eq:mass}) at the EW minimum:
\begin{eqnarray}
0&=&\left.\frac{\partial\Delta V^{\rm (1)}_{T=0}}{\partial \phi_0}\right|_{\begin{subarray}{l}\phi_0=v_{\rm EW}\\S=0\end{subarray}}=\left.\frac{\partial\Delta V^{(1)}_{T=0}}{\partial S}\right|_{\begin{subarray}{l}\phi_0=v_{\rm EW}\\S=0\end{subarray}}\label{eq:vevren}\\
0&=&\left.\frac{\partial^2 \Delta V^{(1)}_{T=0}}{\partial\phi_0^2}\right|_{\begin{subarray}{l}\phi_0=v_{\rm EW}\\S=0\end{subarray}}=\left.\frac{\partial^2\Delta V^{(1)}_{T=0}}{\partial S^2}\right|_{\begin{subarray}{l}\phi_0=v_{\rm EW}\\S=0\end{subarray}}=\left.\frac{\partial^2\Delta V^{(1)}_{T=0}}{\partial\phi_0\partial S}\right|_{\begin{subarray}{l}\phi_0=v_{\rm EW}\\S=0\end{subarray}}.\label{eq:massren}
\end{eqnarray}
However, the trilinears in Eqs.~(\ref{eq:Trilin},\ref{eq:lam112}) will have loop corrections. The full zero temperature potential is then
\begin{eqnarray}
V^{(1)}_{T=0}(\phi_0,S)=V_{\rm tree}((0,\phi_0/\sqrt{2})^{\rm T},S)+\Delta V_{T=0}^{(1)}(\phi_0,S).\label{eq:PotZeroT}
\end{eqnarray}

In principle, the on-shell renormalization conditions of Eq.~(\ref{eq:massren}) introduce infrared (IR) divergences for massless particles (such as Goldstone bosons) included in the sum over particles in the effective potential. These IR divergences can be traced to the fact that the effective potential is evaluated with zero external momentum. The IR divergences can be tamed by properly shifting the mass renormalization conditions to the Higgs pole mass, see  Refs.~\cite{Casas:1994us,Delaunay:2007wb}. While the Goldstones do introduce IR divergences in the renormalization of the scalar mass, their effects after shifting the renormalization conditions are small. Hence, we use the form of the on-shell effective potential in Eq.~(\ref{eq:VeffOS}) and neglect the Goldstone bosons when calculating the one-loop scalar trilinear couplings.

\subsection{Theory Constraints}
\label{sec:TheoryConstraints}
There are many theory constraints on the parameters of this model. For example, parameters should be perturbative, the scalar potential should be bounded from below, and the global minimum of the scalar potential should be $(\langle \Phi^\dagger\Phi\rangle, S)=(v_{\rm EW}^2/2,0)$. In this section we give an overview of these constraints.

\subsubsection{Perturbativity}
Perturbative unitarity applied to two-to-two scattering essentially places perturbative bounds on the quartics. The matrix element for two-to-two scattering has the partial wave decomposition:
\begin{eqnarray}
\mathcal{M}=16\pi \sum_k (2\,k+1)a_k P_k(\cos\theta).
\end{eqnarray}
The unitarity upper bound on matrix elements implies that in the complex plane the coefficients $a_k$ must lie within a circle centered at $a_k=i/2$ with a radius of $1/2$. Projecting this onto the real axis results in the perturbative unitarity upper limit 
\begin{eqnarray}
|{\rm Re}(a_k)|\leq \frac{1}{2}.\label{eq:pertunit}
\end{eqnarray}
This bound is applied order-by-order in perturbation theory. When it is saturated, there is a minimum of a $41\%$ higher order correction to return $a_k$ back into the unitarity circle~\cite{Schuessler:2007av}. Using Eq.~(\ref{eq:pertunit}) on the scattering processes $hh\rightarrow hh$, $SS\rightarrow SS$, and $Sh\rightarrow Sh$,  we find the upper limits on the quartic couplings
\begin{eqnarray}
|\lambda|\leq \frac{4\pi}{3}, \quad |b_4| \leq \frac{4\pi}{3}, \quad {\rm and}\quad |a_2|\leq 8\pi,\label{eq:pertunitUpp}
\end{eqnarray}
respectively.  To ensure perturbativity, we further constrain $a_2$:
\begin{eqnarray}
|a_2|\leq 4\pi.\label{eq:a2pert}
\end{eqnarray}

The trilinear terms in the scalar potential in Eq.~(\ref{eq:PotRen}) also have perturbative upper limits.  Roughly, a triangle loop correction to a trilinear would scale as $\sim\kappa^3/(16\pi^2\,m_i^2)$ where $\kappa$ is a trilinear coupling entering the loop and $m_i$ is the relevant mass scale. Since loops correcting $a_1$ or $b_3$ involve the scalar $S$, we choose $m_2$ as the relevant mass scale. Requiring that the loop corrections be smaller than the tree-level contributions leads to the rough upper bounds of
\begin{eqnarray}
|a_1|\lesssim 4\pi m_2,\quad |b_3|\lesssim 4\pi m_2.\label{eq:tribound}
\end{eqnarray}

Finally, we place perturbative upper bounds on the Wilson coefficients in Eq.~(\ref{eq:PotNonRen}).  Throughout the paper, we will consider two cases of upper bounds:
\begin{eqnarray}
&{\rm Weakly~coupled}:\quad &|a_{3,4,5,6}|,\,|b_{5,6}|,\,|C_H|\leq 1\label{eq:weak}\\
&{\rm Strongly~coupled}:\quad & |a_{3,4,5,6}|,\,|b_{5,6}|,\,|C_H|\leq 4\pi.\label{eq:strong}
\end{eqnarray}
The strongly coupled theory is at the limit of a perturbative theory, but we use it as a benchmark to determine if a strongly coupled theory is necessary to have large phenomenological effects.

\subsubsection{Bounded From Below/Global Minimum Constraints}
To prevent runaway directions in field space, the potential must be bounded from below. At the renormalizable level, this requirement is typically enforced for arbitrarily large field values. In that case, the dominant contributions to the renormalizable scalar potential are the quartics. The bounded from below condition on Eq.~(\ref{eq:PotRen})  is then:
\begin{eqnarray}
0\leq \frac{\lambda}{4}\phi^4_0+\frac{a_2}{4}S^2\phi_0^2+\frac{b_4}{4}S^4 = \frac{1}{4}\left(\lambda -\frac{a_2^2}{4\,b_4}\right)\phi_0^4+\frac{b_4}{4}\left(S^2+\frac{1}{2}\frac{a_2}{b_4}\phi^2_0\right)^2.
\end{eqnarray}
Hence, we have the global minimum constraints:
\begin{eqnarray}
0\leq\lambda,\quad 0\leq b_4,\quad{\rm and}\quad -2\sqrt{\lambda\,b_4}\leq a_2.\label{eq:BoundRen}
\end{eqnarray}

Including the non-renormalizable interactions in Eq.~(\ref{eq:PotNonRen}) relaxes these constraints. Since this is an EFT, we only require that the theory be defined up to the cutoff scale: $|\phi_0|,|S|\leq \Lambda$. Hence, we enforce a boundedness condition at the boundaries of the allowed field space: 
\begin{eqnarray}
V(\phi_0=\pm \Lambda,S=\pm \Lambda)-V(0,0)\geq 0.\label{eq:Bound}
\end{eqnarray}
It is assumed that whatever new physics exists above the cutoff scale will stabilize the potential to arbitrarily large field values.

Since the new scalar $S$ is a gauge singlet, it does not contribute to the gauge boson or fermion masses at the renormalizable level. Hence, $(\langle \Phi^\dagger\Phi\rangle, S)=(v_{\rm EW}^2/2,0)$ must be a global minimum. In the renormalizable theory, this must be true for all field values:
\begin{eqnarray}
V_{\rm ren}((0,\phi_0/\sqrt{2})^{\rm T},S)\geq V_{\rm ren}((0,v_{\rm EW}/\sqrt{2})^{\rm T},0),\quad {\rm for}\quad |\phi_0|,|S|< \infty.\label{eq:GlobalRen}
\end{eqnarray}
The renormalizable potential $V_{\rm ren}$ has six extrema, whose analytical formulae can be found in Refs.~\cite{Chen:2014ask,Lewis:2017dme}.  Hence, this condition can be checked explicitly.

Once the effective operators are considered, the global minimum constraint only needs to be valid to field values at the cutoff scale.  The global minimum constraint is then
\begin{eqnarray}
V(\phi_0,S)\geq V(\phi_0=v_{\rm EW},S=0),\quad{\rm for}\quad |\phi_0|,|S|\leq \Lambda.\label{eq:Global}
\end{eqnarray}

\subsubsection{Numerical Results}

\begin{figure}
\begin{center}
\subfigure[]{\includegraphics[width=0.45\textwidth,clip]{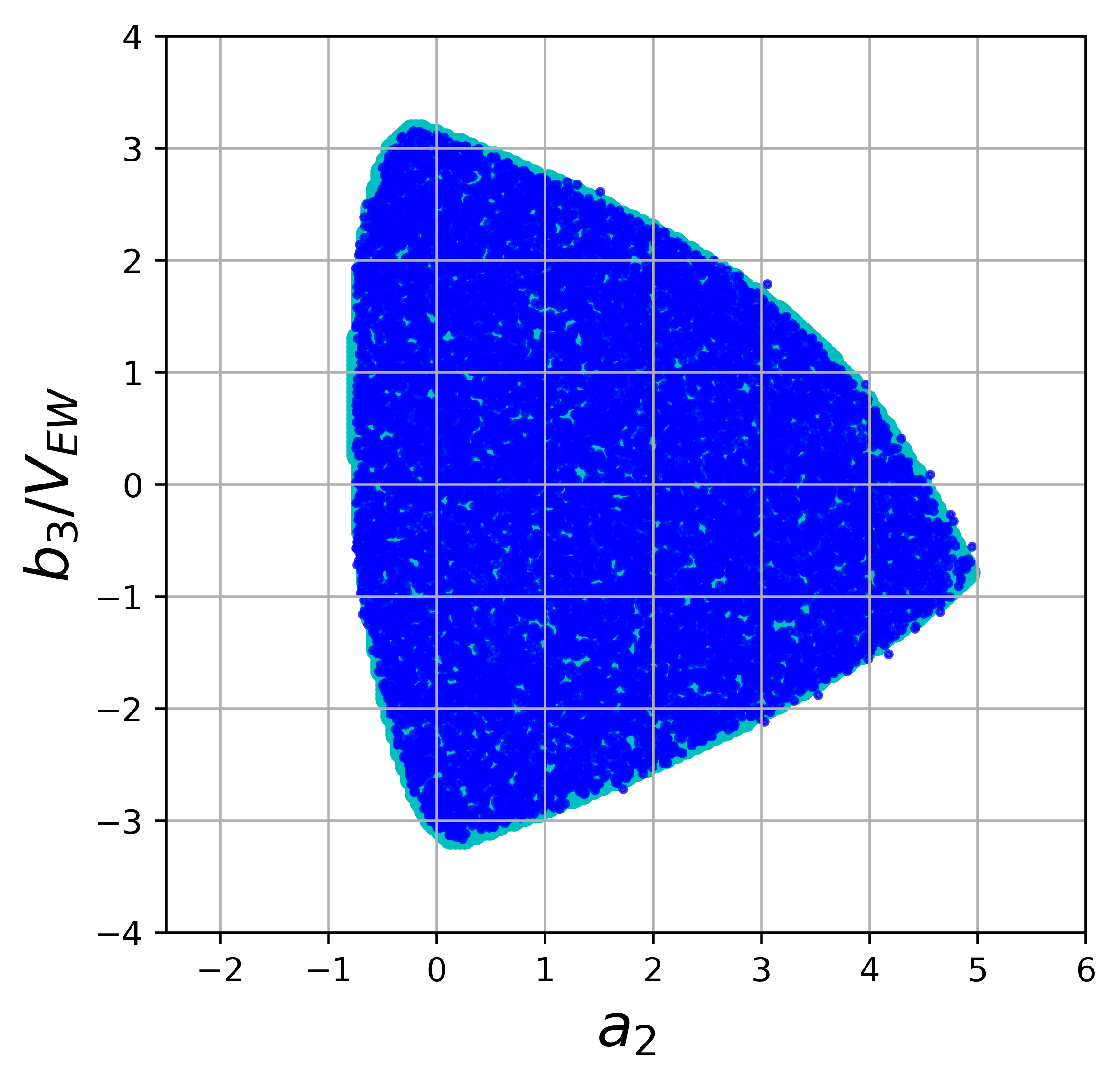}}
\subfigure[]{\includegraphics[width=0.45\textwidth,clip]{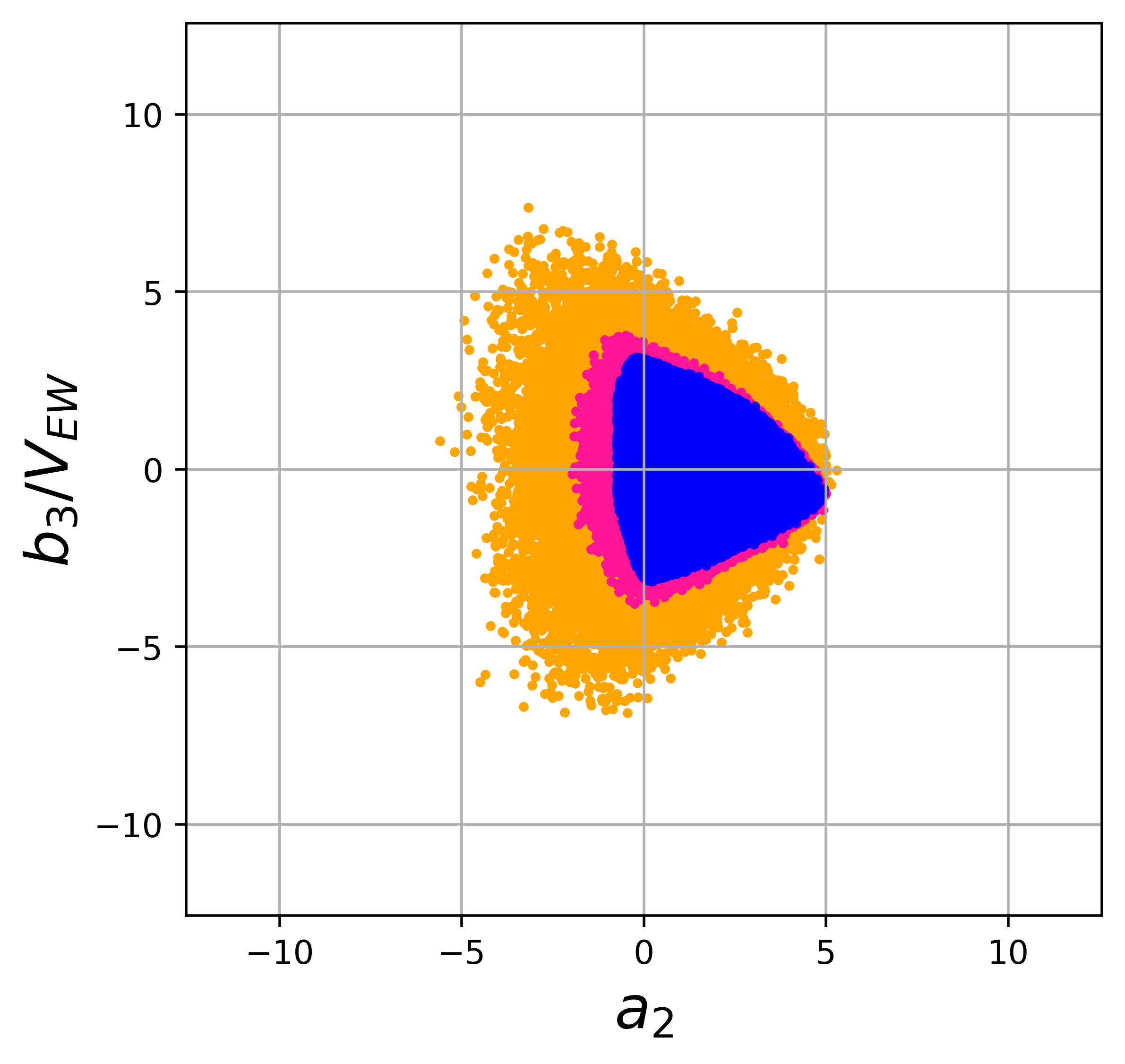}}
\subfigure[]{\includegraphics[width=0.45\textwidth,clip]{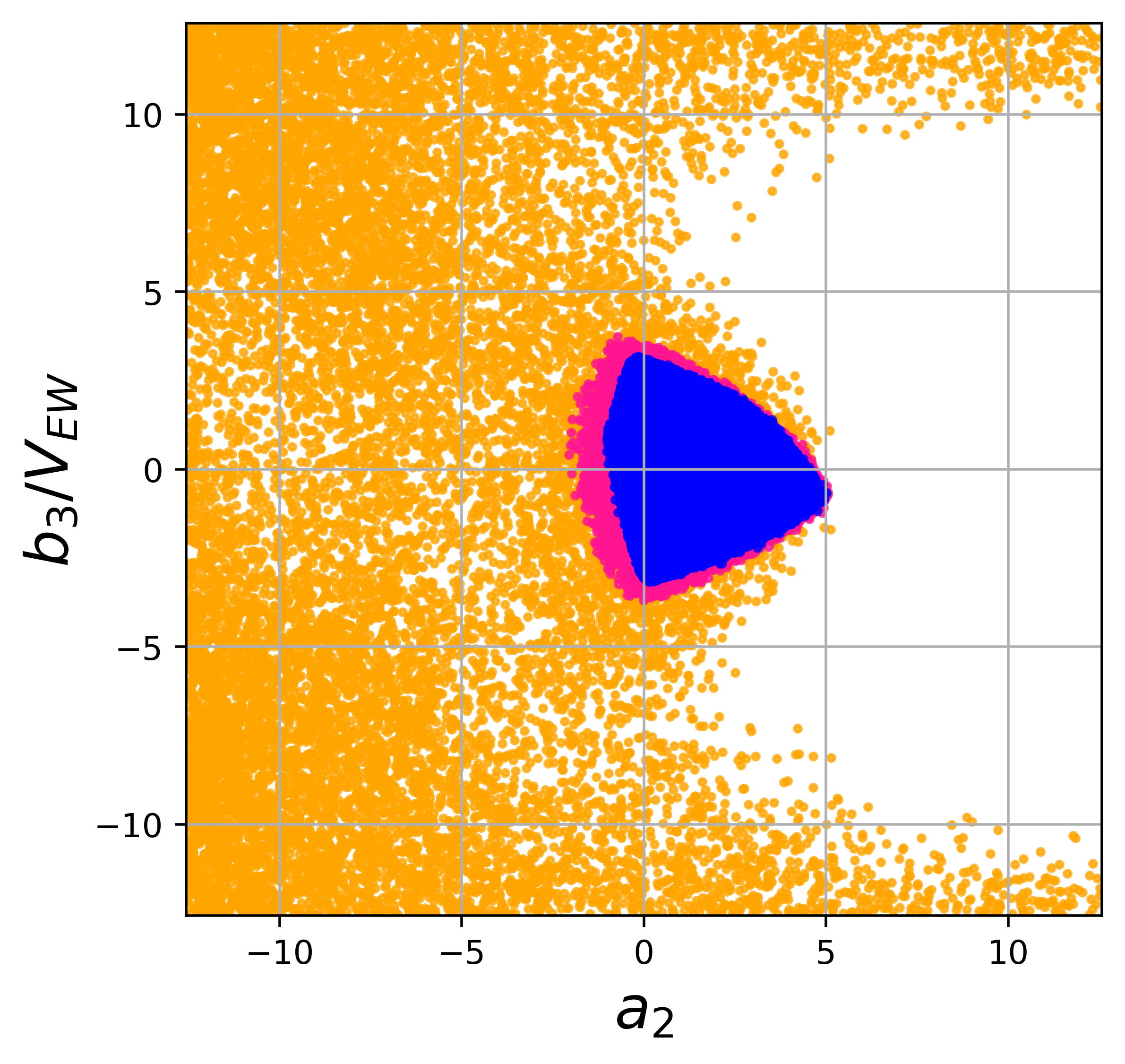}}
\end{center}
\caption{\label{fig:theory} In (a) we show the allowed parameter region for the tree level renormalizable potential $V_{\rm ren}$ in Eq.~(\ref{eq:PotRen})  with global minimum and boundedness constraints applied (cyan) to all field values [Eqs.~(\ref{eq:BoundRen},\ref{eq:GlobalRen})] and (blue) up to field values of 1 TeV [Eqs.~(\ref{eq:Bound},\ref{eq:Global})].  In (b) we show the allowed parameter space with global minimum and boundedness constraints applied to tree level potential out to field values of 1~TeV for (blue) $V_{\rm ren}$, (pink) the non-renormalizable potential $V_{\rm tree}$ in Eq.~(\ref{eq:PotTree}) with Wilson coefficients in the weakly coupled limit [Eq.~(\ref{eq:weak})], and (orange) $V_{\rm tree}$ with Wilson coefficients in the strongly coupled limit [Eq.~(\ref{eq:strong})]. (c) is the same as (b), using the one-loop potential. In all plots, $m_2=370$~GeV, $b_4=1$ and $\sin\theta=0.1$. For the EFT we set $\Lambda=1$~TeV. }
\end{figure}

We now explore the parameter space allowed by the various theoretical limits. 
We will explore three cases: 
\begin{eqnarray}
&{\rm renormalizable}: \quad & V(\phi_0,S)=V_{\rm ren}((0,\phi_0/\sqrt{2})^{\rm T},S)\quad{\rm [Eq.~(\ref{eq:PotRen})]}\\
&{\rm tree~level}:\quad & V(\phi_0,S)=V_{\rm tree}((0,\phi_0/\sqrt{2})^{\rm T},S)\quad{\rm [Eq.~(\ref{eq:PotTree})]},\\
&{\rm one-loop}: \quad & V(\phi_0,S)=V^{(1)}_{T=0}(\phi_0,S)\quad{\rm [Eq.~(\ref{eq:PotZeroT})]}.
\end{eqnarray}
In all cases, the perturbative bounds on the quartics and trilinears in Eqs.~(\ref{eq:pertunitUpp},\ref{eq:a2pert}) and (\ref{eq:tribound}), respectively, are applied.

Figure~\ref{fig:theory}(a) shows the theoretically allowed parameter space considering (cyan) the boundedness and global minimum constraints on the tree level renormalizable theory in Eqs.~(\ref{eq:BoundRen},\ref{eq:GlobalRen}) and (blue) those bounds only applied up to field values of $1$~TeV in Eqs.~(\ref{eq:Bound},\ref{eq:Global}).  As can be seen, there is no appreciable difference in allowed parameter spaces whether the constraints are applied to arbitrarily high field values or for field values up to the cutoff: $|\phi_0|,|S|< \Lambda$.

In Fig.~\ref{fig:theory}(b) we show the allowed parameter space considering boundedness and global minimum constraints up to field values of 1 TeV [Eqs.~(\ref{eq:Bound},\ref{eq:Global})] for tree level (blue) the renormalizable model, (pink) EFT with weakly coupled Wilson coefficients in Eq.~(\ref{eq:weak}), and (orange) the EFT with strongly coupled Wilson coefficients in Eq.~(\ref{eq:strong}). When the EFT operators are added, the allowed parameter space is increased compared to the renormalizable model. Unsurprisingly, when the Wilson coefficients are allowed to be as large as $4\pi$ the allowed parameter space greatly expands.

Figure~\ref{fig:theory}(c) is the same as Fig.~\ref{fig:theory}(b), but with the one-loop, zero temperature effective potential as described in Eq.~(\ref{eq:PotZeroT}). 
As can be seen, similarly to the tree level results, adding the weakly coupled EFT slightly expands the allowed region.  However, the strongly coupled EFT greatly expands the allowed parameter space at one loop. Indeed, in the strongly coupled case, the allowed parameter space often saturates the perturbative bounds on $a_2$ and $b_3$ {\it except} in the region $|b_3|\lesssim 10 v_{\rm EW}$ and $0\lesssim a_2\lesssim 5$. Comparing the bounds in the tree level and one-loop strongly interacting EFT indicates that the strongly interacting EFT is non-perturbative. We will neglect the strongly interacting theory until we discuss the GW signal, where we will show that the EFT will need to be strongly interacting to effect the GW phenomenology.

\subsection{Experimental Constraints}
\label{sec:ExpConst}

In the real singlet model, after rotating to the mass basis, the couplings of the scalar $h_1$ are suppressed by $\cos\theta$ relative to the SM Higgs boson.  Hence, all $h_1$ rates are suppressed by $\cos^2\theta$ relative to the SM predictions:
\begin{eqnarray}
\sigma_i(pp\rightarrow h_1)=\cos^2\theta\, \sigma_{i,{\rm SM}}(pp\rightarrow h_1),\quad \Gamma(h_1\rightarrow f_{\rm SM})=\cos^2\theta \,\Gamma_{\rm SM}(h_1\rightarrow f_{\rm SM}),
\end{eqnarray}
where the subscript ${\rm SM}$ on rates indicates the predictions of a SM Higgs at $m_1=125$~GeV and $f_{\rm SM}$ indicates SM gauge boson and fermion final states. Hence, assuming $h_2$ is massive enough to forbid $h_1\rightarrow h_2h_2$, there is a universal signal strength for $h_1$:
\begin{eqnarray}
\mu_i^f=\frac{\sigma_i(pp\rightarrow h_1){\rm BR}(h_1\rightarrow f_{\rm SM})}{\sigma_{i,{\rm SM}}(pp\rightarrow h_1){\rm BR}_{\rm SM}(h_1\rightarrow f_{\rm SM})}=\cos^2\theta.
\end{eqnarray}
Bounds from Higgs precision data then put a bound on the scalar mixing angle. The bounds are found via a $\chi^2$ fit:
\begin{eqnarray}
\chi^2_{h_1}=\sum_{i,f}\frac{(\mu_i^f-\hat{\mu}_i^f)^2}{(\delta_i^f)^2},\label{eq:h1chi2}
\end{eqnarray}
where $\hat{\mu}_i^f$ is the measured signal strength and $\delta_i^f$ are the 1$\sigma$ uncertainties on $\hat{\mu}_i^f$. A list of signal strengths included in the fit is given in App.~\ref{app:HiggsPrec}. We find the following 95\% CL upper limits from Higgs precision measurements:
\begin{eqnarray}
|\sin\theta|<0.25.
\end{eqnarray}

There are many searches at the LHC for scalar resonances. The coupling of $h_2$ to SM fermions and gauge bosons is suppressed by $\sin\theta$ relative to the SM Higgs couplings.  Hence, the production rate at the LHC and decay rates into SM fermions and gauge bosons are
\begin{eqnarray}
\sigma_i(pp\rightarrow h_2)=\sin^2\theta\,\sigma_{i,{\rm SM}}(pp\rightarrow h_2),\quad \Gamma(h_2\rightarrow f_{\rm SM})=\sin^2\theta \,\Gamma_{\rm SM}(h_2 \rightarrow f_{\rm SM}),
\end{eqnarray}
where the subscript ${\rm SM}$ indicates a rate calculated for a SM Higgs boson at a mass of $m_2$. For the SM-rates we use the results from the LHC Higgs Cross Section Working Group~\cite{LHCHiggsCrossSectionWorkingGroup:2016ypw}.

\begin{figure}[tb]
\begin{center}
\subfigure[]{\includegraphics[width=0.45\textwidth,clip]
{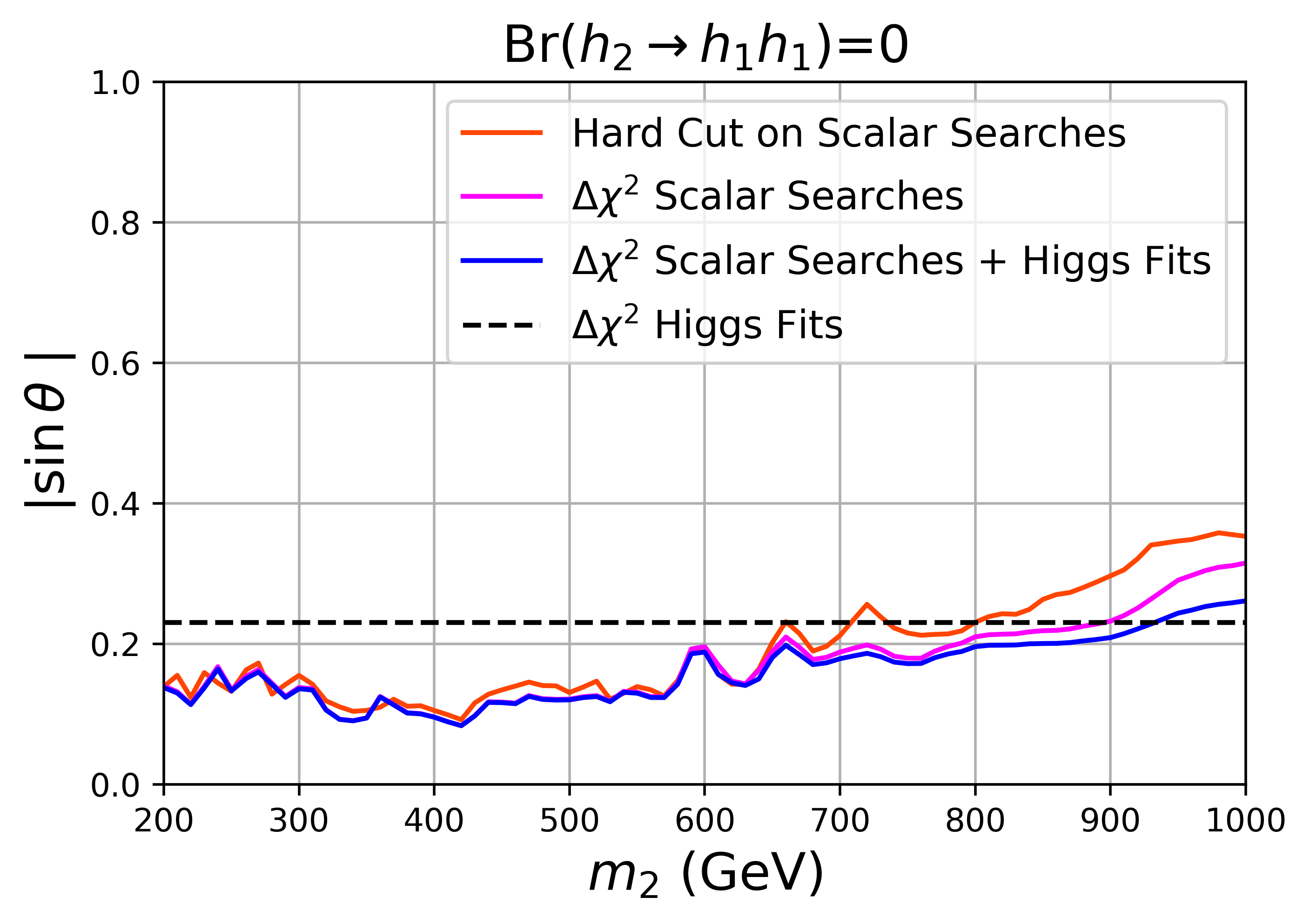}}
\subfigure[]{\includegraphics[width=0.45\textwidth,clip]
{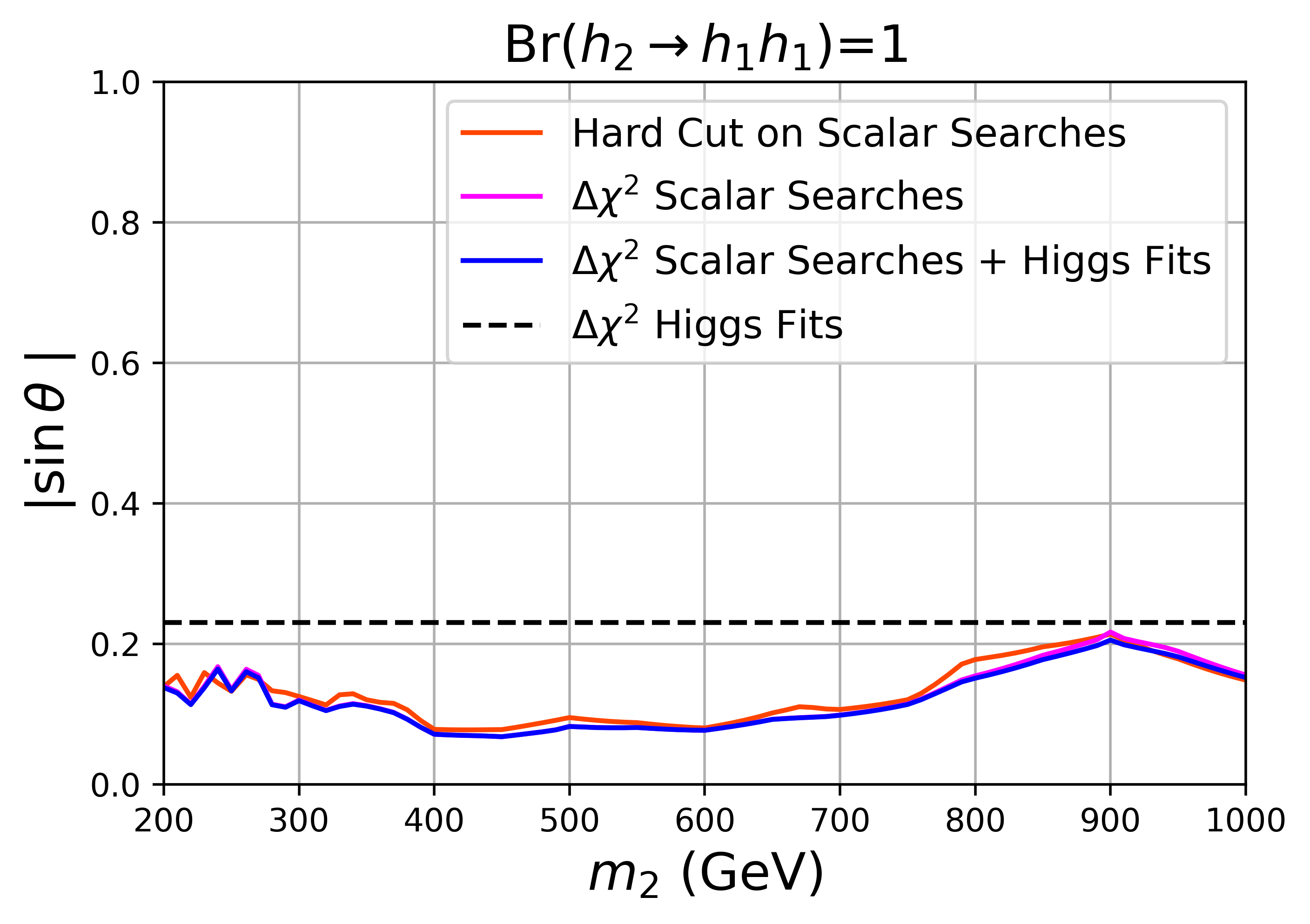}}
\subfigure[]{\includegraphics[width=0.45\textwidth,clip]
{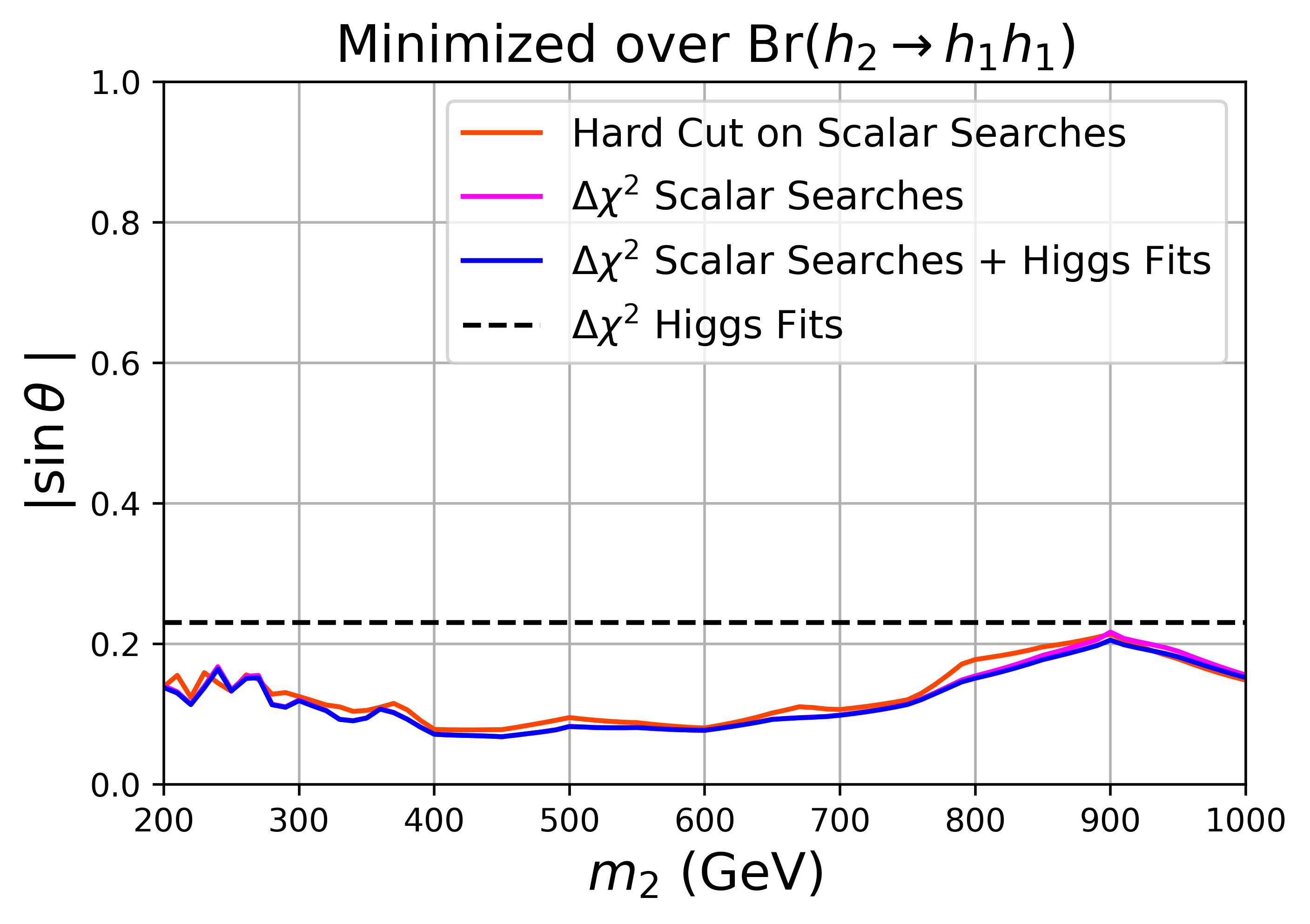}}
\end{center}
\caption{\label{fig:sth} Bounds on $|\sin\theta|$ for (a) ${\rm BR}(h_2\rightarrow h_1h_1)=0$, (b) ${\rm BR}(h_2\rightarrow h_1 h_1)=1$, and (c) minimized over ${\rm BR}(h_2\rightarrow h_1h_1)$. All plots show constraints from (orange solid) the hard cut method for scalar searches, (magenta solid) the $\Delta \chi^2$ method for scalar searches, (blue solid) the combination of $\Delta \chi^2$ method for scalar searches and Higgs precision, and (black dashed) $\Delta\chi_{h_1}^2$ fits from Eq.~(\ref{eq:h1chi2}).}
\end{figure}

We will keep $\sin\theta$, $m_2$, and ${\rm BR}(h_2\rightarrow h_1h_1)$ as free parameters. With this choice, the branching ratio of $h_2$ into SM fermions and gauge bosons can be expressed as
\begin{eqnarray}
{\rm BR}(h_2\rightarrow f_{\rm SM})={\rm BR}_{\rm SM}(h_2\rightarrow f_{\rm SM})\left[1-{\rm BR}(h_2\rightarrow h_1h_1)\right].
\end{eqnarray}
The experimental constraints are then applied using the narrow width approximation
\begin{eqnarray}
\sigma_i(pp\rightarrow h_2\rightarrow f)\approx \sigma_i(pp\rightarrow h_2){\rm BR}(h_2\rightarrow f),
\end{eqnarray}
where $f$ can be any kinematically allowed decay product. A list of experimental results included in the fits is given in App.~\ref{app:ScalSearch}.  

Bounds from scalar searches are implemented in two ways: a hard cut or a $\Delta\chi^2$ method. The hard cut method accepts a parameter point only if  all predicted production rates are below the observed limits:
\begin{eqnarray}
\sigma_{i}(pp\rightarrow h_2\rightarrow f)<\hat{\sigma}^f_{i,{\rm Obs}},\label{eq:HardCut}
\end{eqnarray}
where $\hat{\sigma}^f_{i,{\rm Obs}}$ is the 95\% CL observed upper limit on the cross section for initial state $i$ and final state $f$.
This is the traditional method used in tools such as \texttt{HiggsTools}~\cite{Bahl:2022igd}. The $\Delta \chi^2$ method was developed in Ref.~\cite{Adhikari:2020vqo}:
\begin{eqnarray}
\chi^2_{h_2}&=&\sum_{i,f}\left(\chi^f_{i,h_2}\right)^2\\
\left(\chi^f_{i,h_2}\right)^2&=&\displaystyle\begin{cases}
\displaystyle\left(\frac{\sigma_i(pp\rightarrow h_2\rightarrow f)+\hat{\sigma}^f_{i,{\rm Exp}}-\hat{\sigma}^f_{i,{\rm Obs}}}{\hat{\sigma}^f_{i,{\rm Exp}}/1.96}\right)^2\quad& {\rm if}~\hat{\sigma}_{i,{\rm Obs}}^f\geq \hat{\sigma}_{i,{\rm Exp}}^f\\
\displaystyle\left(\frac{\sigma_i(pp\rightarrow h_2\rightarrow f)}{\hat{\sigma}_{i,{\rm Obs}}^f/1.96}\right)^2&{\rm if}~\hat{\sigma}_{i,{\rm Obs}}^f< \hat{\sigma}_{i,{\rm Exp}}^f
\end{cases}\nonumber
\end{eqnarray}
For a single measurement, applying a $\Delta\chi^2_{h_2}$ fit will reproduce the hard cut upper bound in Eq.~(\ref{eq:HardCut}). The $\Delta\chi^2_{h_2}$ allows for a consistent combination of all searches together with Higgs precision measurements.

Fig.~\ref{fig:sth} shows the bounds on the scalar mixing angle $|\sin\theta|$ as a function of $h_2$ mass for (a) ${\rm BR}(h_2\rightarrow h_1h_1)=0$, (b) ${\rm BR}(h_2 \rightarrow h_1h_1)=1$, and (c) most stringent for all ${\rm BR}(h_2\rightarrow h_1h_1)$.  As can be seen in all the plots, the hard cut and $\Delta\chi^2$ methods for employing scalar searches give similar results while the $\Delta\chi^2$ combination of scalar searches and Higgs data is slightly stronger. When kinematically allowed, the bounds assuming ${\rm BR}(h_2\rightarrow h_1h_1)=1$ are stronger than those assuming ${\rm BR}(h_2\rightarrow h_1h_1)=0$ for all $m_2$ masses except $320~{\rm GeV}\lesssim m_2\lesssim 355-365~{\rm GeV}$. For all masses, the bounds from scalar searches with ${\rm BR}(h_2\rightarrow h_1h_1)=1$ are stronger than Higgs precision measurements.

%% file: EWPT.tex
The main results of this paper will be how the dimension-5 and 6 non-renormalizable operators in the scalar potential [Eq.~(\ref{eq:PotNonRen})] affect the parameter space that accommodates a SFOEWPT in the real scalar singlet model, as well as the signatures of the model.  Here we give an overview of the finite temperature calculation we use and report on the parameter space that allows for a SFOEWPT.

At finite temperatures, the effective scalar potential has additional contributions at one loop\footnote{There is work on higher order calculations of the finite temperature effective potential~\cite{Niemi:2021qvp,Niemi:2024axp,Bahl:2024ykv,Ekstedt:2022zro}. To compare the EFTs, we use the one loop result.}:
\begin{eqnarray}
V_{eff}(\phi_0,S, T)&=&V^{(1)}_{T=0}(\phi_0,S) + V_T(\phi_0,S,T)
\end{eqnarray}
where $V^{(1)}_{T=0}(\phi_0,S)$ is the zero temperature one-loop effective potential in Eq.~(\ref{eq:PotZeroT}) and the finite temperature contributions, $V_T$, are given as 
\begin{eqnarray}
    V_T(\phi_0, S,T) &=& \sum_{i} \frac{n_i T^4}{2\pi^2}\int_{0}^{\infty } dy\,y^2 \log\left( 1 \pm e^{-\sqrt{y^2 + m_i^2(\phi_0,S)/T^2}}\right),
\end{eqnarray}
where the sum is over all fields, $+$ is for fermions, and $-$ is for bosons. 

When calculating the temperature dependent effective potential, there are well-known infrared divergences in the form of ``daisy'' diagrams~\cite{Parwani:1991gq,Arnold:1992rz,Pilaftsis:2013xna,Pilaftsis:2015cka,Funakubo:2023cyv,Funakubo:2023eic,Boyd:1993tz,Curtin:2016urg,Curtin:2022ovx,Bahl:2024ykv,Bittar:2025lcr} where perturbation theory begins to fail when $m\ll T$, where $m$ is a typical mass scale. For more accurate and stable predictions, these contributions need to be resummed. We will use the Parwani scheme~\cite{Parwani:1991gq}\footnote{ Other schemes to resum infrared contributions to the finite-temperature effective potential involve solving a gap equation~\cite{Boyd:1993tz,Curtin:2016urg,Curtin:2022ovx,Bahl:2024ykv,Bittar:2025lcr} and the Arnold-Espinosa method~\cite{Arnold:1992rz}. At very high temperatures, the Arnold-Espinosa scheme can introduce symmetry non-restoration that is not reproduced with other methods~\cite{Bahl:2024ykv}.
}. In this scheme, all the boson masses are replaced by their high temperature thermally corrected masses in both the zero temperature Coleman-Weinberg potential $\Delta V_{T=0}^{(1)}$ and the finite temperature contributions $V_T$:
\begin{eqnarray}
    m_i^2(\phi_0,S) \rightarrow \mathcal{M}^2_i(\phi_0,S,T)&=& m_i^2(\phi_0, S) + \Pi_i(T)
\end{eqnarray}
where $\Pi_i(T)$ are the high temperature one-particle irreducible self-energies. The $\Pi_i(T)$ that we use are given in App.~\ref{app:TempMasses}.

For sphalerons to turn off sufficiently fast to freeze in the baryon asymmetry, the phase transition must be strong and first order. The first order phase transition occurs when a barrier forms between two degenerate minima of the effective potential at a critical temperature $T_c$. Bubbles will nucleate when the vacuum decay rate matches the Hubble rate at the nucleation temperature $T_n$. For temperatures around the EW scale, nucleation occurs when $S_3/T_n\approx 140$, where $S_3$ is the 3D Euclidean action. We use \texttt{CosmoTransitions}~\cite{Wainwright:2011kj} to solve the bounce equations and determine the nucleation temperature. 

The first order phase transition is said to be strong where $\phi_n/T_n\gtrsim1$~\cite{Kuzmin:1985mm,Arnold:1987mh,Shaposhnikov:1987tw,Cohen:1993nk} such that the sphalerons fall out of equilibrium sufficiently quickly, where $\phi_n$ is the location of the EW symmetry breaking minimum at the nucleation temperature. There are uncertainties in this relation due to the gauge dependence of the finite temperature effective potential~\cite{Dolan:1973qd,Dolan:1974gu,Morrissey:2012db,Garny:2012cg,Patel:2011th,Arunasalam:2021zrs,Lofgren:2021ogg} which introduces gauge dependence in both $\phi_n$ and $T_n$.  This can be traced to the Nielsen identities~\cite{Nielsen:1975fs,Fukuda:1975di} not being implemented order-by-order in the $\hbar$ expansion. While there are proposals for gauge invariant criteria for a SFOEWPT~\cite{Patel:2011th,Arunasalam:2021zrs,Lofgren:2021ogg}, for our purposes, we will use the na\"{i}ve estimation of $\phi_n/T_n\gtrsim1$ to compare the predictions of the renormalizable and non-renormalizable theories.

\begin{figure}[tb]
\begin{center}
\subfigure[]{\includegraphics[width=0.45\textwidth,clip]%
{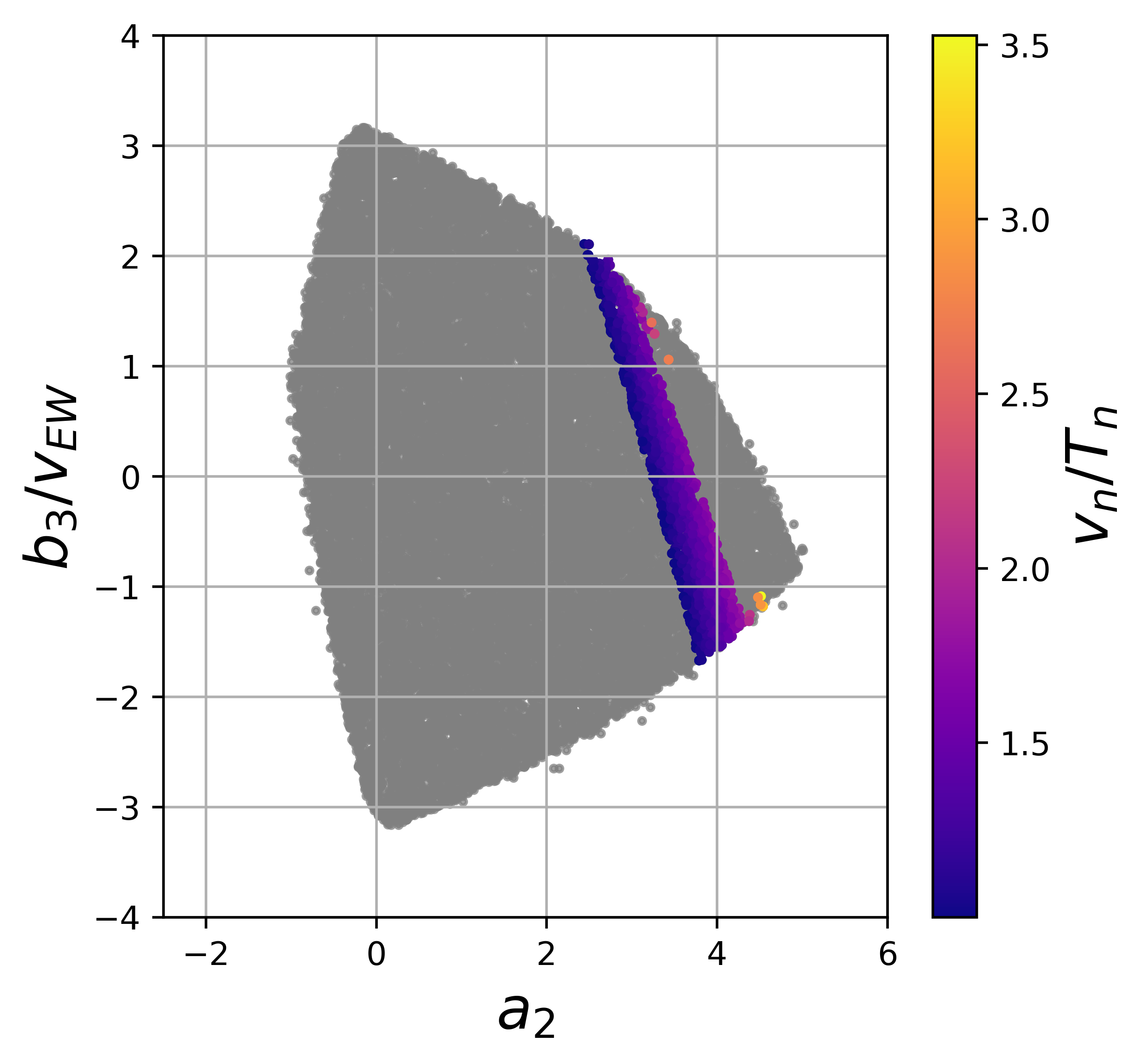}}
\subfigure[]{\includegraphics[width=0.45\textwidth,clip]%
{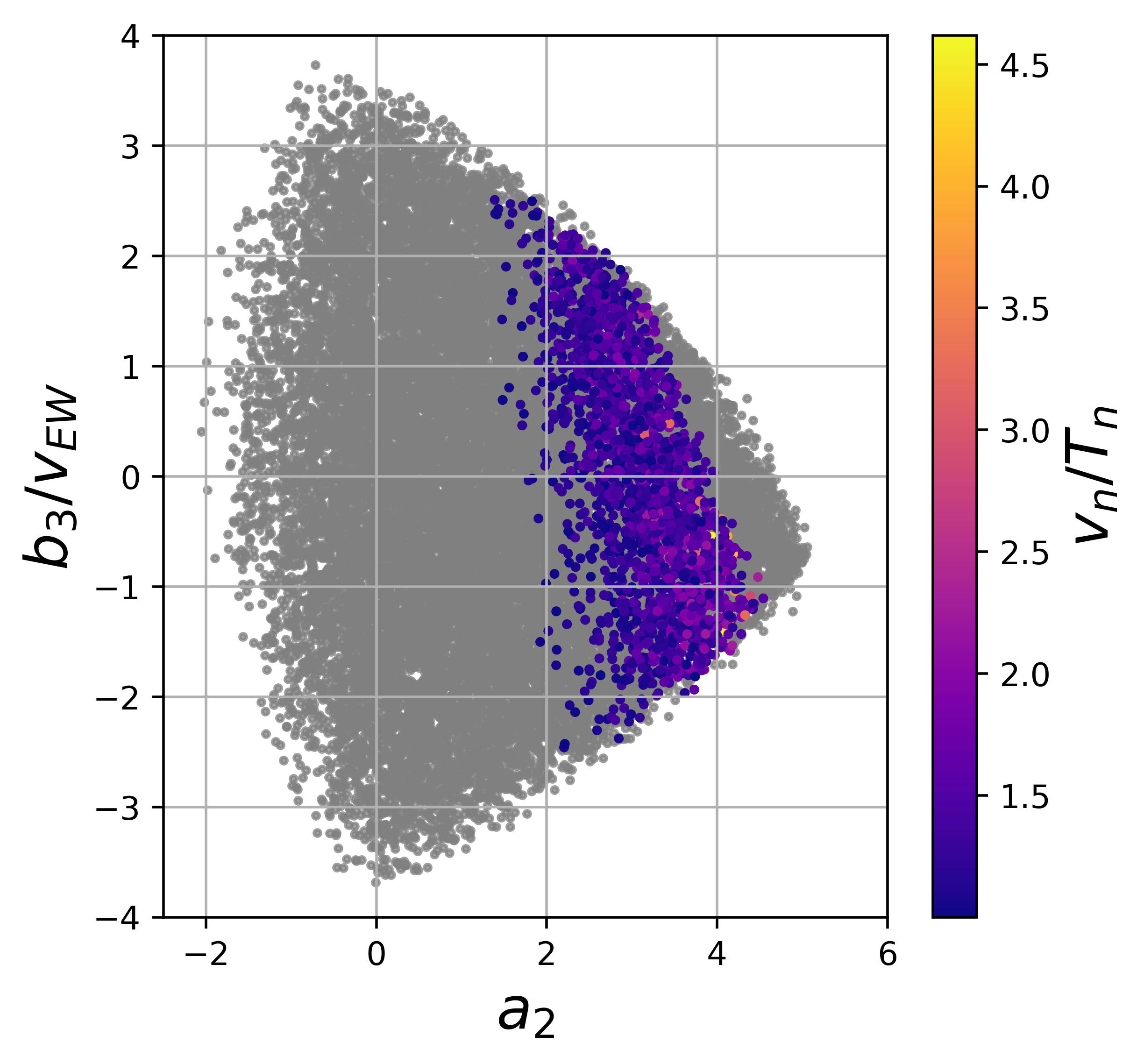}}
\end{center}
\caption{\label{fig:EWPT} 
Comparison of the parameter space with a SFOEWPT for (a) the renormalizable model and (b) the weakly coupled non-renormalizable model. Grey regions show the allowed parameter space at zero temperature while the colored heat map shows the strength of the phase transition. We fix $\sin\theta=0.1$, $m_2 = 370$ GeV, $b_4=1$, and $\Lambda=1$~TeV.
}
\end{figure}

In Fig.~\ref{fig:EWPT} we show the allowed parameter space for a SFOEWPT in the (a) renormalizable model and (b) the weakly coupled non-renormalizable model defined in Eq.~(\ref{eq:weak}). We use a cutoff of $\Lambda=1$~TeV and apply the boundedness and global minimum criteria of Eqs.~(\ref{eq:Bound},\ref{eq:Global}) up to the cutoff. In the renormalizable model, we find that only a relatively small band of allowed parameter space allows for a SFOEWPT. This is similar to results for the $Z_2$ symmetric singlet model~\cite{Kurup:2017dzf}. We do note that the allowed parameter space consistent with a SFOEWPT is smaller than in Ref.~\cite{Chen:2017qcz}. This is due to Ref.~\cite{Chen:2017qcz} using the condition $\phi_c/T_c\gtrsim 1$ as the SFOEWPT condition, where $\phi_c$ is the location of the second minimum at the critical temperature. The condition $\phi_c/T_c\gtrsim 1$ is weaker than $\phi_n/T_n\gtrsim 1$. Even if $\phi_c/T_c\gtrsim 1$, it is possible for the field to be stuck in the electroweak symmetric vacuum and fail to nucleate.  As shown in Fig.~\ref{fig:EWPT}(b), the weakly coupled EFT expands the region that nucleates while still requiring $a_2\gtrsim 1.5$.

%% file: pheno.tex
We now discuss the phenomenology of the SFOEWPT in the singlet extended model with non-renormalizable operators in the potential.   All constraints discussed in Secs.~\ref{sec:TheoryConstraints} and \ref{sec:ExpConst} are applied for $|\phi_0|,|S|<\Lambda$ using the one-loop effective potential,  as well as requiring $(\lambda_{112}^{\rm eff})^2>0$. Throughout this section we use $\Lambda=1$~TeV, as in the previous sections.

\subsection{Collider Phenomenology}
\begin{figure}[ht!]
\begin{center}
\includegraphics[width=0.99\textwidth]{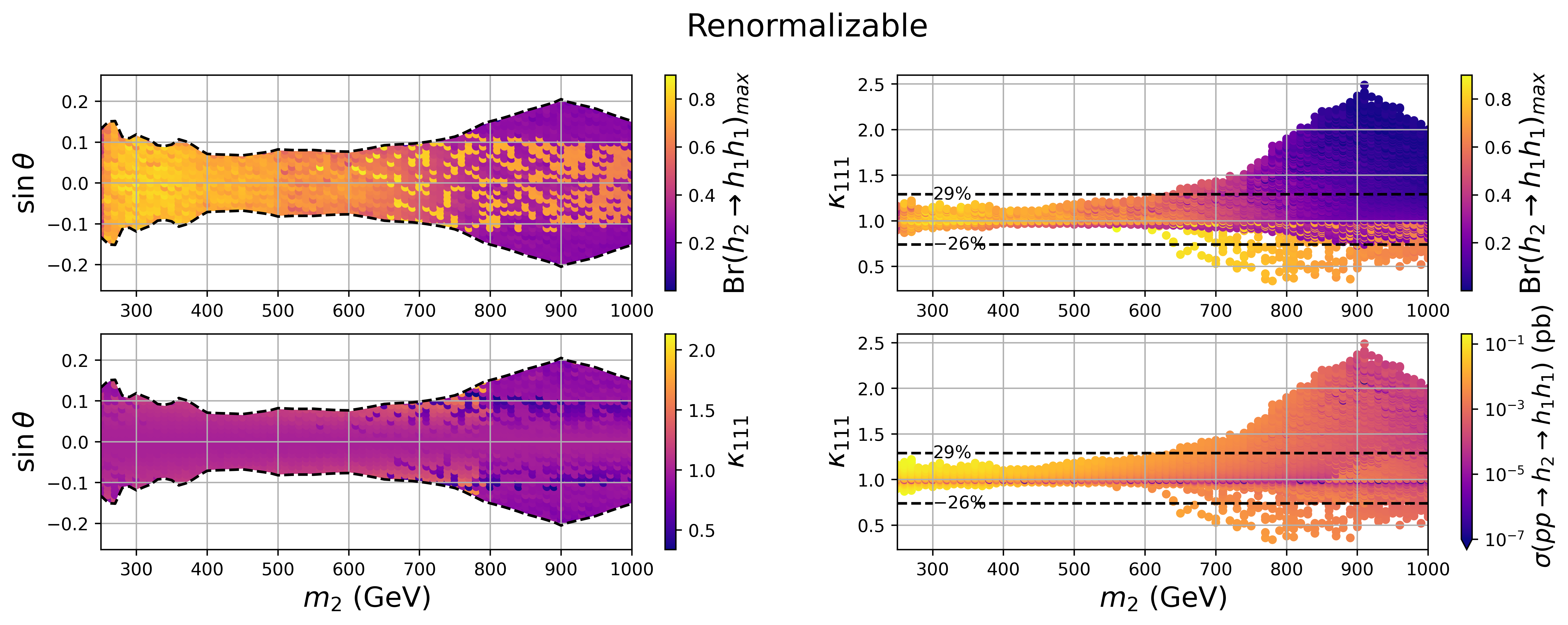}\\
\includegraphics[width=0.99\textwidth]{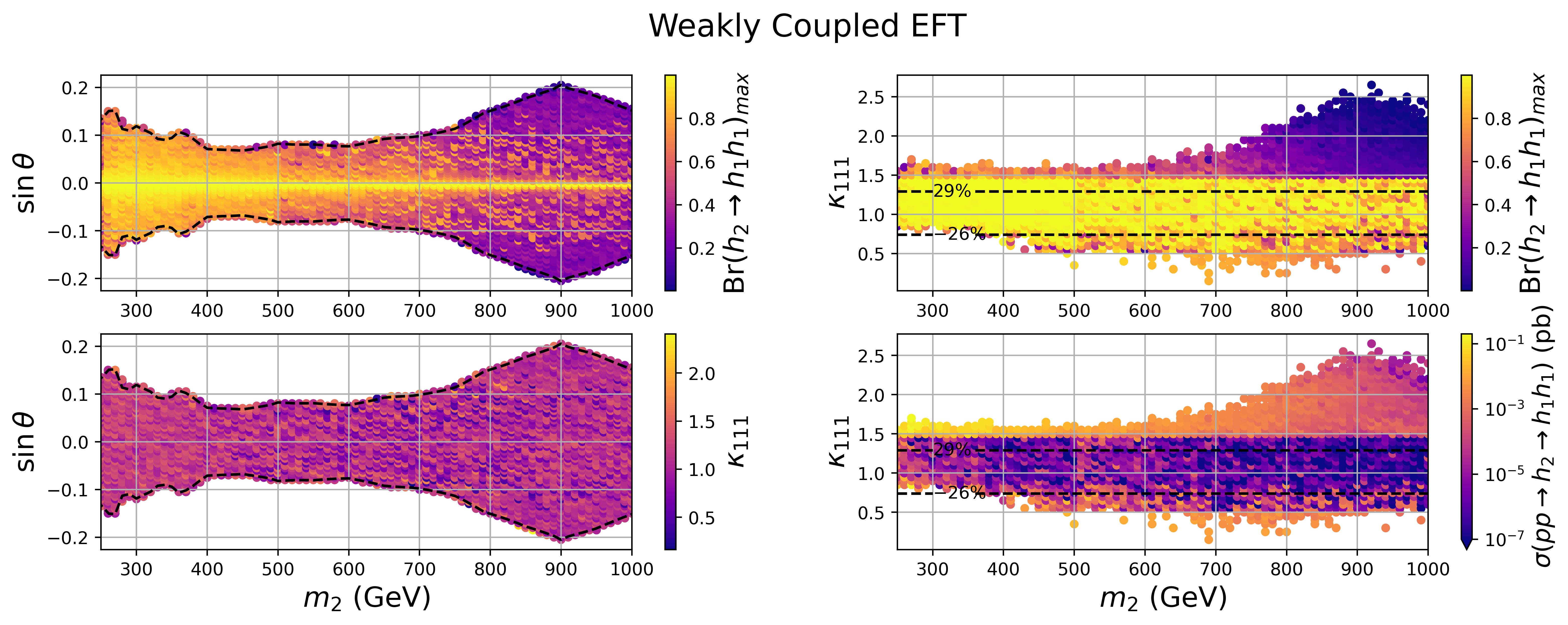}
\end{center}
\caption{\label{fig:collider_tree} Plots are paired vertically to show (left) the maximum ${\rm BR}(h_2\rightarrow h_1h_1)$ in the $\sin\theta-m_2$ plane and directly below the corresponding variation in $\kappa_{111}$, and (right) the maximum ${\rm BR}(h_2\rightarrow h_1h_1)$ in the $\kappa_{111}-m_2$ plane and directly below the corresponding resonant cross section $\sigma(pp\rightarrow h_2\rightarrow h_1h_1)$. The top two plots show the renormalizable model and the bottom two show the weakly coupled EFT.  All parameters are scanned over. All trilinear couplings are evaluated at tree level and we use $\lambda_{112}^{\rm eff}$. Dashed lines on the right correspond to projected HL-LHC bounds $0.74\leq \kappa_{111}\leq1.29$~\cite{ATLAS:2025eii}.
}

\end{figure}

\begin{figure}[t!]
\begin{center}
\includegraphics[width=0.99\textwidth]{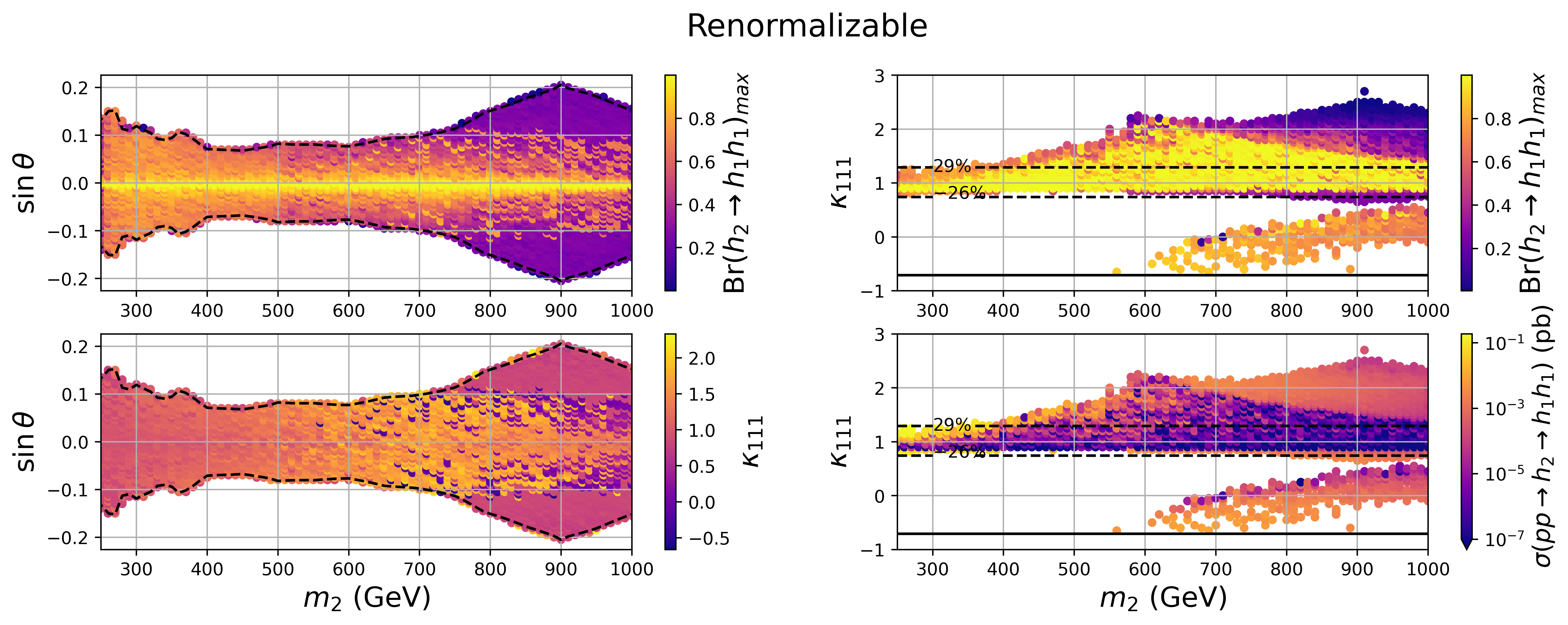}\\
\includegraphics[width=0.99\textwidth]{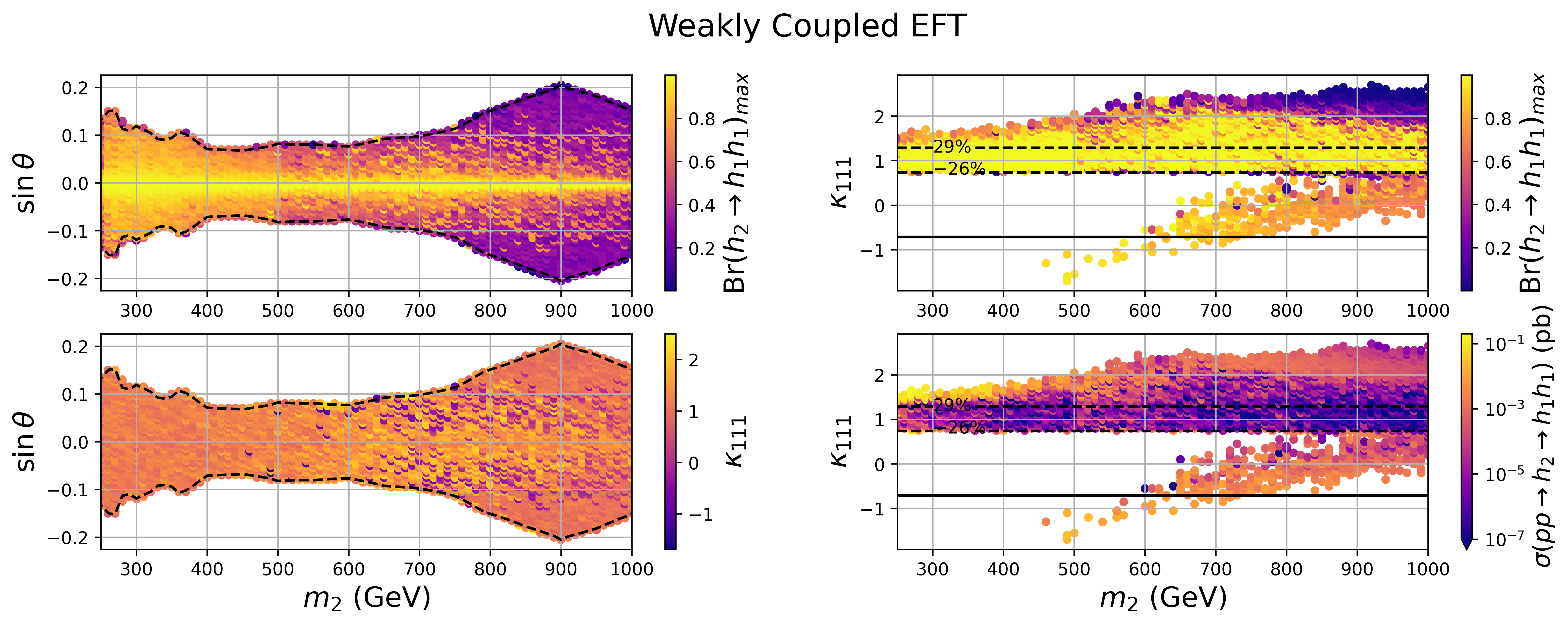}
\end{center}
\caption{\label{fig:collider_CW}  Same as Fig.~\ref{fig:collider_tree} with Coleman Weinberg definition of trilinears. Solid black lines on the right correspond to current LHC bounds $-0.71\leq \kappa_{111}\leq6.1$~\cite{CMS:2026nuu}.}
\end{figure}

\begin{figure}[tb]
\begin{center}
\includegraphics[width=0.99\textwidth]{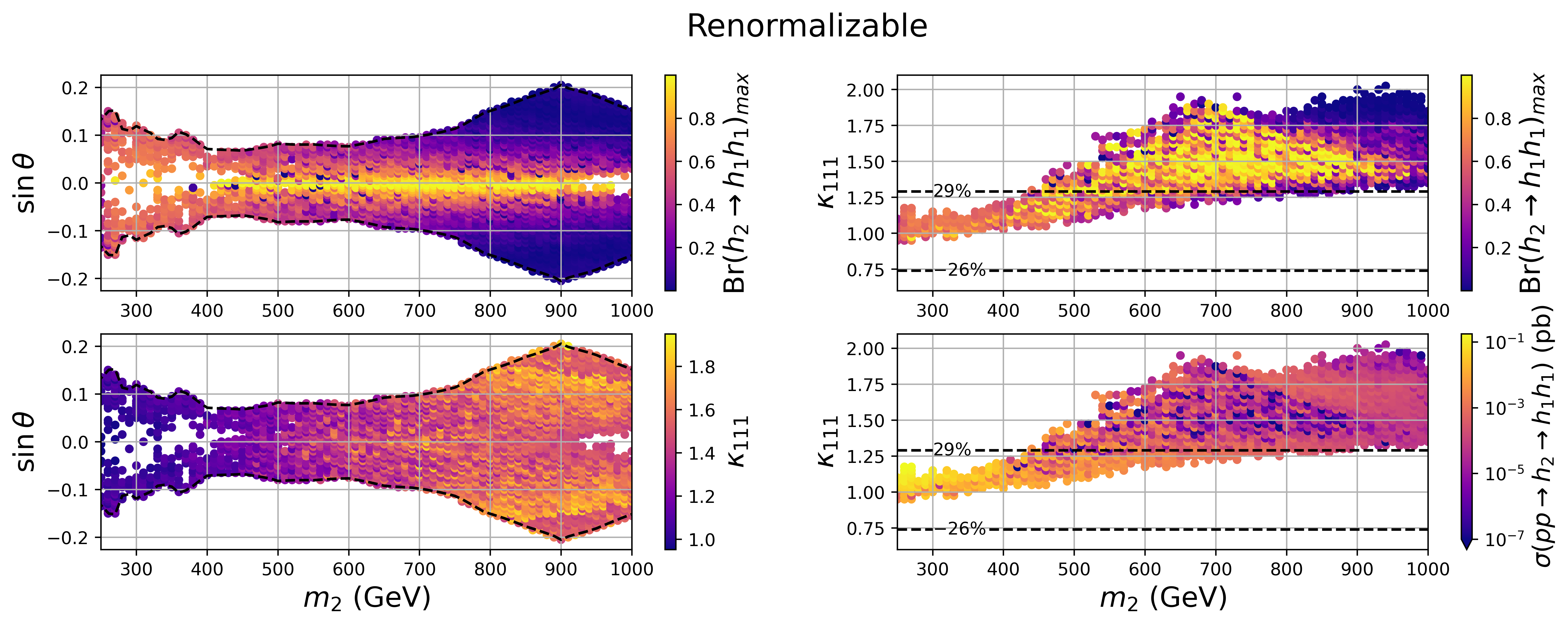}\\
\includegraphics[width=0.99\textwidth]{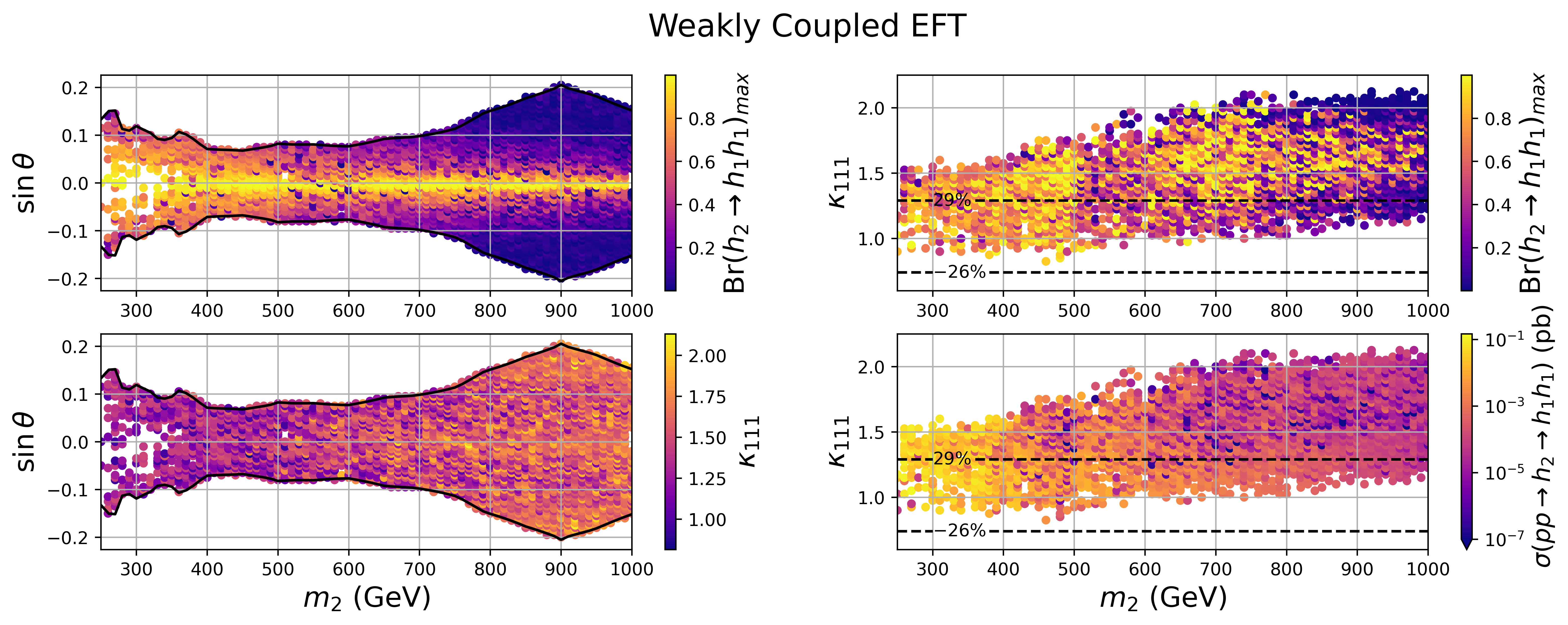}
\end{center}
\caption{\label{fig:collider_EWPT_oneloop}Same as Fig.~\ref{fig:collider_CW} but now requiring a strong first order EWPT. 
}
\end{figure}

Since we focused on the effective operators in the scalar potential, the main effects of the EFT at a collider are to change the maximal $pp\rightarrow h_2\rightarrow h_1h_1$ rate and the variation of the Higgs trilinear coupling 
\begin{eqnarray}
\kappa_{111}=\frac{\lambda_{111}}{\lambda_{111,{\rm SM}}},
\end{eqnarray}
where the subscript SM indicates the prediction within the Standard Model. In Figs.~\ref{fig:collider_tree},~\ref{fig:collider_CW}, and  \ref{fig:collider_EWPT_oneloop}, we compare the maximal ${\rm BR}(h_2\rightarrow h_1h_1)$ in the (left) $\sin\theta-m_2$ plane and (right) $\kappa_{111}-m_2$ plane in the  (top two) renormalizable model and (bottom two) weakly coupled EFT. All free parameters are scanned over. On the left the plots are paired such that when ${\rm BR}(h_2\rightarrow h_1h_1)$ is maximized the corresponding $\lambda_{111}$ variation is shown directly below. Similarly, on the right, we pair the maximal ${\rm BR}(h_2\rightarrow h_1h_1)$ with the corresponding production rate $\sigma_{ggF}(pp\rightarrow h_2\rightarrow h_1h_1)$ in the gluon fusion channel in the narrow width approximation directly below.  The branching ratio of $h_2\rightarrow h_1h_1$ is calculated consistently using $\lambda_{112}^{\rm eff}$ defined in Eq.~(\ref{eq:lam112eff}).

\begin{figure}[ht!]
\begin{center}
\subfigure[] {\includegraphics[width=0.45\textwidth]{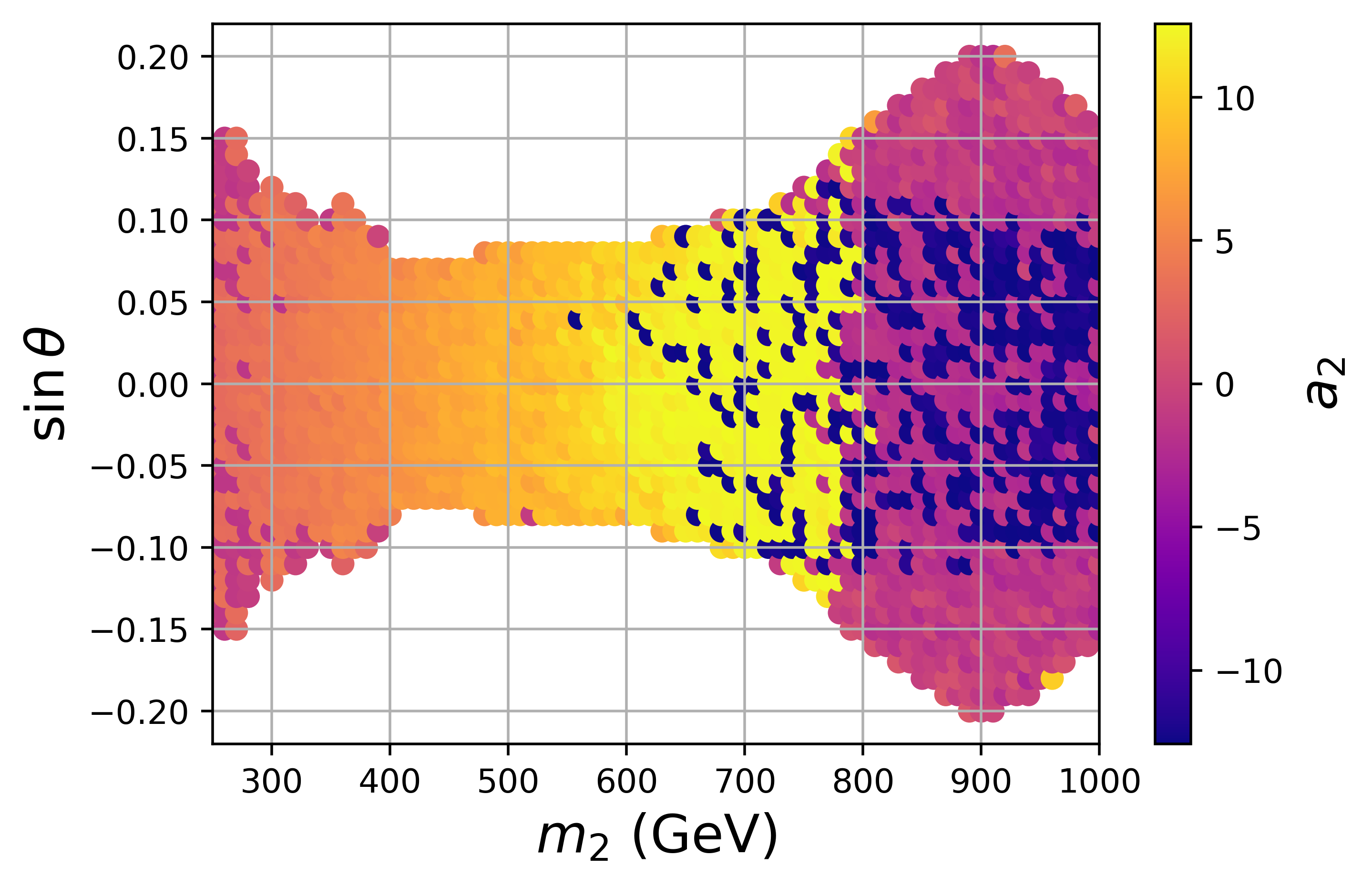}} 
\subfigure[]{\includegraphics[width=0.45\textwidth]{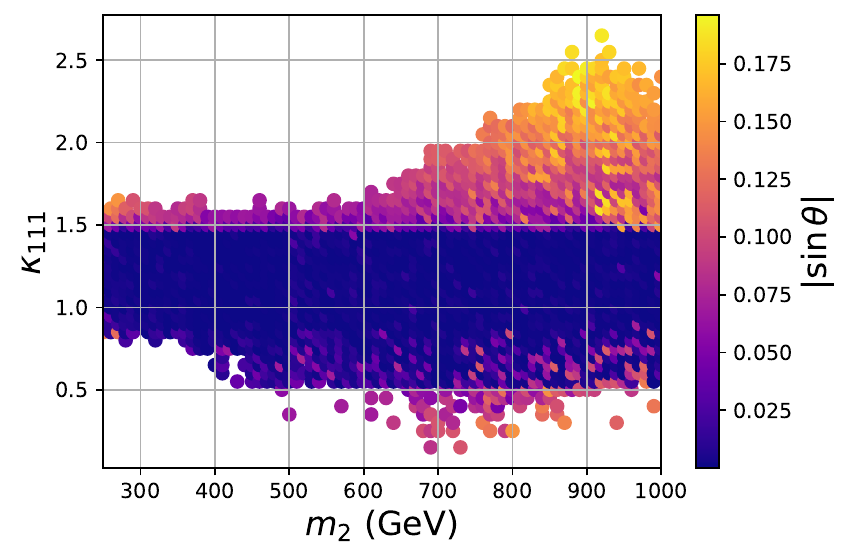}}
\subfigure[]{\includegraphics[width=0.45\textwidth]{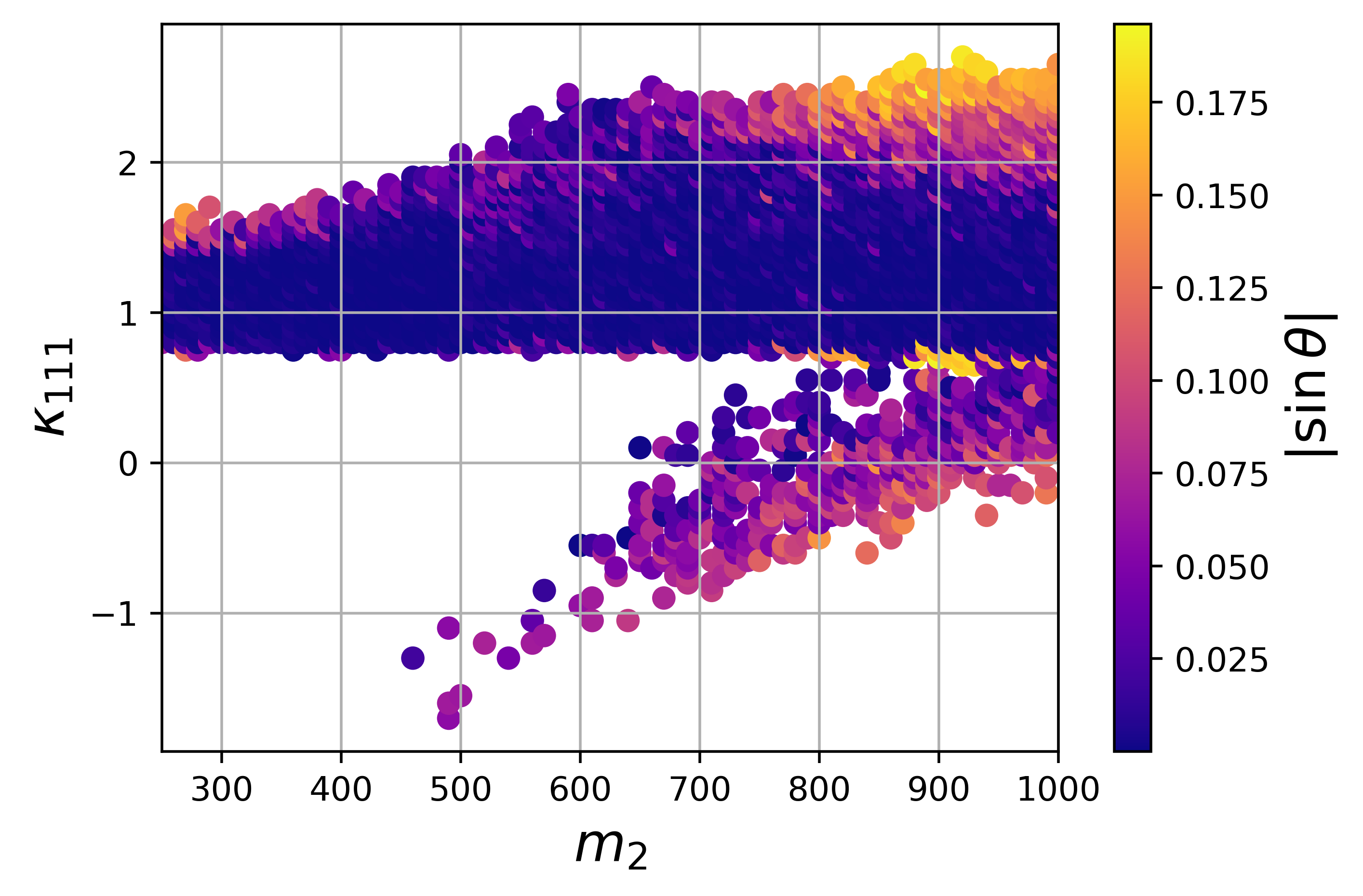}}
\end{center}
\caption{\label{fig:clarification} (a) Heat map of $a_2$ that maximizes ${\rm BR}(h_2\rightarrow h_1h_1)$ at tree level in the renormalizable model [Fig.~\ref{fig:collider_tree}] in the $\sin\theta-m_2$ plane. (b,c) Heat map of $\sin\theta$ that maximizes BR($h_2\rightarrow h_1 h_1$) for the weakly coupled EFT in (b) Fig.~\ref{fig:collider_tree} and (c) Fig.~\ref{fig:collider_CW}. }
\end{figure}

Fig.~\ref{fig:collider_tree} shows these results when the trilinears are evaluated at tree level, while Fig.~\ref{fig:collider_CW} is the same but with the trilinear couplings evaluated at the one-loop level calculated with the zero temperature effective potential. In the renormalizable model, at tree level the maximal ${\rm BR}(h_2\rightarrow h_1h_1)$ can reach as high as $\sim70\%$ over the entire mass range. However, for $m_2\gtrsim 750$~GeV the largest branching ratios occur for $|\sin\theta|\lesssim 0.1$. This can be understood numerically from Fig.~\ref{fig:clarification}(a), which shows the corresponding values of $a_2$. For $m_2\lesssim 750$~GeV, the $h_2\rightarrow h_1h_1$ branching ratio can be enhanced by large positive $a_2$, as seen in Ref.~\cite{Lewis:2024yvj}. For $m_2\gtrsim750$~GeV, the large branching ratios occur for $a_2$ that are large and negative. When boundedness is applied to all field values, as in Eq.~(\ref{eq:BoundRen}), $a_2$ cannot be as large and negative as when the boundedness condition is applied only to field values out to $\Lambda=1$~TeV. Those large and negative $a_2$ only occur for $|\sin\theta|\lesssim0.1$.

Comparing Figs.~\ref{fig:collider_tree} and~\ref{fig:collider_CW} the importance of loop level corrections is apparent. With the loop corrections, the ${\rm BR}(h_2\rightarrow h_1h_1)$ can be $\sim100\%$
across the entire $m_2$ range in the renormalizable model. In particular, as shown in App.~\ref{app:V1_cutoff}, 1-loop corrections to the $h_1-h_1-h_2$ coupling in the renormalizable model persist as $|\sin\theta|\rightarrow 0$:
\begin{eqnarray}
\lambda_{112}=\frac{a_2^2b_3}{16\pi^2}\frac{v_{\rm EW}^2}{m_2^2}+\mathcal{O}\left(\sin\theta,\Lambda^{-1}\right).
\end{eqnarray}
Hence, in the small mixing angle limit the $h_2\rightarrow h_1h_1$ partial width can be enhanced over the $h_2\rightarrow WW,ZZ,tt$ partial widths. This also explains why the resonant di-Higgs rate is smaller at larger ${\rm BR}(h_2\rightarrow h_1h_1)$ at the one-loop trilinear level, since the $h_2$ production rate is suppressed by $\sin^2\theta$.

Loop corrections are similarly important for $\kappa_{111}$. At tree level in the renormalizable model, the value of $\kappa_{111}$ can be as large as $2$ or as small as zero for $m_2\gtrsim 800$~GeV\footnote{Finding large suppression of $\kappa_{111}$ would seem to contradict the results in Ref.~\cite{Lewis:2024yvj}. However, Ref.~\cite{Lewis:2024yvj} applied the boundedness and global minimum constraints at tree level for all field values, limiting how large and negative $a_2$ can be.}.  With the loop corrections, $\kappa_{111}$ can be larger than $2$ for $m_2\gtrsim 550$~GeV.  Additionally, the loop corrections open up a region of parameter space where $\kappa_{111}$ can be substantially lower than $1$, even negative. That is, in the renormalizable model loop corrections show that there is a substantially larger region of parameter space with larger $\kappa_{111}$ variations than indicated by a tree level analysis. These results are similar to those found in Refs.~\cite{Kanemura:2002vm,Bahl:2023eau,Arco:2025pgx,Braathen:2026jfg,Kanemura:2016lkz}.

When the non-renormalizable operators are turned on, these results change considerably. In the weakly coupled EFT, the tree and one-loop level $h_2\rightarrow h_1h_1$ branching ratio can reach as high as $\sim 100\%$ across the entire mass range. In the small mixing angle limit, the partial widths $h_2\rightarrow WW,ZZ,tt$ are suppressed, while the Wilson coefficient $a_4$ can enhance  the partial width of $h_2\rightarrow h_1h_1$ at tree level. This can be seen clearly via the $h_1-h_1-h_2$ coupling in Eq.~(\ref{eq:lam112}). The role of effective operators in enhancing ${\rm BR}(h_2\rightarrow h_1h_1)$ and other multi-Higgs resonant channels has recently gained attention in the literature~\cite{Roy:2026kud,Li:2026kqk}, and we show the importance of these effective operators at one loop as well. 

In the weakly coupled EFT, $\kappa_{111}$ can generically be further from the SM predictions for both the tree and loop level, with larger variations at loop level. Unlike the renormalizable model, the tree level $\kappa_{111}$ can be as large as $1.5-1.6$ across the entire mass range, and can be lower than $0.5$ for $m_2\gtrsim 400$~GeV.
A striking feature of the weakly coupled EFT at tree level is that for $\kappa_{111}\lesssim1.5$ the branching ratio of $h_2\rightarrow h_1h_1$ is $\sim100\%$ across the entire parameter space, while for $\kappa_{111}\gtrsim1.5$ the $h_2\rightarrow h_1h_1$ branching ratio is  smaller. This can be understood from Fig.~\ref{fig:clarification}(b). For $0.5\lesssim \kappa_{111}\lesssim 1.5$, the $\sin\theta$ that maximizes ${\rm BR}(h_2\rightarrow h_1h_1)$ is very small. In this limit, the ratio of the Higgs trilinear at tree level is
\begin{eqnarray}
\kappa_{111}=1+2\frac{C_Hv_{\rm EW}^4}{\Lambda^2m_{1}^2}+\mathcal{O}(\sin\theta,\Lambda^{-3}).
\end{eqnarray}
Hence, $0.5\lesssim \kappa_{111}\lesssim 1.5$ for $-1<C_H<1$, small $\sin\theta$, and $\Lambda=1$~TeV. For $\kappa_{111}$ to be outside of that range, a larger $|\sin\theta|$ is needed, resulting in a smaller $h_2\rightarrow h_1h_1$ branching ratio. 
 However, the cross sections at maximum branching ratio are enhanced for $\kappa_{111}\gtrsim 1.5$ due to a larger mixing angle. As shown in Fig.~\ref{fig:clarification}(c), at one loop $\kappa_{111}$ can be larger with smaller $|\sin\theta|$. Hence, the range of $\kappa_{111}$ with larger $h_2\rightarrow h_1h_1$ is increased.  However, again, there is an anti-correlation between maximum branching ratios and corresponding rates due to large branching ratios occurring mostly at small mixing angle. 

The HL-LHC is expected to measure the Higgs trilinear coupling to within $-26\%$ and $+29\%$ of the SM expectations~\cite{ATLAS:2025eii}. Currently, the LHC places a bound of $-0.71\leq\kappa_{111}\leq6.1$~\cite{CMS:2026nuu}. In Figs.~\ref{fig:collider_tree} and~\ref{fig:collider_CW}, HL-LHC projected limits are shown as dashed black lines and current LHC sensitivities as solid black lines. As is clear, loop corrections to the Higgs trilinear are necessary to fully determine the reach of the LHC. Current LHC limits on $\kappa_{111}$ already rule out some of the weakly coupled EFT space. The HL-LHC will be sensitive to the renormalizable model and weakly coupled EFT.

\begin{figure}[tb]
\begin{center}
\subfigure[]{\includegraphics[width=0.45\textwidth,clip]{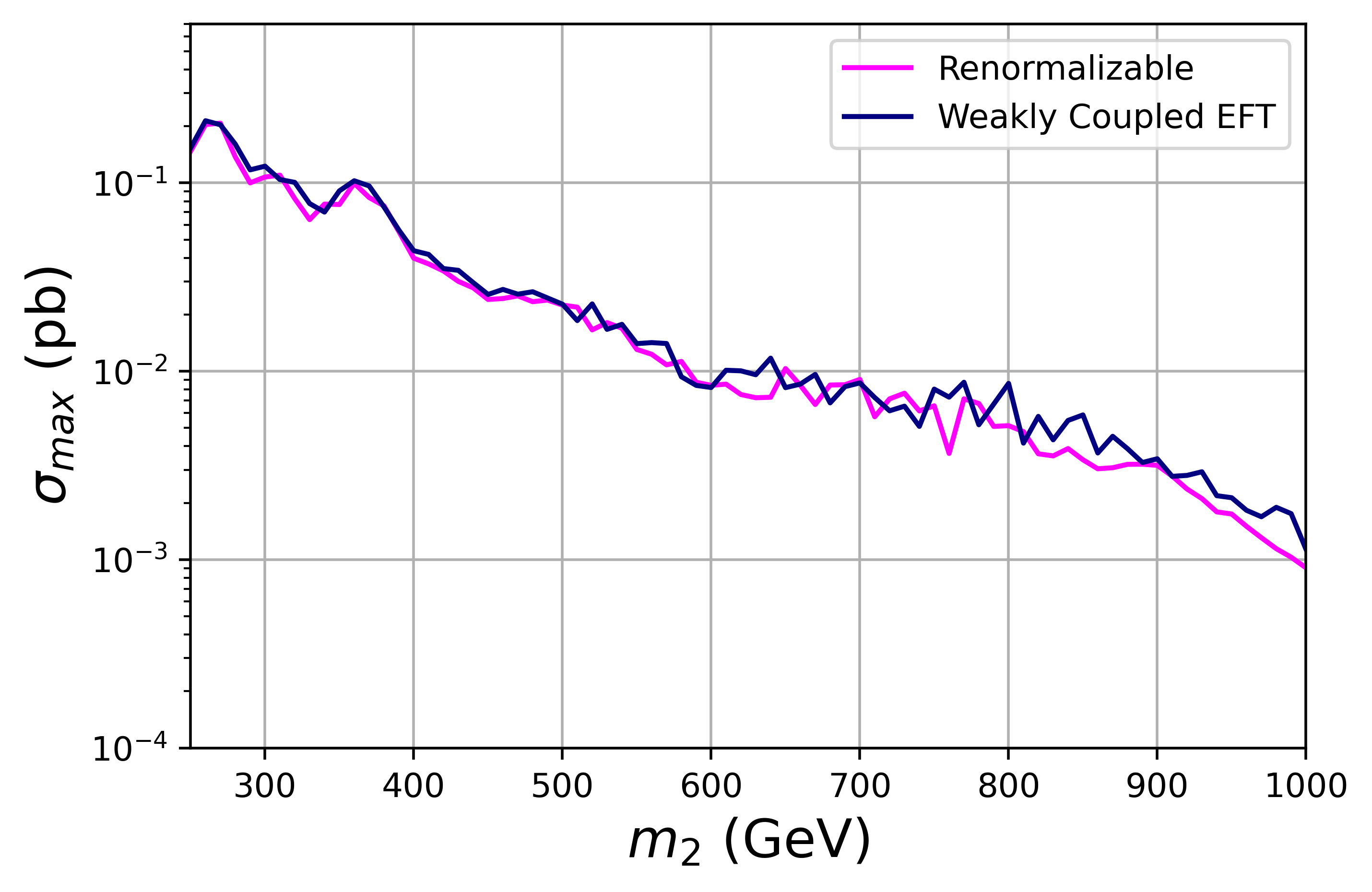}}
\subfigure[]{\includegraphics[width=0.45\textwidth,clip]{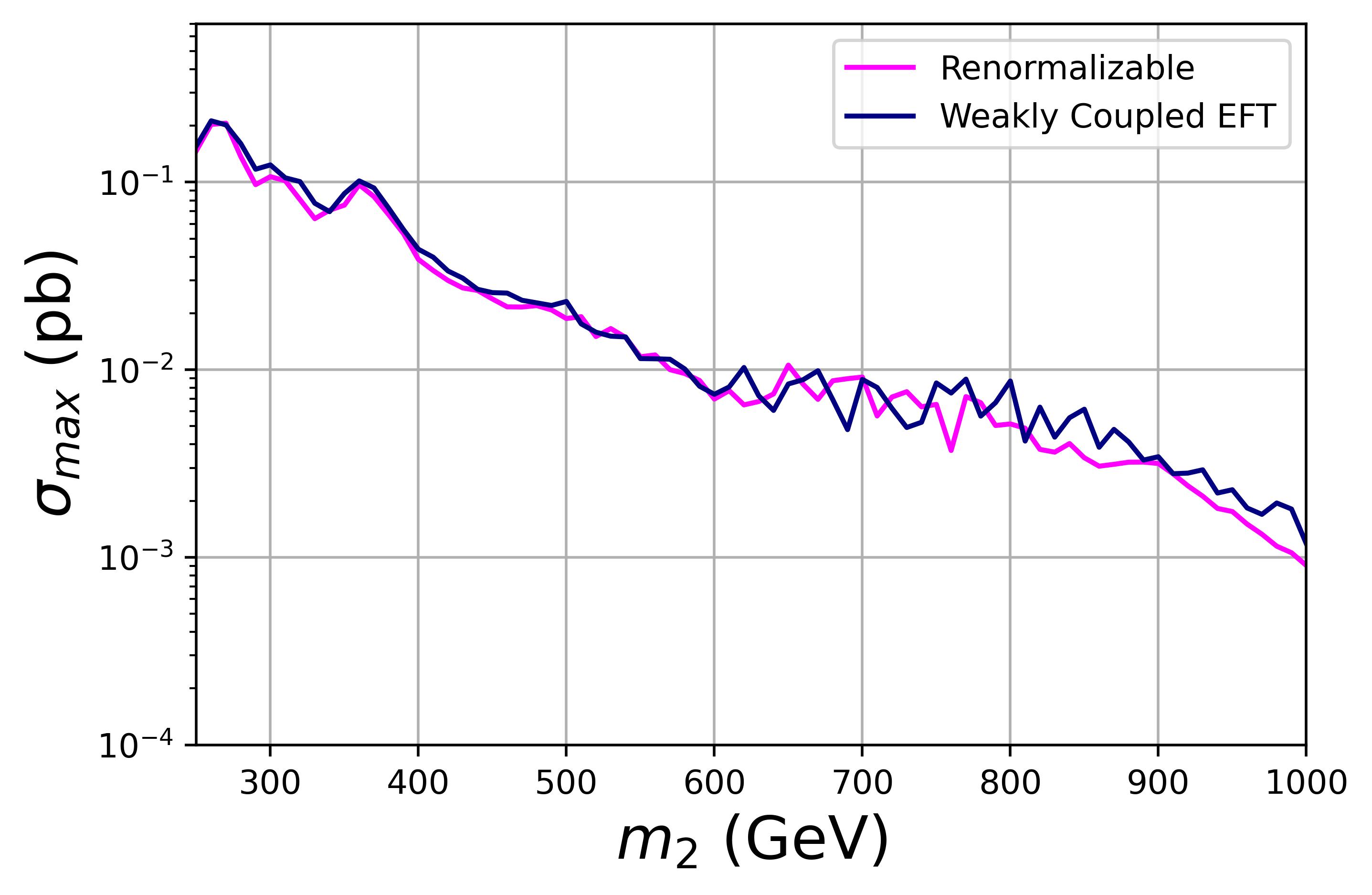}}
\subfigure[]{\includegraphics[width=0.45\textwidth,clip]{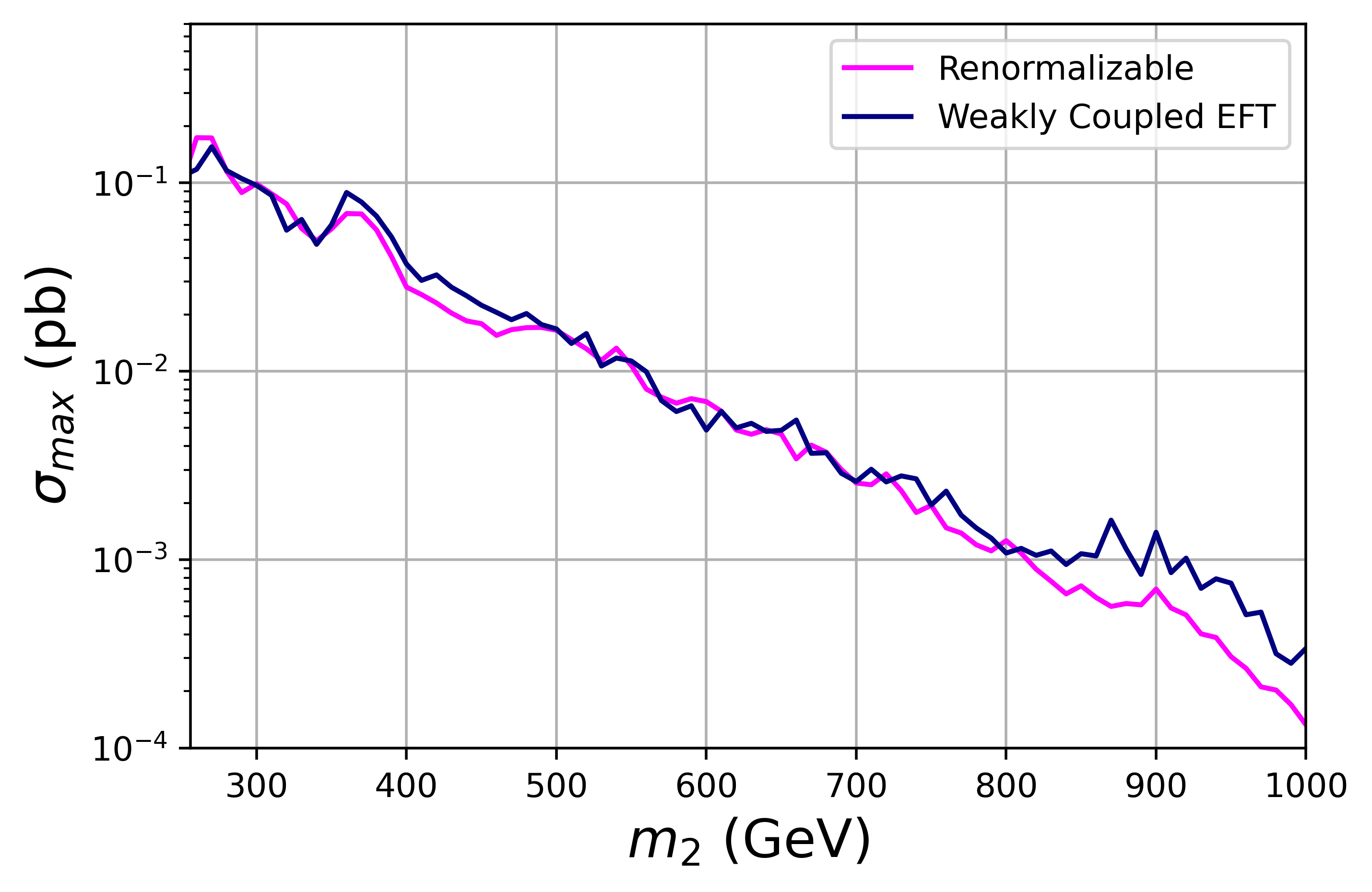}}
\end{center}
\caption{\label{fig:xsect}Maximum allowed rate $\sigma(pp\rightarrow h_2\rightarrow h_1h_1)$ vs. $m_2$ for (a) trilinears evaluated at tree level, (b) trilinears defined with Coleman-Weinberg potential, and (c) after applying EWPT constraints for (magenta) renormalizable and (navy) weakly coupled EFT. All constraints, including current bounds on the Higgs trilinear, are applied.}
\end{figure}

Fig.~\ref{fig:collider_EWPT_oneloop} is the same as Fig.~\ref{fig:collider_CW} but now requiring a strong first order EW phase transition. 
For all models, there are fewer points that have the branching ratio of  $h_2\rightarrow h_1h_1$ above $90\%$, and the variation of $\kappa_{111}$ is smaller.  At the renormalizable level, requiring that there be a SFOEWPT forces the Higgs trilinear to be larger than the SM prediction for $m_2\gtrsim 400$~GeV. From Fig.~\ref{fig:EWPT}(a), a SFOEWPT is driven by an $a_2$ that is greater than zero. As shown in App.~\ref{app:V1_cutoff}, in the small $\sin\theta$ limit, the loop corrected Higgs trilinear in the renormalizable model is~\cite{Curtin:2014jma}\footnote{The full one-loop $\lambda_{111}$ including EFT operators in the $\sin\theta\rightarrow 0$ limit can be found in App.~\ref{app:V1_cutoff}.}:
\begin{eqnarray}
\kappa_{111}=1+\frac{a_2^3}{32\pi^2}\frac{v_{\rm EW}^3}{\lambda_{111,{\rm SM}}m_2^2}+\mathcal{O}\left(\sin\theta,\Lambda^{-1}\right),
\end{eqnarray}
where $\lambda_{111,{\rm SM}}=0.71\,v_{\rm EW}$ at one loop. Hence, positive $a_2$ raises $\kappa_{111}$ above one. As the mass $m_2$ increases, the lower bound on $a_2$ also increases, which in turn increases the one-loop correction to $\lambda_{111}$. Indeed, we find that the HL-LHC measurements of the Higgs trilinear will be sensitive to the majority of parameter space of the renormalizable model consistent with a SFOEWPT for $m_2\gtrsim 450$~GeV. Similar results show that a SFOEWPT requires a non-SM Higgs trilinear~\cite{Curtin:2014jma,Braathen:2026wib,Kanemura:2004ch,Reichert:2017puo,Goncalves:2021egx,Dorsch:2017nza,Basler:2017uxn,Biekotter:2022kgf,Braathen:2025qxf} for heavier scalars $\gtrsim 400-500$~GeV.

The situation in the EFTs is different. In the weakly coupled EFT, after requiring a SFOEWPT, the current LHC bounds on $\kappa_{111}$ no longer rule out any parameter points.  While the HL-LHC is sensitive to a large portion of the parameter space in the weakly coupled EFT, there are points consistent with a strong first order EWPT that will survive across the entire $h_2$ mass range, in contrast to the renormalizable model.  It should be noted that as $m_2$ approaches $1$~TeV, it is approaching the cutoff and the validity comes into question. However, even for $m_2$ far below the cutoff scale, the weakly coupled EFT still has many more points that would evade HL-LHC Higgs trilinear measurements than would survive in the renormalizable model.

Finally, in Fig.~\ref{fig:xsect} we show the maximum allowed di-Higgs rate for (a) trilinears evaluated at tree level, (b) trilinears evaluated via the one-loop Coleman-Weinberg potential, and (c) requiring a strong first order EWPT. We only show points that pass current LHC bounds on the Higgs trilinear coupling. In all three cases (a,b,c), the maximum resonant di-Higgs cross section is similar in the renormalizable and weakly coupled EFT. This is true when effective operators are added only to the potential and the largest ${\rm BR}(h_2\rightarrow h_1h_1)$ occur for small $\sin\theta$. However, if new physics or effective operators alter the $h_2$ production mechanism, the resonant di-Higgs rate can be enhanced~\cite{Roy:2026kud}. Additionally, the maximum rates for the trilinear evaluated at tree or one loop level are similar. Once a SFOEWPT is required, the maximum rate for the renormalizable and weakly coupled EFT is decreased. Even though the di-Higgs rate is decreased, as we showed earlier, measurements of the Higgs trilinear are still promising at large masses.

\subsection{Gravitational Waves}

A SFOEWPT offers a direct cosmological probe of physics beyond the SM through the production of a stochastic GW background. The transition proceeds by the nucleation of bubbles of the EW symmetry breaking phase, followed by their expansion in the primordial plasma. The released vacuum energy is converted into collective plasma motion, including sound waves and turbulence, as well as bubble wall collisions, which act as sources of gravitational radiation. For transitions occurring near the EW scale, the peak frequency of the resulting spectrum typically lies in the millihertz band, allowing space based interferometers to probe this scenario. In this work, we compute the GW signal associated with the SFOEWPT in the singlet-extended SM with higher-dimensional effective operators. In addition to the theoretical and experimental constraints discussed previously, in this section we also apply the current LHC bounds on the Higgs trilinear.

\begin{figure}[tb]
\begin{center}
\subfigure[]{\includegraphics[width=0.45\textwidth,clip]{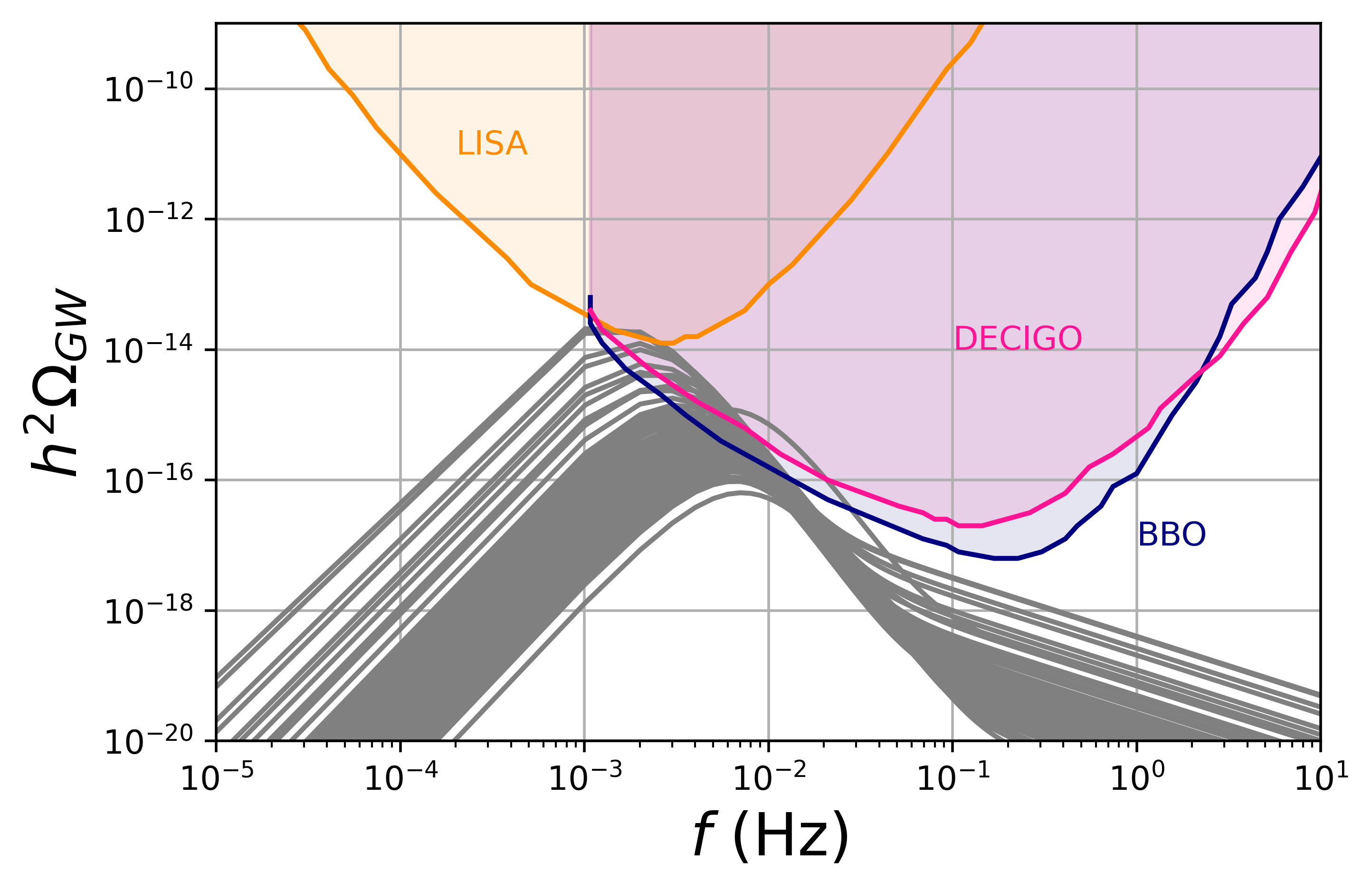}}
\subfigure[]{\includegraphics[width=0.45\textwidth,clip]{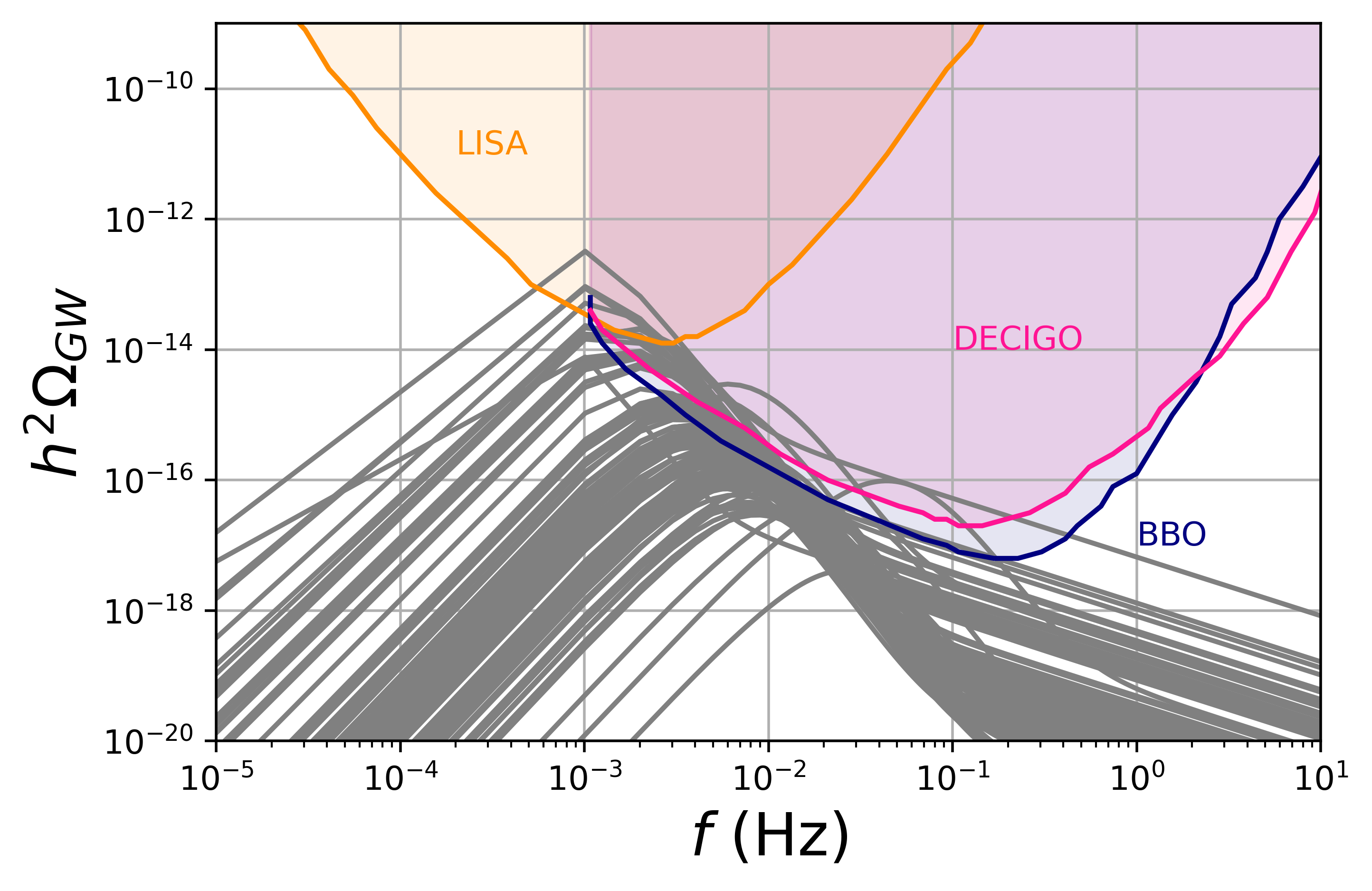}}
\subfigure[]{\includegraphics[width=0.45\textwidth,clip]{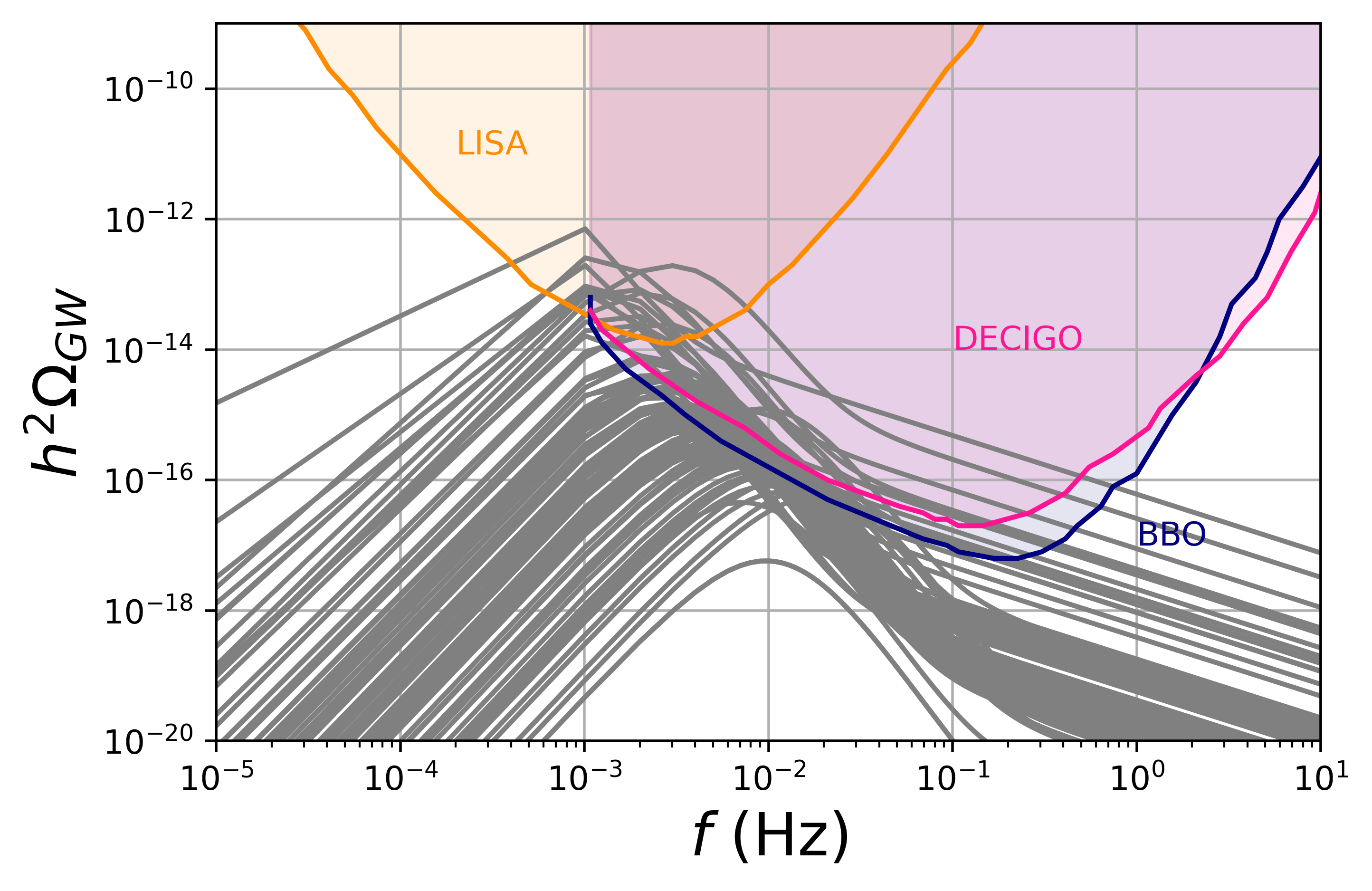}}
\end{center}
\caption{\label{fig:grav_wave}
GW spectrum sourced by bubble collisions and sound waves for 100 randomly selected points with $\sqrt{\Delta V_{\text{eff}}/\alpha\rho_r} \,>\, v_J$ for (a) the renormalizable model, (b) weakly coupled EFT, and (c) the strongly coupled EFT. The sensitivity curves are taken from ~\cite{Schmitz:2020syl} for BBO~\cite{Cutler:2005qq}, DECIGO~\cite{Kawamura:2020pcg}, and LISA~\cite{LISA:2024hlh}.
}
\end{figure}

The GW spectrum is primarily characterized by four macroscopic parameters: the transition strength $\alpha$, the inverse duration of the phase transition normalized by the Hubble rate, $\beta/H$, the bubble wall velocity $v_w$, and the characteristic transition temperature $T_*$, which we set to the nucleation temperature $T_n$. The parameter $\alpha$ characterizes the amount of energy released during the transition relative to the radiation energy density.
\begin{eqnarray}
   \alpha = \frac{\displaystyle \Delta \left(- V_{\text{eff}} + \frac{T_n}{4} \frac{dV_{ \text{eff} }}{dT}\right)}{\rho_r}
\end{eqnarray}
where
\begin{eqnarray}
    \rho_r = \frac{\pi^2}{30}g_* T_n^4 
\end{eqnarray}
is the radiation energy density and $\Delta$ denotes taking the difference between the true and false vacuum. Our scans result in $ \alpha \lesssim 10^{-1}$ for the renormalizable model and weakly coupled EFT, while for the strongly coupled EFT we find $\alpha \lesssim3\times10^{-1}.$ We calculate the inverse duration of the phase transition as
\begin{eqnarray}
\frac{\beta}{H} &=& T_n \frac{d }{dT} \left(\frac{S_3}T \right) \Big|_{T=T_n}
\end{eqnarray}
and find  $10^{2} \lesssim \beta/H \lesssim 6\times10^{4} $.
Determining the wall velocity is an involved calculation that requires solving the coupled scalar field equations of motion and fluid equations for the plasma including the effects of hydrodynamic backreaction and, in general, deviations from local thermal equilibrium~\cite{Carena:2025flp,vandeVis:2025plm,Ekstedt:2024fyq}. 
A full calculation is beyond the scope of the present analysis. We instead adopt the simple analytic approximation proposed in Ref.~\cite{Lewicki:2021pgr}
\begin{eqnarray}
v_{w} = \begin{cases} \displaystyle
\sqrt{\frac{\Delta V_{\rm eff}}{\alpha \rho_r}} & \displaystyle\text{for} \,\,\,\, \sqrt{\frac{\Delta V_{\rm eff}}{\alpha \rho_r}} < v_J \\
1 & \displaystyle\text{for} \,\,\,\, \sqrt{\frac{\Delta V_{\rm eff}}{\alpha \rho_r}} \geq v_J
\end{cases}
\end{eqnarray}
where $\Delta V \equiv V_{\rm false}-V_{\rm true}$ denotes the free energy difference between the false and true vacua. The Jouguet velocity is given by
\begin{eqnarray}
    v_J = \displaystyle\frac{1}{\sqrt{3}} \frac{1 + \sqrt{3 \alpha^2 + 2 \alpha}}{1+\alpha}.
\end{eqnarray}
For the sub-Jouguet branch, the expression of $v_w$ follows from the steady state momentum flux matching condition in the local thermodynamic equilibrium approximation. It provides a useful estimate of the wall velocity when deviations from thermal equilibrium are small. For transitions sufficiently strong that this estimate exceeds $v_J$, we follow Ref.~\cite{Lewicki:2021pgr} and take the wall to be highly relativistic, setting $v_w=1$.

For parameter points with $\sqrt{\frac{\Delta V_{\rm eff}}{\alpha\rho_r}} > v_J$ we calculate the contributions to the GW spectrum from sound waves and bubble collisions. We use \texttt{PTPlot}~\cite{Caprini:2015zlo,Caprini:2019egz} to calculate the contributions from sound waves to the GW spectrum. App.~\ref{app:gw} contains the details of the GW calculations for both sound waves and bubble collisions. We show the spectrum in Fig.~\ref{fig:grav_wave} for a subset of 100 randomly selected points. The peak is set by sound waves
and lies at $f\sim10^{-3}$--$10^{-2}$~Hz. In the renormalizable model and the weakly coupled EFT
the peak amplitudes are $h^2\Omega_{\rm GW}\lesssim10^{-13}$, at or below the LISA
sensitivity, with a handful of points within reach of DECIGO and BBO. In the strongly
coupled EFT, where $\alpha$ reaches $0.3$, the amplitudes are larger and a few points
exceed the LISA sensitivity. The peak of the spectrum is dominated by the sound waves, while the change in the power spectrum at higher frequencies is due to the dominance of bubble wall collisions at those frequencies. 
\begin{figure}[tb]
\begin{center}
\subfigure[]{\includegraphics[width=0.45\textwidth,clip]{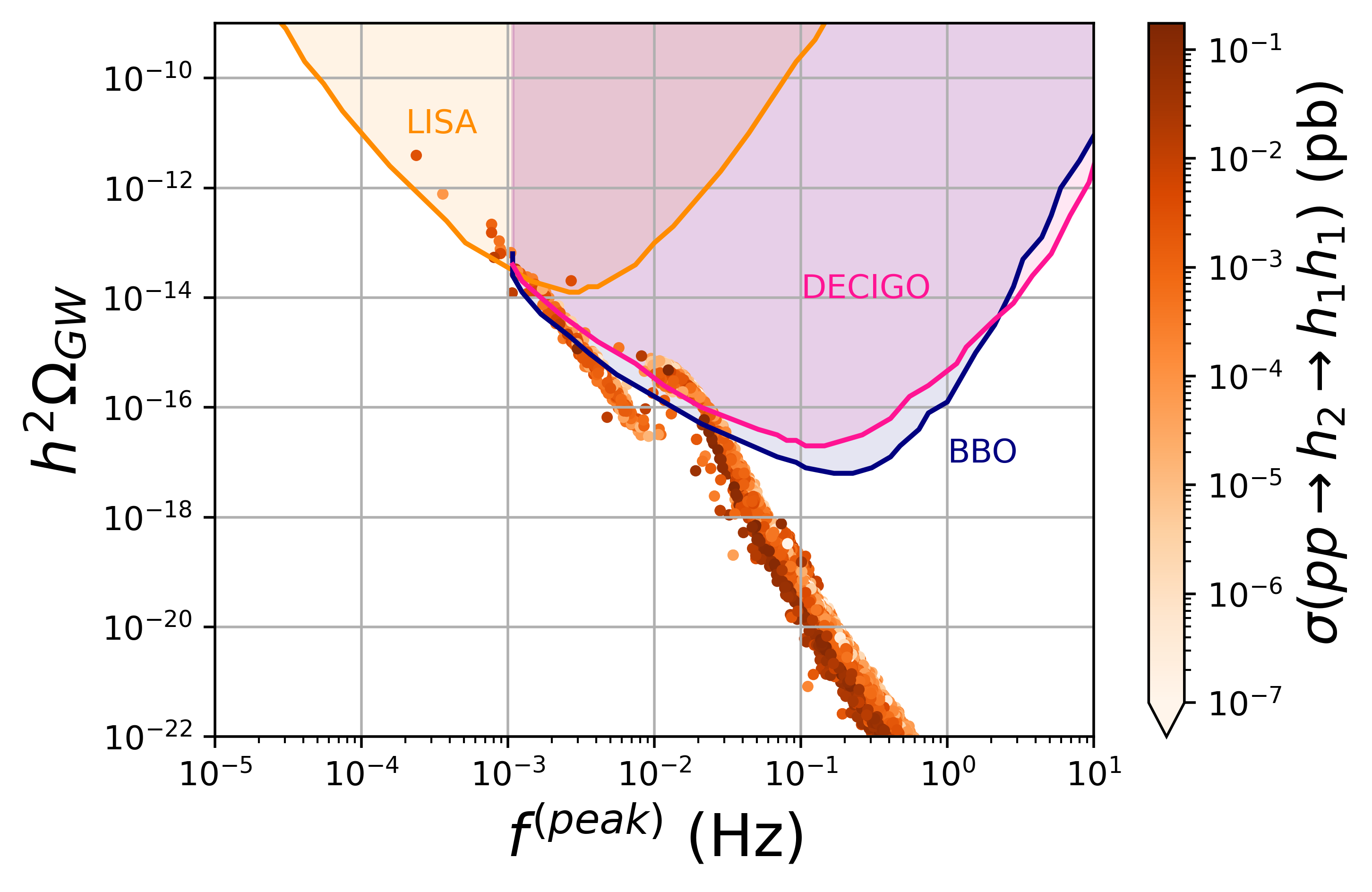}}
\subfigure[]{\includegraphics[width=0.45\textwidth,clip]{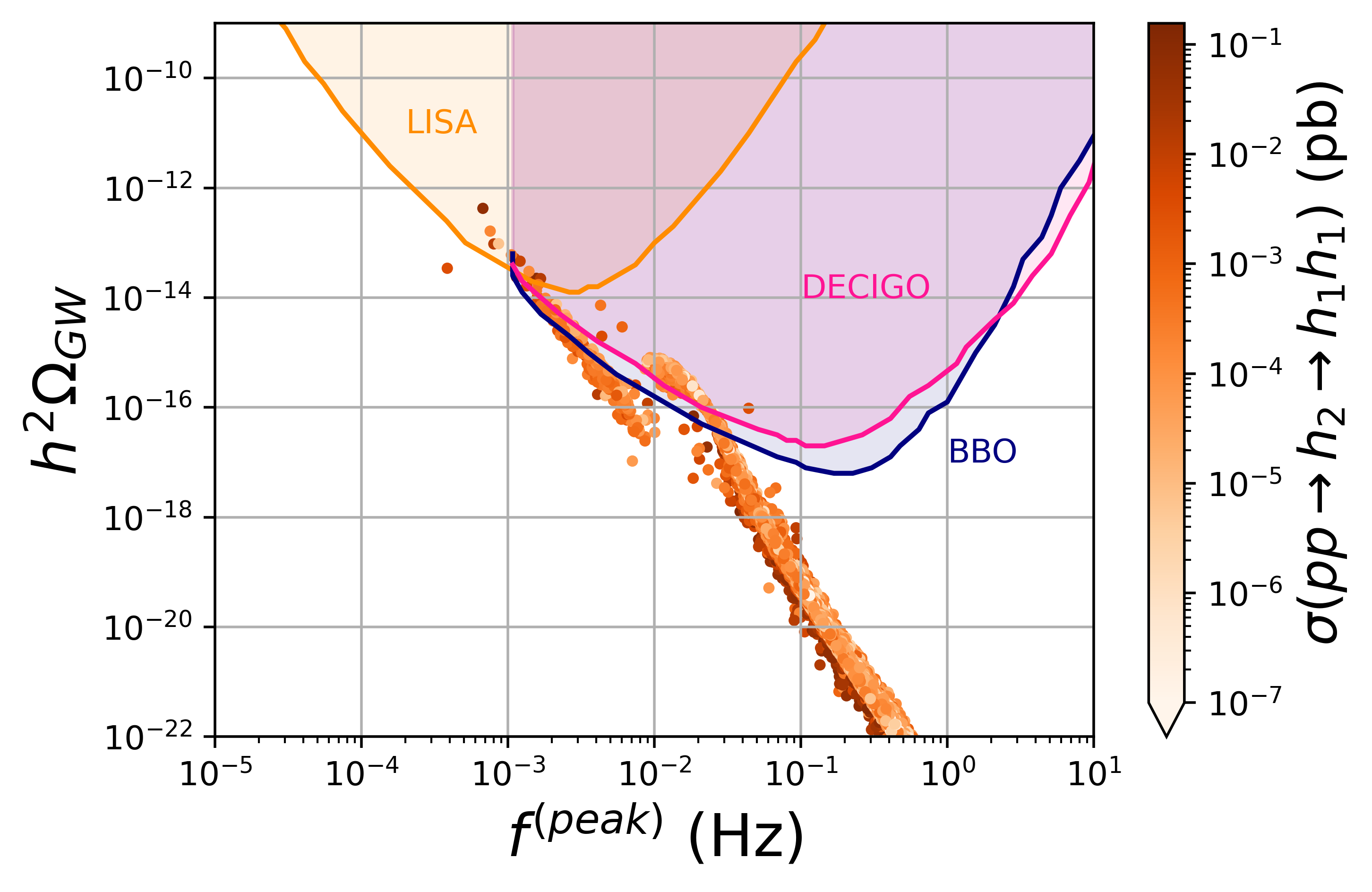}}
\subfigure[]{\includegraphics[width=0.45\textwidth,clip]{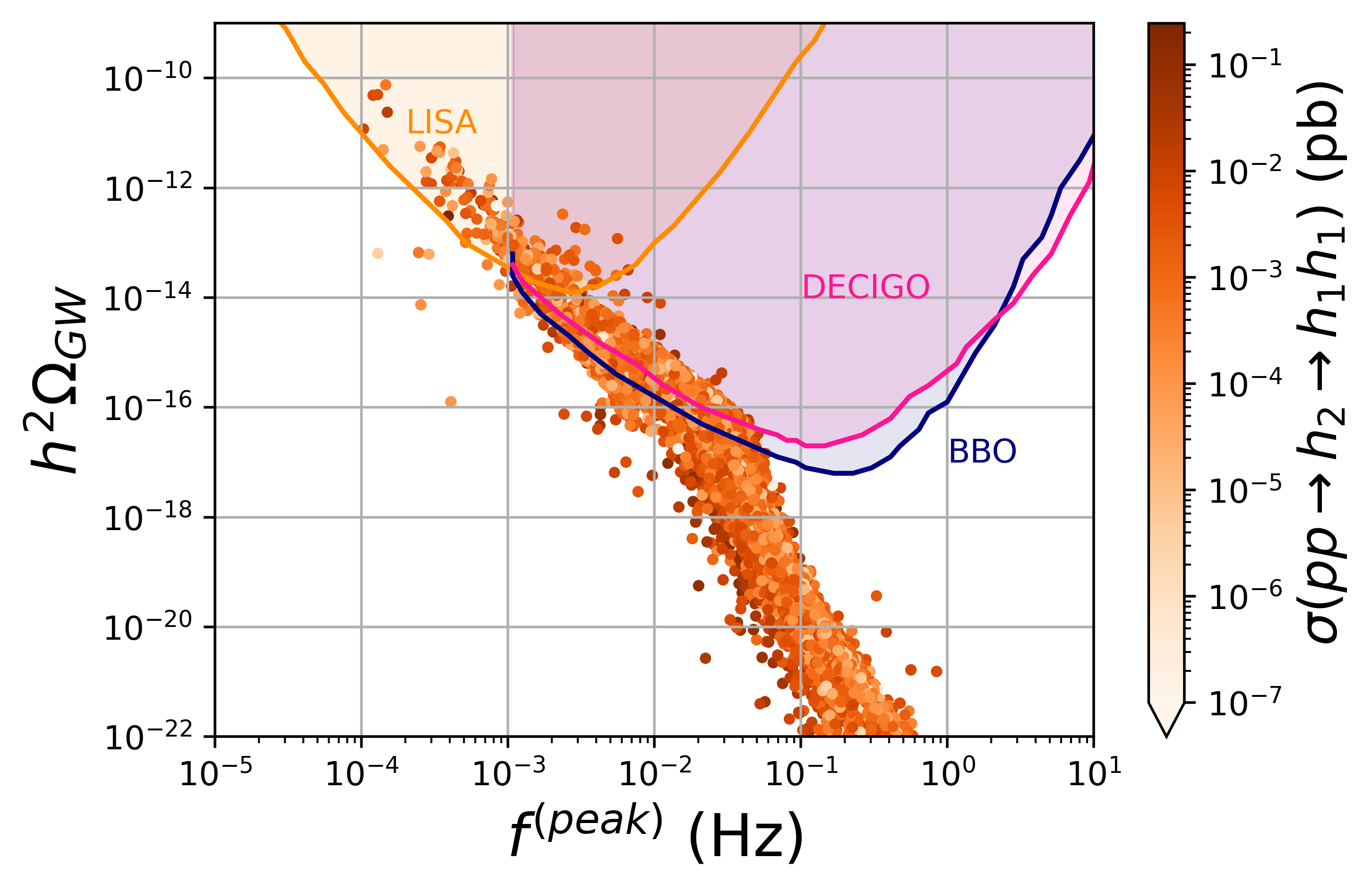}}
\end{center}
\caption{\label{fig:grav_wave2}
Peak amplitudes versus frequency for GW sourced by sound waves for (a) the renormalizable model (b) weakly coupled EFT, and (c) strongly coupled EFT. The color of each point shows the corresponding $\sigma(pp\rightarrow h_2\rightarrow h_1 h_1)$.}
\end{figure}

\begin{figure}[tb]
\begin{center}
\subfigure[]{\includegraphics[width=0.45\textwidth,clip]{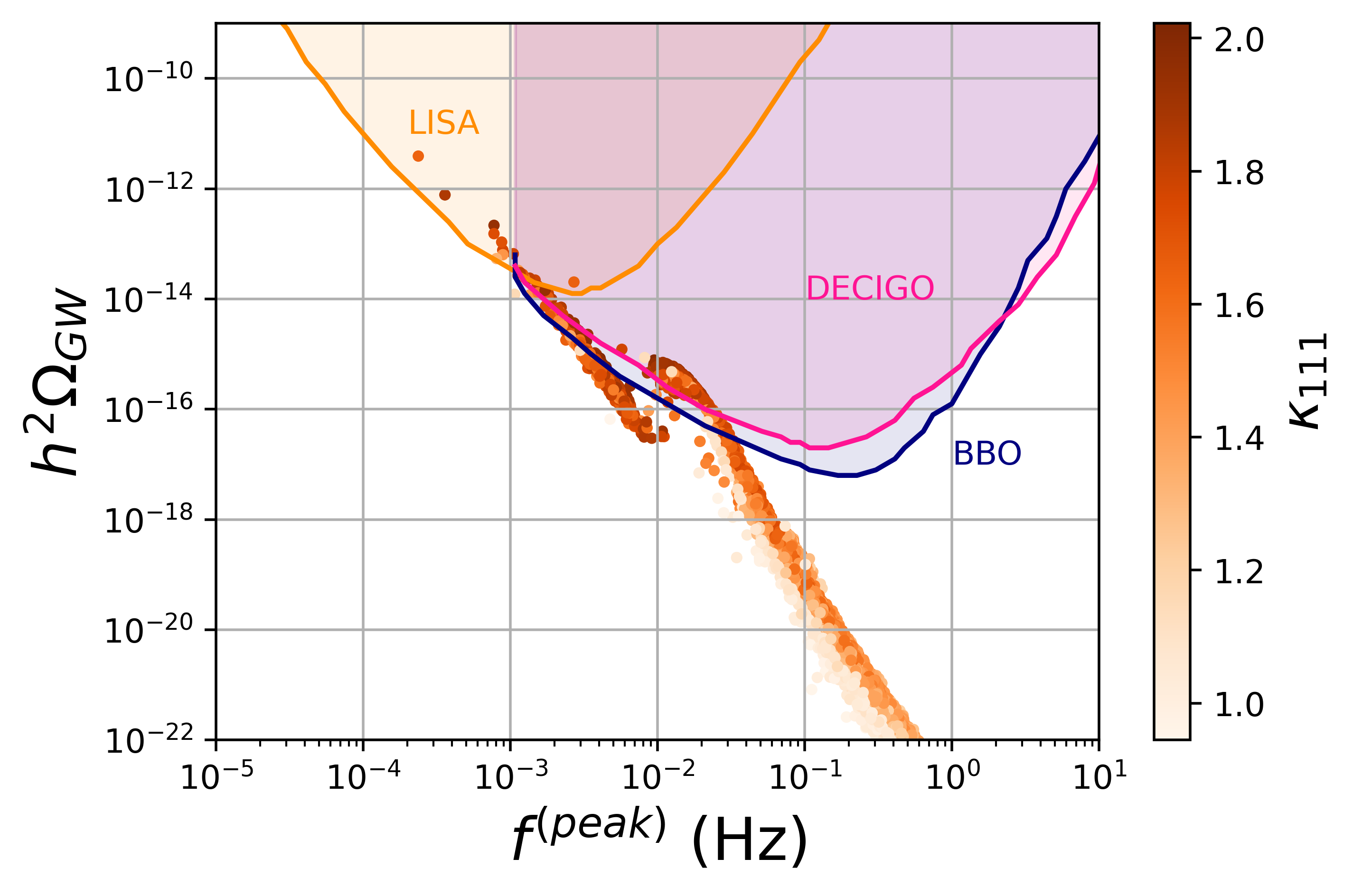}}
\subfigure[]{\includegraphics[width=0.45\textwidth,clip]{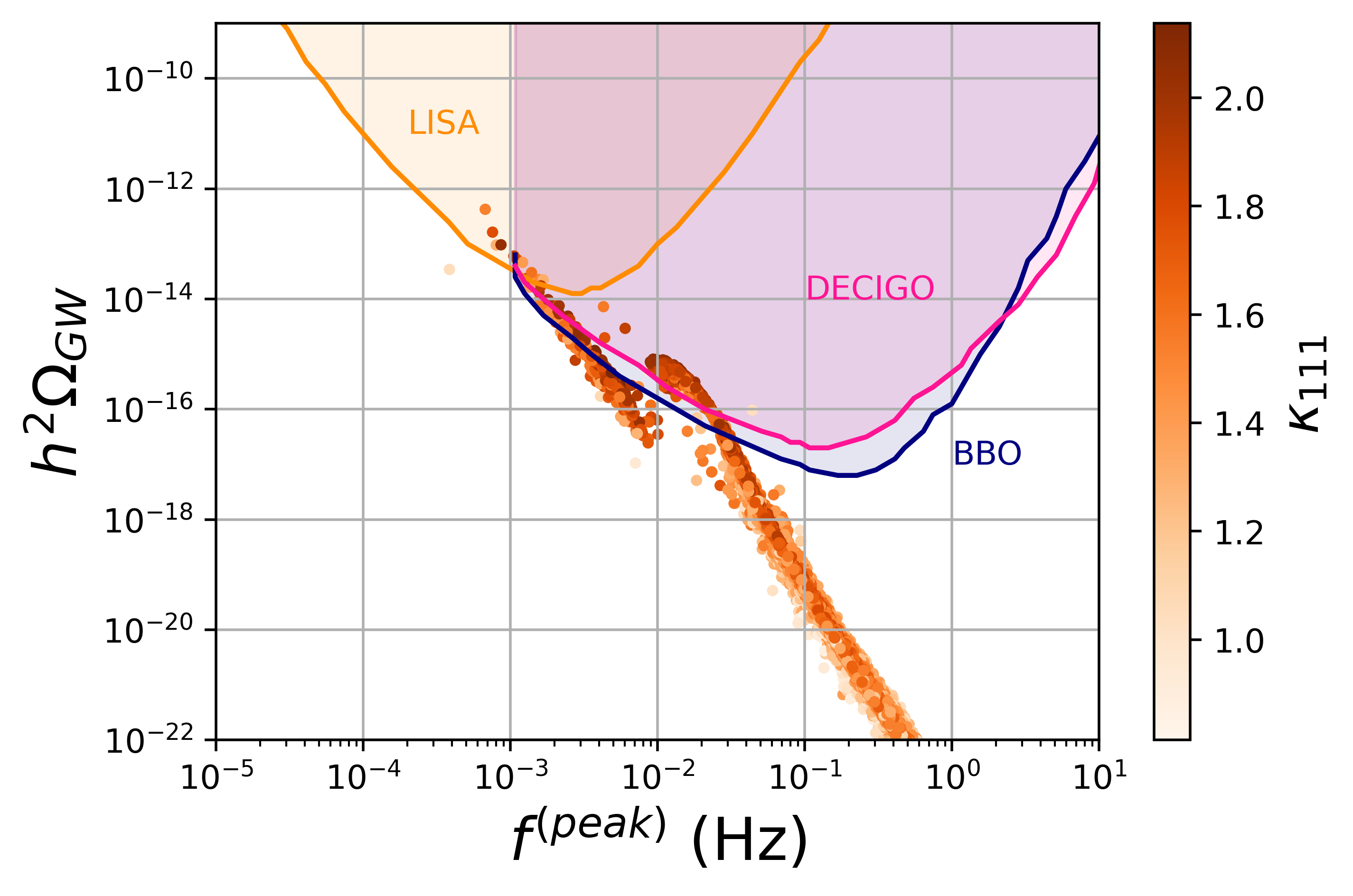}}
\subfigure[]{\includegraphics[width=0.45\textwidth,clip]{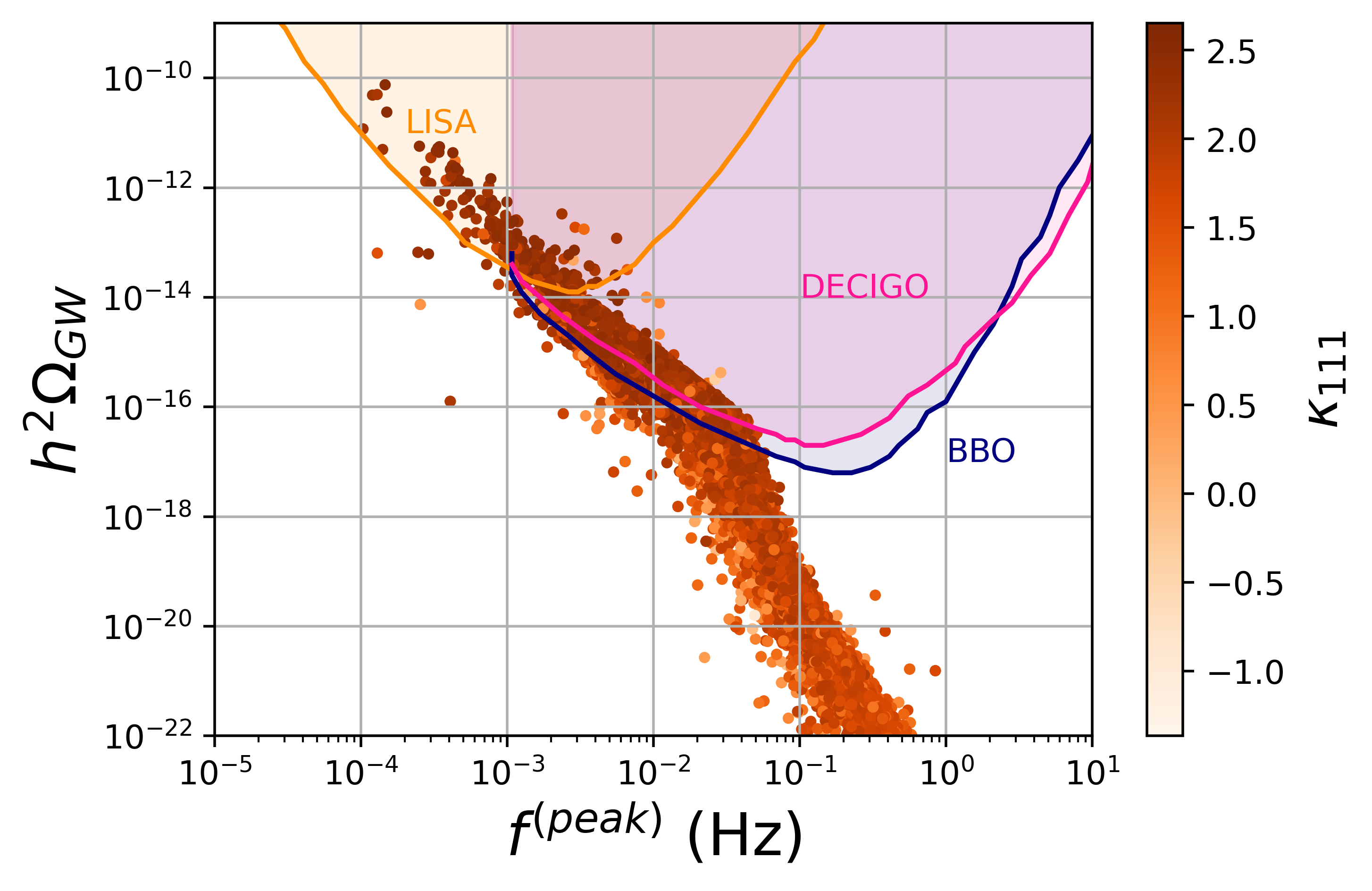}}
\end{center}
\caption{\label{fig:grav_wave3}
Peak amplitudes versus frequency for GW sourced by sound waves for (a) the renormalizable model (b) weakly coupled EFT, and (c) strongly coupled EFT. The color of each point shows the corresponding value of $\kappa_{111}$.}
\end{figure}

In Figs.~\ref{fig:grav_wave2} and~\ref{fig:grav_wave3} we show the peak frequency of the GW spectrum as a heat map in the resonant di-Higgs cross section and $\kappa_{111}$, respectively. This is shown for the (a) renormalizable model, (b) weakly interacting EFT, and (c) strongly interacting EFT. The prospects for GW observation are similar in the renormalizable and weakly coupled EFT, while many more points in the strongly coupled EFT have detection prospects at GW detectors. This shows that to have a significant change in GW prospects away from that in the renormalizable model, either there should be a strongly interacting theory at a TeV or new physics below a TeV beyond that in the real scalar singlet model.

\begin{figure}[tb]
\begin{center}
\subfigure[]{\includegraphics[width=0.45\textwidth,clip]{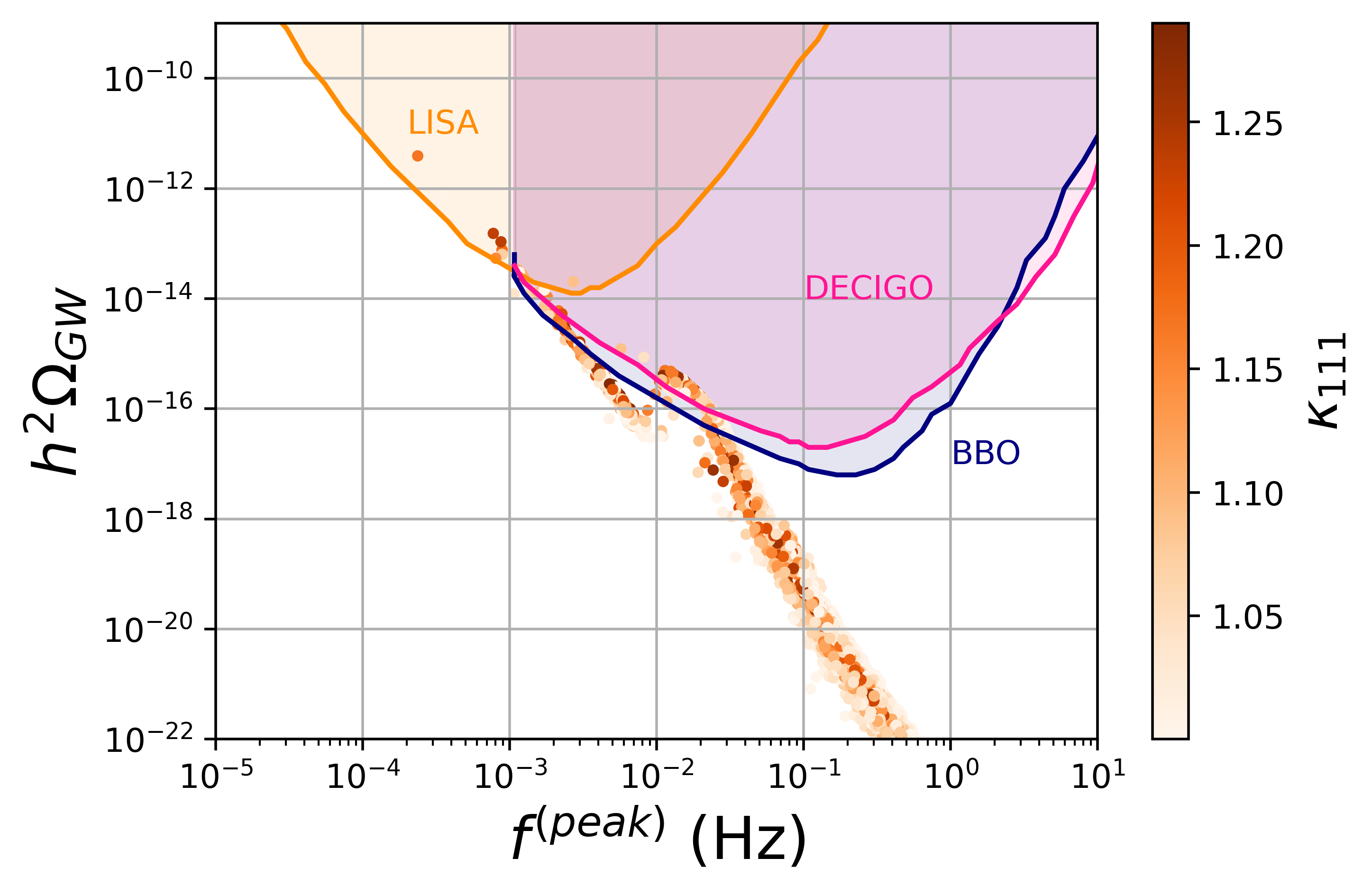}}
\subfigure[]{\includegraphics[width=0.45\textwidth,clip]{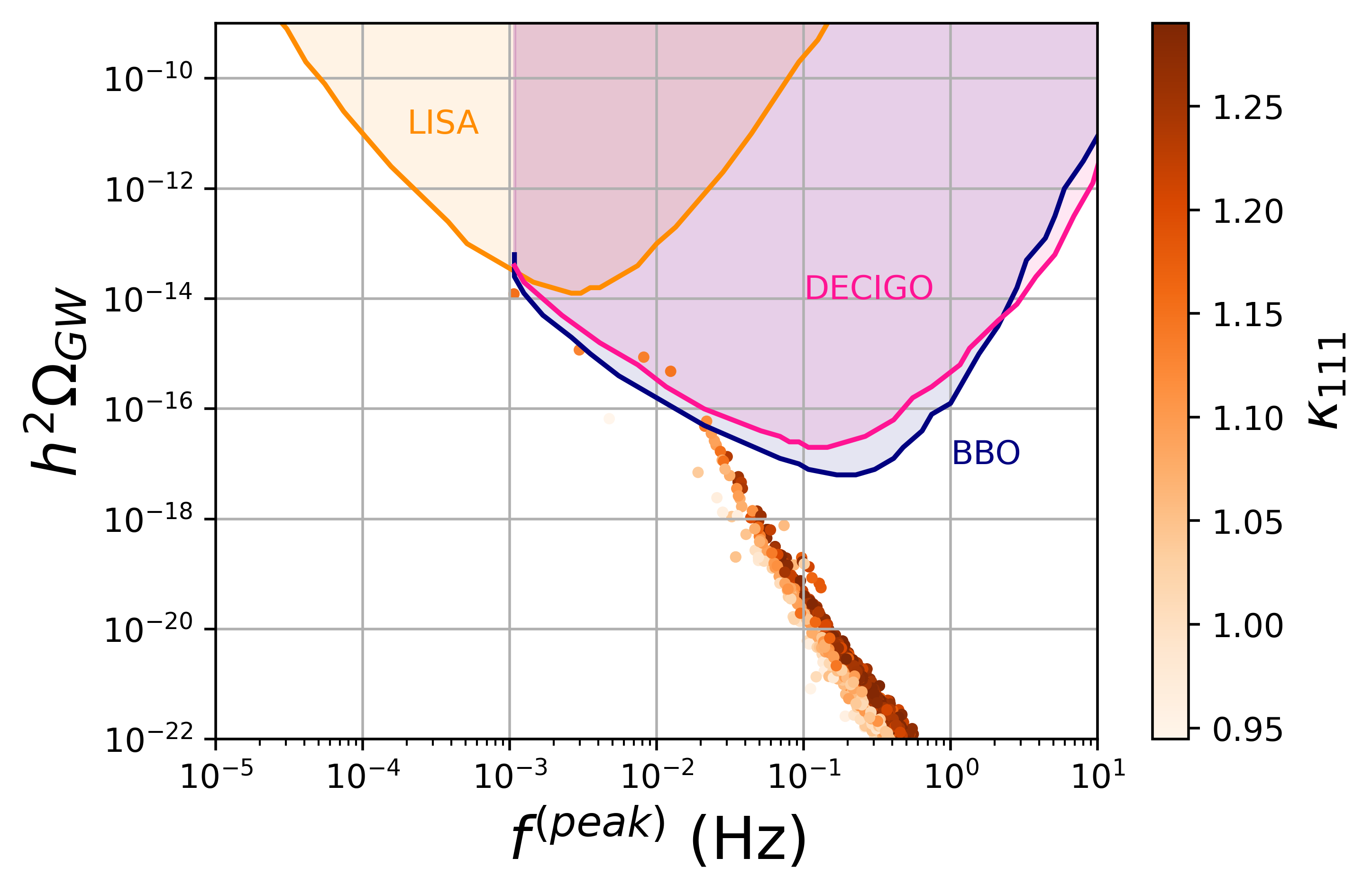}}
\subfigure[]{\includegraphics[width=0.45\textwidth,clip]{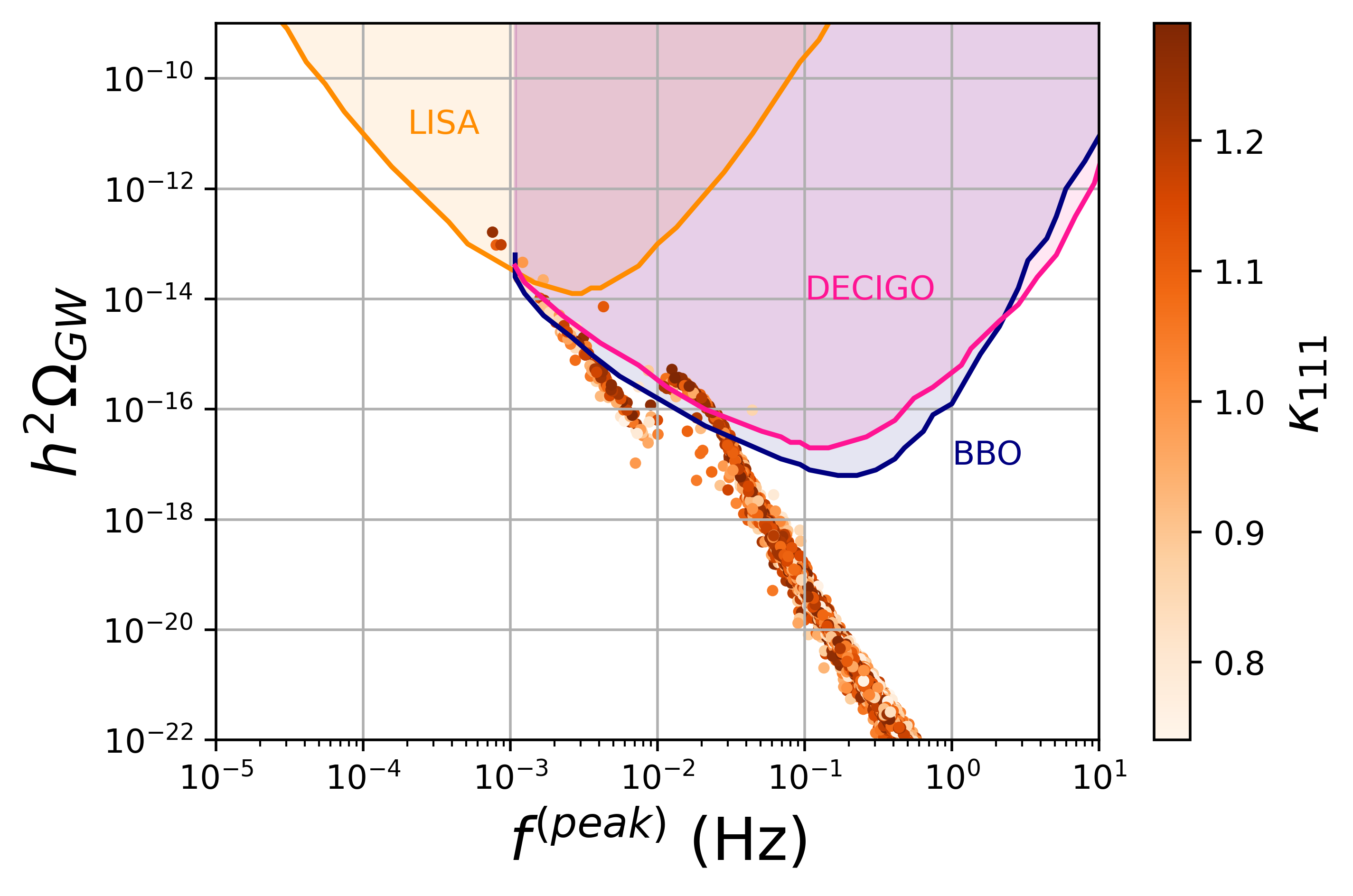}}
\subfigure[]{\includegraphics[width=0.45\textwidth,clip]{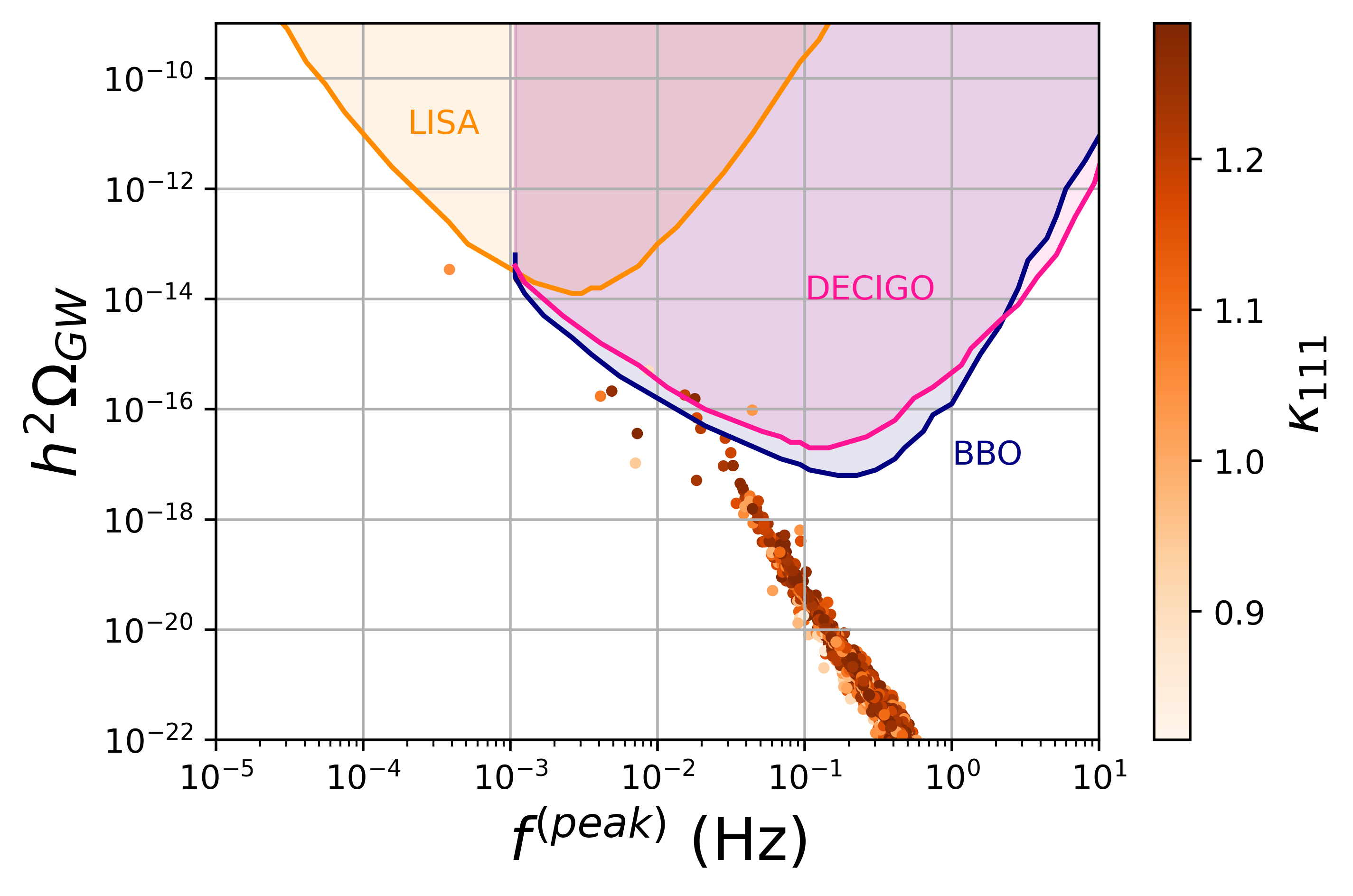}}
\subfigure[]{\includegraphics[width=0.45\textwidth,clip]{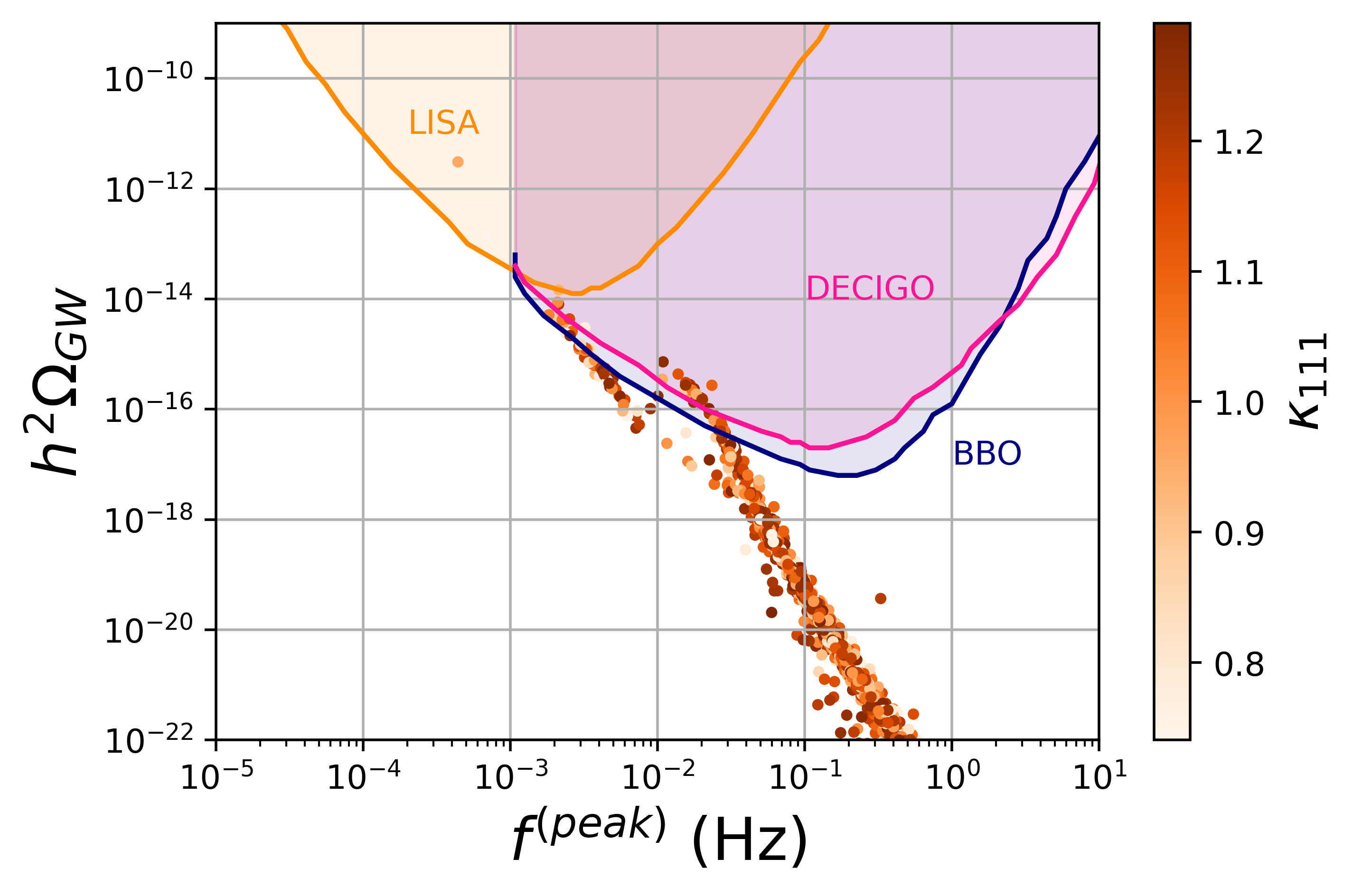}}
\subfigure[]{\includegraphics[width=0.45\textwidth,clip]{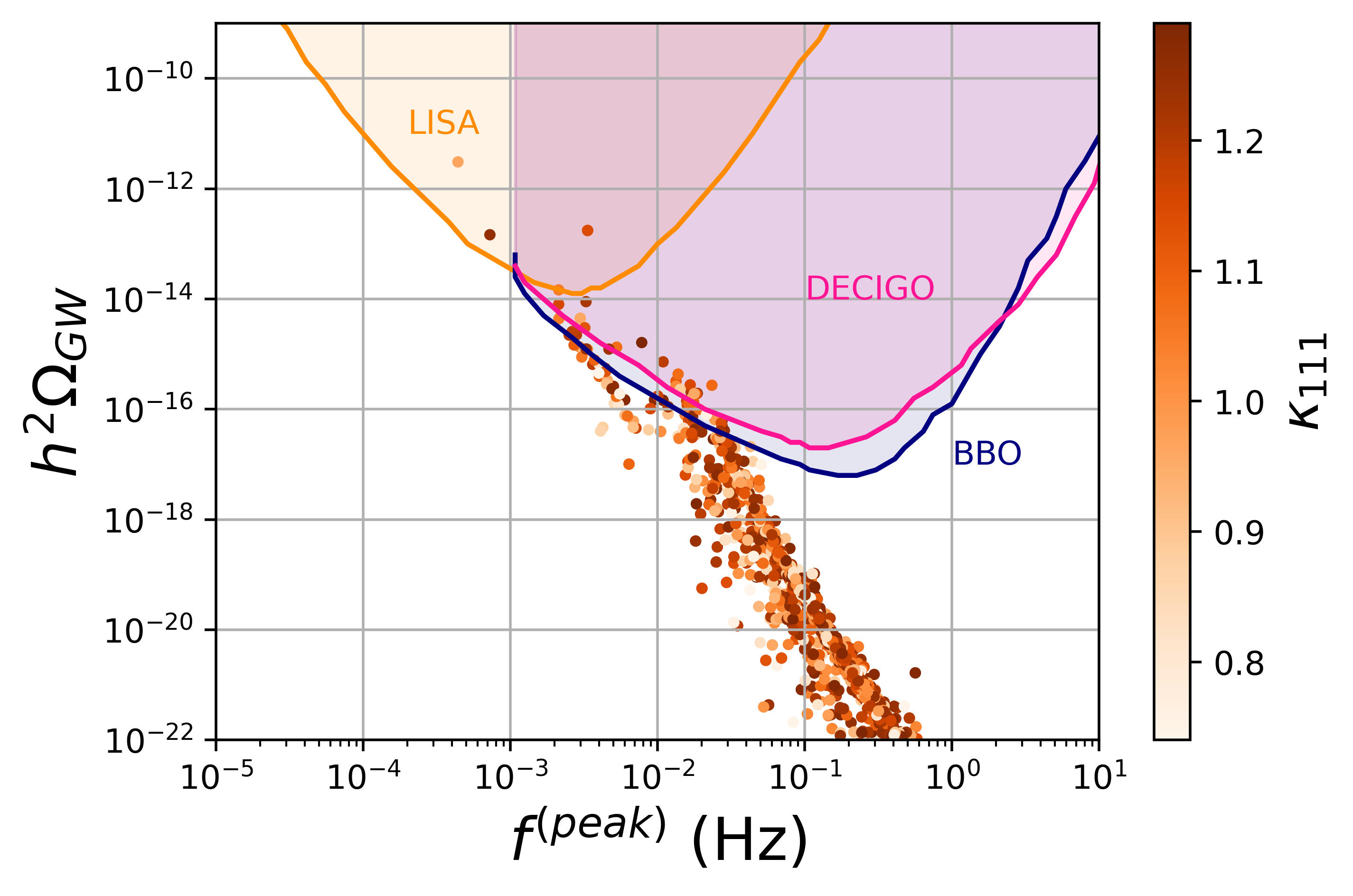}}
\end{center}
\caption{\label{fig:grav_wave4}
Peak amplitudes versus frequency for GW sourced by sound waves for (a,b) the renormalizable model (c,d) weakly coupled EFT, and (e,f) strongly coupled EFT. The color of each point shows the corresponding value of $\kappa_{111}$ after imposing $0.74\lesssim \kappa_{111}\lesssim 1.29$. The trilinear scalar couplings are evaluated at (a,c,e) tree and (b,d,f) loop level.}
\end{figure}

As can be seen in Fig.~\ref{fig:grav_wave2}, there  is little to no correlation between the GW signal and resonant di-Higgs production, indicating a complementarity between the two signals. In contrast, in Fig.~\ref{fig:grav_wave4} we show the peak frequency of the GW spectrum while applying HL-LHC projected bounds $0.74\lesssim \kappa_{111}\lesssim 1.29$. We show this for (a,b) the renormalizable model, (c,d) the weakly coupled EFT, and (e,f) the strongly coupled EFT. For completeness, we show the results when the trilinear is evaluated at (a,c,e) tree level and (b,d,f) loop level. As can be seen at loop level, applying the bounds on $\kappa_{111}$ at one loop level eliminates most of the parameter space with a detectable GW spectrum. This is not the case when the trilinears are evaluated at tree level. As discussed above, loop corrections raise the value of $\kappa_{111}$ at high $h_2$ masses.  In Ref.~\cite{Braathen:2025svl} it was found that in the low mass region there exist points with a GW signature and almost no signature at the LHC in the di-Higgs final state. Our bounds on $\sin\theta$ in Fig.~\ref{fig:sth}(c) are in conflict with most of that parameter space. Hence, we find only a few points that pass future Higgs trilinear bounds that still have a potential GW signature in the renormalizable or weakly coupled EFT. While more points pass in the strongly interacting EFT, this theory is non-perturbative and its predictions are unreliable, as mentioned previously.

%% file: conc.tex
While the real singlet model is a much studied benchmark model for a strong first order EWPT necessary for EWBG, it is not sufficient to generate the matter-antimatter asymmetry. That is, there is expected to be additional new physics at the TeV scale. In this paper we studied the potential effects of this new physics on the EWPT using EFT methods and extending the scalar potential to include effective operators up to dimension six.  Both a weakly coupled EFT, with Wilson coefficients bounded by $|C_i|\leq 1$, and a strongly coupled theory $|C_i|\leq 4\pi$ were studied. We found that requiring global minimum and bounded from below constraints on the scalar potential for field values up to $1$ TeV increased the allowed parameter space of the renormalizable model. Unsurprisingly, the allowed parameter space was increased further in the EFTs, even after the requirement of a strong first order electroweak phase transition.

Similar to previous literature, we found the one loop corrections to scalar trilinear couplings effected the phenomenology of these models considerably. At the renormalizable level, one loop corrections to the Higgs trilinear significantly increased the deviations away from SM predictions. Once a strong first order EWPT was required, the HL-LHC measurement of the Higgs trilinear would be sensitive to the majority of the parameter space for heavy scalar masses above $\sim450$~GeV. For the LHC phenomenology, we presented results for the weakly coupled EFT only, since the one-loop effective potential with the strongly coupled EFT indicated it was non-perturbative or nearly non-perturbative. The weakly coupled EFT increases the allowed ranges of the Higgs trilinear significantly. There were parameter points for all scalar masses that could evade HL-LHC Higgs trilinear measurements, in contrast to the renormalizable case where the HL-LHC would be sensitive to the large majority of parameter points at high scalar masses. We also found that in the renormalizable and weakly coupled EFT requiring a SFOEWPT suppressed the maximum resonant di-Higgs rate relative to when the SFOEWPT was not required. The maximum resonant di-Higgs rates were similar in the renormalizable and weakly coupled EFT. This is true when effective operators are added only to the potential and the largest ${\rm BR}(h_2\rightarrow h_1h_1)$ occur for small $\sin\theta$. However, if new physics or effective operators alter the $h_2$ production mechanism, the resonant di-Higgs rate can be enhanced~\cite{Roy:2026kud}.

For the gravitational wave signals, the results in both the renormalizable model and weakly coupled theory were similar. Only a few points had peak frequencies that would be observable at future gravitational wave detectors. That is, the predictions of the GW signal in the real singlet model are robust against additional weakly coupled physics at the TeV scale. The strongly coupled EFT had many more points that are potentially observable. This indicates that to enhance the GW signature in the real singlet model, either there should be a strongly interacting theory in the TeV range or the theory needs additional contributions to the phase transition below a TeV.  Additionally, there is a correlation between the one-loop Higgs trilinear and gravitational wave signals. When HL-LHC constraints were applied, most of the points in the renormalizable model and weakly coupled EFT with stronger gravitational wave signals are eliminated. This is not the case when the trilinear coupling is evaluated at tree level. In the strongly coupled EFT, there were still many points with an observable GW signal even after the HL-LHC constraints.

Finally, our analysis of the one-loop trilinear scalar couplings used the zero temperature effective potential. However, the effective potential is evaluated when the external momenta are zero, which is not the case in measurements of the Higgs trilinear coupling and the decay $h_2\rightarrow h_1h_1$. Indeed, there are loops that can contribute when the external momenta are non-zero which do not contribute when the external momenta are zero. Our work shows the importance of these loop corrections at the renormalizable and EFT level, as has been found in other studies on the renormalizable level~\cite{Curtin:2014jma,Braathen:2026wib,Kanemura:2004ch,Reichert:2017puo,Goncalves:2021egx,Dorsch:2017nza,Basler:2017uxn,Biekotter:2022kgf}. We leave a more careful calculation of the one-loop corrections to trilinear scalar couplings~\cite{Kanemura:2017gbi,Aiko:2023xui,Kanemura:2019slf,Bahl:2023eau} to future work.

%% file: App_Exp.tex
In this appendix are detailed the Higgs signal strengths [App.~\ref{app:HiggsPrec}] and scalar searches [App.~\ref{app:ScalSearch}] included in the $\sin\theta$ bounds as presented in Sec.~\ref{sec:ExpConst}.
\subsection{Higgs Precision Measurements}
\label{app:HiggsPrec}
Both ATLAS~\cite{ATLAS:2026pdi} and CMS~\cite{CMS:2026nce} have released combinations of Higgs signal strengths measured during Run 2 at 13 TeV. Since the renormalizable singlet model has a universal Higgs signal strength we use the combined global signal strengths:
\begin{eqnarray}
&\hat{\mu}_{\rm ATLAS}=0.990^{+0.053}_{-0.051}\quad&{\rm ATLAS~Combination}\label{eq:Combo}\\
&\hat{\mu}_{\rm CMS}=1.014^{+0.055}_{-0.053}\quad&{\rm CMS~Combination}.\nonumber
\end{eqnarray}
Not included in the CMS combination are Run 2 searches for a boosted Higgs decaying to hadronic or semi-hadronic $WW^{*}$~\cite{CMS:2026yrb} or boosted $Vh$ production with hadronic $V$ and $h\rightarrow bb$~\cite{CMS:2026fsx}.  Hence, we also include those signal strengths in our fit:
\begin{eqnarray}
\hat{\mu}^{WW}_{\rm CMS}&=&-0.19^{+0.48}_{-0.46}\label{eq:CMS13}\\
\hat{\mu}^{bb}_{Vh,{\rm CMS}}&=&0.72^{+0.75}_{-0.71}.\nonumber
\end{eqnarray}

Both ATLAS and CMS have released several Run 3 measurements at 13.6~TeV.  ATLAS has released Run 3 measurements of $h\rightarrow Z\gamma$ with 165 fb$^{-1}$ of data~\cite{ATLAS:2025aip}, $h\rightarrow ZZ$ with 164 fb$^{-1}$~\cite{ATLAS:2026hsl}, $h\rightarrow \mu\mu$ with 165 fb$^{-1}$~\cite{ATLAS:2025coj}, and $h\rightarrow \gamma\gamma$ with 31.4 fb$^{-1}$~\cite{ATLAS:2023tnc}:
\begin{eqnarray}
\hat{\mu}^{Z\gamma}_{{\rm ATLAS}} &=& 0.9^{+0.7}_{-0.6}\label{eq:ATLAS136}\\
\hat{\mu}^{ZZ}_{{\rm ATLAS}} &=&0.99\pm0.13\nonumber\\
\hat{\mu}^{\mu\mu}_{{\rm ATLAS}}&=&1.4\pm 0.4\nonumber\\
\hat{\mu}^{\gamma\gamma}_{{\rm ATLAS}}&=&1.12\pm0.21\nonumber.
\end{eqnarray}
CMS has released Run 3 measurements of $h\rightarrow\gamma\gamma$ with 34.7 fb$^{-1}$~\cite{CMS:2025fwn} and $h\rightarrow ZZ$ with 171 fb$^{-1}$~\cite{CMS-PAS-HIG-25-015}:
\begin{eqnarray}
\hat{\mu}^{\gamma\gamma}_{\rm CMS}&=&1.09\pm0.19\label{eq:CMS136}\\
\hat{\mu}^{ZZ}_{\rm CMS}&=&1.02\pm0.11\nonumber.
\end{eqnarray}
All the signal strengths in Eqs.~(\ref{eq:Combo}-\ref{eq:CMS136}) are included in the $\chi^2$ fit for $h_1$ in Eq.~(\ref{eq:h1chi2}).

\subsection{Scalar Searches}
\label{app:ScalSearch}
Table~\ref{tab:ScalSearch} lists all the scalar searches included in the $\sin\theta$ bounds that are presented in Sec.~\ref{sec:ExpConst}.  Since the couplings of $h_2$ to SM fermions and gauge bosons are the same as a SM Higgs at the same mass, $h_2$ is expected to decay like a SM Higgs at mass $m_2$ with the additional decay channel of $h_2\rightarrow h_1h_1$. That is, when kinematically allowed, $h_2$ predominantly decays into $WW,ZZ,$ and $h_1h_1$. Hence, those are the searches included in our fits. If two searches from a collaboration have overlapping data sets, the most constraining of the two searches was used.

\begin{table}[tb]
\resizebox{\textwidth}{!}{
\begin{tabular}{|c|c|c|c|c|}\hline\hline
Experiment & Search Channel & Luminosity & $h_2$ Mass Range & Ref. \\\hline\hline
\multirow{2}{*}{ATLAS} & \multirow{2}{*}{$h_2\rightarrow WW \rightarrow e\nu\mu\nu$ in ggF and VBF} & \multirow{2}{*}{139 fb$^{-1}$} & 200-3800 GeV in ggF & \multirow{2}{*}{\cite{ATLAS-CONF-2022-066}}\\
& & & 200-4000 GeV in VBF& \\\hline
ATLAS & $h_2\rightarrow ZZ\rightarrow 4\ell/2\ell 2\nu$ in ggF and VBF & 139 fb$^{-1}$ & 210-2000 GeV& \cite{ATLAS:2020tlo}\\\hline
ATLAS & $h_2\rightarrow WW+ZZ\rightarrow 2\ell2q/\ell\nu2q$ in ggF and VBF & 139 fb$^{-1}$ & 300-5000 GeV & \cite{ATLAS:2020fry}\\\hline
ATLAS & $h_2\rightarrow h_1h_1\rightarrow 4b/2b2\gamma/2b2\tau$ in ggF & 126-136 fb$^{-1}$ & 250-5000 GeV& \cite{ATLAS:2023vdy}\\\hline
\multirow{2}{*}{ATLAS} & \multirow{2}{*}{$h_2\rightarrow (h_1\rightarrow 2b)(S\rightarrow 2\gamma)$ in ggF} & 140 fb$^{-1}$ at 13 TeV& \multirow{2}{*}{275-1000 GeV}& \multirow{2}{*}{\cite{ATLAS:2025nda}}\\
& & +59 fb$^{-1}$ at 13.6 TeV  & & \\\hline
ATLAS & $h_2\rightarrow h_1h_1\rightarrow 4W$ in ggF & 36.1 fb$^{-1}$ & 250-500 GeV & \cite{ATLAS:2018ili}\\\hline
ATLAS & $h_2\rightarrow h_1h_1 \rightarrow 2b2W$ in ggF & 36.1 fb$^{-1}$ & 500-3000 GeV & \cite{ATLAS:2018fpd}\\\hline
ATLAS &$h_2\rightarrow h_1h_1\rightarrow 2\gamma2W$ in ggF & 36.1 fb$^{-1}$ & 250-500 GeV & \cite{ATLAS:2018hqk}\\\hline
ATLAS & $h_2\rightarrow h_1h_1\rightarrow 4b$ in VBF & 126 fb$^{-1}$ & 260-1000 GeV & \cite{ATLAS:2020jgy}\\\hline
CMS & $h_2\rightarrow ZZ\rightarrow 4\ell$ in ggF and VBF & 138 fb$^{-1}$ & 130-3000 GeV & \cite{CMS:2026xbb}\\\hline
CMS & $h_2\rightarrow ZZ\rightarrow 2\ell2q$ in ggF & 138 fb$^{-1}$ & 450-1800 GeV & \cite{CMS:2021xor}\\\hline
CMS & $h_2\rightarrow WW\rightarrow 2\ell2\nu$ in ggF and VBF combined & 138 fb$^{-1}$ & 180-5000 GeV & \cite{CMS-PAS-HIG-20-016}\\\hline
CMS & $h_2\rightarrow WW\rightarrow 2\ell 2\nu/\ell\nu2q$ in ggF and VBF combined & 35.9 fb$^{-1}$ & 200-3000 GeV & \cite{CMS:2019bnu}\\\hline
CMS & $h_2\rightarrow h_1h_1\rightarrow 4W/4\tau/2W2\tau/2b2W/2b2\tau/2b2\gamma/4b$ in ggF & 138 fb$^{-1}$ & 250-4000 GeV & \cite{CMS:2024phk}\\\hline
CMS & $h_2\rightarrow h_1h_1\rightarrow 4b$ in ggF  & 138 fb$^{-1}$ & 400-1400 GeV & 
\cite{CMS:2026mwf}\\\hline
CMS & $h_2\rightarrow (h_1\rightarrow 2b)(S\rightarrow 2\gamma)$ in ggF & 132 fb$^{-1}$ & 280-1000 GeV & \cite{CMS:2025qit}\\\hline
CMS & $h_2\rightarrow (h_1\rightarrow 2\gamma)(S\rightarrow 2b)$ in ggF & 138 fb$^{-1}$ & 260-1000 GeV & \cite{CMS:2023boe}\\\hline
CMS & $h_2\rightarrow h_1 h_1 \rightarrow 2\tau2\gamma$ & 138 fb$^{-1}$ & 300-1000 GeV & \cite{CMS:2025tqi}\\\hline
CMS & $h_2\rightarrow h_1h_1\rightarrow 2b2Z$ in ggF & 138 fb$^{-1}$ & 260-4000 GeV & \cite{CMS:2026fqd}\\\hline
CMS & $h_2\rightarrow h_1h_1\rightarrow 2b2\tau$ in ggF & 138 fb$^{-1}$ & 250-3000 GeV & \cite{CMS-PAS-B2G-24-011}\\\hline
\end{tabular}}
\caption{\label{tab:ScalSearch}The scalar searches included in the bounds on $\sin\theta$. All searches at 13 TeV unless otherwise noted.}
\end{table}

%% file: App_Coup.tex
The trilinear and quartic scalar couplings are defined as
{\allowdisplaybreaks
\begin{eqnarray}
V(\Phi,S)&\supset& \frac{1}{3!}\lambda_{111}h_1^3+\frac{1}{2}\lambda_{112}h_1^2h_2+\frac{1}{2}\lambda_{122}h_1h_2^2+\frac{1}{3!}\lambda_{222}h_2^3+\frac{1}{4!}\lambda_{1111}h_1^4\\
&&+\frac{1}{3!}\lambda_{1112}h_1^3h_2+\frac{1}{4}\lambda_{1122}h_1^2h_2^2+\frac{1}{3!}\lambda_{1222}h_1h_2^3+\frac{1}{4!}\lambda_{2222}h_2^4\nonumber
\end{eqnarray}
The trilinear couplings $\lambda_{111}$ and $\lambda_{112}$ are already reported in Eqs.~(\ref{eq:Trilin},\ref{eq:lam112}), respectively. The remaining trilinear and quartic couplings are:
\begin{eqnarray}
\lambda_{122}&=&\frac{2\,m_2^2+m_1^2}{v_{\rm EW}}\cos\theta\sin^2\theta-2\left(a_2\,v_{\rm EW}\left(1-\frac{3}{2}\cos^2\theta\right)-b_3\sin\theta\,\cos\theta\right)\cos\theta\nonumber\\
&&+\frac{v_{\rm EW}^2}{\Lambda}\sin\theta\left(\frac{3}{2}a_3\cos^2\theta+a_4\left(1-3\,\cos^2\theta\right)\right)\nonumber\\
&&+6\frac{v_{\rm EW}^3}{\Lambda^2}\cos\theta\left(C_H\sin^2\theta-\frac{1}{3}a_6\left(1-\frac{3}{2}\cos^2\theta\right)\right)+\mathcal{O}(\Lambda^{-3})\\
\lambda_{222}&=&-3\frac{m_2^2}{v_{\rm EW}}\sin^3\theta+\left(-3\,a_2\,v_{\rm EW}\sin\theta+2\,b_3\cos\theta\right)\cos^2\theta+3\frac{v_{\rm EW}^2}{\Lambda}\cos\theta\left(\frac{1}{2}a_3\cos^2\theta+a_4\sin^2\theta\right)\nonumber\\
&&-6\frac{v_{EW}^3}{\Lambda^2}\sin\theta\left(C_H\sin^2\theta+\frac{1}{2}a_6\cos^2\theta\right)+\mathcal{O}(\Lambda^{-3})\\
\lambda_{1111}&=&3\frac{m_1^2\cos^2\theta+m_2^2\sin^2\theta}{v_{\rm EW}^2}\cos^4\theta+6\left(a_2\cos^2\theta+b_4\sin^2\theta\right)\sin^2\theta\nonumber\\
&&+12\frac{v_{\rm EW}}{\Lambda}\cos\theta\,\sin\theta\left(a_3\sin^2\theta+a_4\cos^2\theta\right)\nonumber\\
&&+36\frac{v_{\rm EW}^2}{\Lambda^2}\left(C_H\cos^4\theta+\frac{1}{6}a_5\sin^4\theta+\frac{1}{2}a_6\cos^2\theta\sin^2\theta\right)+\mathcal{O}(\Lambda^{-3})\\
\lambda_{1112}&=&-3\frac{m_1^2\cos^2\theta+m_2^2\sin^2\theta}{v_{\rm EW}^2}\cos^3\theta\sin\theta+3\left(a_2\left(\cos^2\theta-\sin^2\theta\right)+2\,b_4\,\sin^2\theta\right)\cos\theta\sin\theta\nonumber\\
&&+3\frac{v_{\rm EW}}{\Lambda}\left(3\,a_3\left(1-\frac{4}{3}\sin^2\theta\right)\sin^2\theta+a_4\left(1-4\,\sin^2\theta\right)\cos^2\theta\right)\nonumber\\
&&-36\frac{v_{\rm EW}^2}{\Lambda^2}\cos\theta\sin\theta\left(C_H\cos^2\theta-\frac{1}{6}a_5\sin^2\theta-\frac{1}{4}a_6\left(\cos^2\theta-\sin^2\theta\right)\right)+\mathcal{O}(\Lambda^{-3})\\
\lambda_{1122}&=&3\frac{m_1^2\cos^2\theta+m_2^2\sin^2\theta}{v_{\rm EW}^2}\cos^2\theta\sin^2\theta+a_2\left(1-6\,\cos^2\theta\sin^2\theta\right)+6\,b_4\cos^2\theta\sin^2\theta\nonumber\\
&&+6\frac{v_{\rm EW}}{\Lambda}\left(\cos^2\theta-\sin^2\theta\right)\cos\theta\sin\theta\left(a_3-a_4\right)\nonumber\\
&&+36\frac{v_{\rm EW}^2}{\Lambda^2}\left(C_H\cos^2\theta\sin^2\theta+\frac{1}{6}a_5\cos^2\theta\sin^2\theta+\frac{1}{12}a_6\left(1-6\,\cos^2\theta\sin^2\theta\right)\right)+\mathcal{O}(\Lambda^{-3})\\
\lambda_{1222}&=&-3\frac{m_1^2\cos^2\theta+m_2^2\sin^2\theta}{v_{\rm EW}^2}\cos\theta\sin^3\theta+3\left(a_2\left(\sin^2\theta-\cos^2\theta\right)+2\,b_4\cos^2\theta\right)\sin\theta\cos\theta\nonumber\\
&&-3\frac{v_{\rm EW}}{\Lambda}\left(3\,a_3\left(1-\frac{4}{3}\cos^2\theta\right)\cos^2\theta+a_4\left(1-4\,\cos^2\theta\right)\sin^2\theta\right)\nonumber\\
&&-36\frac{v_{\rm EW}^2}{\Lambda^2}\sin\theta\cos\theta\left(C_H\sin^2\theta-\frac{1}{6}a_5\cos^2\theta-\frac{1}{4}a_6\left(\sin^2\theta-\cos^2\theta\right)\right)+\mathcal{O}(\Lambda^{-3})\\
\lambda_{2222}&=&3\frac{m_1^2\cos^2\theta+m_2^2\sin^2\theta}{v_{\rm EW}^2}\sin^4\theta+6\left(a_2\sin^2\theta+b_4\cos^2\theta\right)\cos^2\theta\nonumber\\
&&-12\frac{v_{\rm EW}}{\Lambda}\cos\theta\sin\theta\left(a_3\cos^2\theta+a_4\sin^2\theta\right)\nonumber\\
&&+36\frac{v_{\rm EW}^2}{\Lambda^2}\left(C_H\sin^4\theta+\frac{1}{6}a_5\cos^4\theta+\frac{1}{2}a_6\cos^2\theta\sin^2\theta\right)+\mathcal{O}(\Lambda^{-3})
\end{eqnarray}
}

%% file: App_V1_cutoff.tex
In this appendix we discuss the expansion of the scalar mass matrix in inverse powers of the new physics scale and the calculation of the scalar contributions to the trilinears $\lambda_{111}$ and $\lambda_{112}$.

The two-by-two scalar mass matrix is organized as
\begin{eqnarray}
    \M &=& \M_0 + \frac{1}{\Lambda}\mathcal{M}_1+ \frac{1}{\Lambda^2}\M_2,\label{eq:ScalMass}
\end{eqnarray}
where the subscript indicates the order of $\Lambda^{-n}$. The eigenvalues are then:
\begin{eqnarray}
m_{\pm}^2=\frac{1}{2}\left({\rm Tr}\,{\M}\pm\sqrt{({\rm Tr}\,\M)^2-4\,{\rm det}\,\M}\right)
\end{eqnarray}
To order $\Lambda^{-2}$, the determinant is:
\begin{eqnarray}
    \text{det}\M &=& A_0 + \frac{1}{\Lambda}A_1 
     +\frac{1}{\Lambda^2}A_2 
\end{eqnarray}
where
\begin{eqnarray}
    A_0 &=&(\M_{0})_{11}(\M_{0})_{22} - [(\M_{0})_{12}]^2  \\
    A_1 &=&  (\M_{0})_{11}(\M_{1})_{22} + (\M_{0})_{22}(\M_{1})_{11} - 2 (\M_{0})_{12}(\M_{1})_{12} \nonumber\\
    A_2 &=& (\M_{2})_{11}(\M_{0})_{22} + (\M_{2})_{22}(\M_{0})_{11} + (\M_{1})_{11}(\M_{1})_{22} - 2(\M_{2})_{12}(\M_{0})_{12} - [(\M_{1})_{12}]^2\nonumber
\end{eqnarray}
The eigenvalues to order $\Lambda^{-2}$ are 
\begin{eqnarray}
m^2_{\pm}(\phi_0,S) &=& m^2_{\pm, 0} + \frac{1}{\Lambda}m_{\pm,1}^2 + \frac{1}{\Lambda^2}m_{\pm, 2}^2
\end{eqnarray}
where 
\begin{eqnarray}
    m_{\pm,0}^2 &=& \frac{1}{2} \left( \text{Tr}(\M_0) \pm \sqrt{ \text{Tr}(\M_0)^2 - 4 A_0 } \right) \\
    m_{\pm, 1}^2 &=& \frac{1}{2} \left( \text{Tr}(\M_1) \pm \frac{\text{Tr}(\M_0)\text{Tr}(\M_1) - 2 A_1 }{\sqrt{\text{Tr}(\M_0)^2 - 4 A_0}} \right)\nonumber \\
    m_{\pm,2}^2 &=& \frac{1}{2}\left(\text{Tr}(\M_2) \pm \left(\frac{2\text{Tr}(\M_0)\text{Tr}(\M_2) + \text{Tr}(\M_1)^2 - 4 A_2}{2\sqrt{\text{Tr}(\M_0)^2 - 4A}} - \frac{\left(\text{Tr}(\M_0)\text{Tr}(\M_1) - 2 A_1 \right)^2}{2(\text{Tr}(\M_0)^2 - 4 A_0)^{3/2}} \right)\right) \nonumber\
\end{eqnarray}
Then for a term in the 1 loop potential,
\begin{eqnarray}
    \Delta V^{(1)}_{T=0, i} = \frac{n_i}{64\pi^2}\left[ m_i^4(\phi_0, S)\left(\log\frac{m_i^2(\phi_0,S)}{m_i^2(v_{EW},0)} - \frac{3}{2}\right) + 2 m_i^2(\phi_0,S)m_i^2(v_{EW},0) \right],
\end{eqnarray}
with a mass that depends on $\Lambda$, the truncated potential is 
\begin{eqnarray}
    \Delta V^{(1)}_{T=0} = \Delta V^{(1)}_{0, \,(T=0)} + \frac{1}{\Lambda} \Delta V^{(1)}_{1, \,(T=0)} + \frac{1}{\Lambda^2}\Delta V^{(1)}_{2, \,(T=0)}
\end{eqnarray}
where 
\begin{eqnarray}
    \Delta V^{(1)}_{0, \,(T=0)} &=& \frac{n_i}{64\pi^2}\left[(m_{\pm,0}^2(\phi_0,S))^2 \left(\log\frac{m_{\pm,0}^2(\phi_0,S)}{m_{\pm}^2(v_{EW},0)} - \frac{3}{2}\right)  
    +2 m_{\pm,0}^2(\phi_0,S) m_{\pm}^2(v_{EW},0)  \right]  \nonumber\ \\
    \Delta V^{(1)}_{1, \,(T=0)} &=& \frac{n_i}{64\pi^2}\left[ 2 m_{\pm,0}^2(\phi_0,S)m_{\pm,1}^2(\phi_0,S)\left( \log\frac{m_{\pm,0}^2(\phi_0,S)}{m_{\pm}^2(v_{EW},0)} - 1\right) + 2m_{\pm,1}^2(\phi_0,S)m_{\pm}^2(v_{EW},0) \right ] \nonumber\ \\
    \Delta V^{(1)}_{2, \,(T=0)} &=& \frac{n_i}{64\pi^2} \left[ 2m_{\pm,0}^2(\phi_0,S)m_{\pm,2}^2(\phi_0,S) \left( \log\frac{m_{\pm,0}^2(\phi_0,S)}{m_{\pm}^2(v_{EW},0)} - 1\right)\right. \\
    &+& \left.(m_{\pm,1}^2(\phi_0,S))^2 \left( \log\frac{m_{\pm,0}^2(\phi_0,S)}{m_{\pm}^2(v_{EW},0)}\right) + 2m_{\pm,2}^2(\phi_0,S) m_{\pm}^2(v_{EW},0)  \right]
\end{eqnarray}

In the $\sin\theta\rightarrow 0$ limit, we find the one-loop correction to the Higgs trilinear including the top, $W$, and $Z$:
\begin{eqnarray}
\lambda_{111}^{(1)}&=\displaystyle\frac{v_{\rm EW}}{4\pi^2}&\left[\frac{3}{v_{\rm EW}^4}\left(-4\,m_t^4+2\,M_W^4+M_Z^4+\frac{9}{8}m_1^4\right)+\frac{81}{4}C_H\frac{m_1^2}{\Lambda^2}\right.\\
&&\left.+\frac{1}{8}\frac{v_{\rm EW}^2}{m_2^2}a_2^2\left(a_2+3\,a_6 \frac{v_{\rm EW}^2}{\Lambda^2}\right)\right]+\mathcal{O}\left(\sin\theta,\Lambda^{-3}\right)\nonumber\\
\lambda_{112}^{(1)}&=\displaystyle\frac{1}{16\pi^2}&\left[\frac{v_{\rm EW}^2}{m_2^2}a_2\left(a_2\,b_3+\frac{3}{4}a_2 a_3 \frac{v_{\rm EW}^2}{\Lambda}+2\,a_6b_3\frac{v_{\rm EW}^2}{\Lambda^2}\right)\right.\\
&&\left.+\frac{9}{2}a_4\frac{m_1^2}{\Lambda}\right]+\mathcal{O}\left(\sin\theta,\Lambda^{-3}\right).\nonumber
\end{eqnarray}

%% file: App_TempMasses.tex
The temperature corrections to the Goldstone boson masses are
\begin{eqnarray}
    \Pi_G(T) &=& \left(\frac{3}{16}g^2 + \frac{1}{16}g'^2 + \frac{1}{4}y_t^2 + \frac{1}{2}\lambda + \frac{1}{24}a_2 \right) T^2.
\end{eqnarray}
For the scalars, the temperature dependent mass corrections are
\begin{eqnarray}
    \Delta\mathcal{M}_T^2 &=&\begin{pmatrix}
        \frac{3}{16}g^2 + \frac{1}{16}g'^2 + \frac{1}{4}y_t^2 + \frac{1}{2}\lambda + \frac{1}{24}a_2 & 0 \\
        0 & \frac{1}{6}a_2 + \frac{1}{4}b_4
    \end{pmatrix} T^2.
\end{eqnarray}
The thermally corrected masses are then the eigenvalues of $\mathcal{M}+\Delta\mathcal{M}^2_T$, where $\mathcal{M}$ is the zero temperature tree level mass matrix defined in Eq.~(\ref{eq:ScalMass}).

For the longitudinal modes of the gauge bosons, the temperature corrections to the masses are
\begin{eqnarray}
    \Pi_{W_L}(T) = \frac{11}{6}g^2T^2 \\
    \mathcal{M}_{Z_L}^2(\phi_0,S,T) &=& \frac{1}{2}\left[ m_z^2(\phi_0,S) + \frac{11}{6}\frac{g^2}{\cos^2\theta_W}T^2 +\Delta(\phi_0,S,T) \right] \\
    \mathcal{M}_{\gamma_L}^2(\phi_0,S,T) &=& \frac{1}{2}\left[ m_z^2(\phi_0,S) + \frac{11}{6}\frac{g^2}{\cos^2\theta_W}T^2 -\Delta(\phi_0,S,T) \right] \\ 
    \Delta(\phi_0,S,T)&=& \sqrt{m_z^4(\phi_0,S) + \frac{11}{3}\frac{g^2 \cos^22\theta_W}{\cos^2\theta_W}\left(m_z^2(\phi_0,S) + \frac{11}{12}\frac{g^2}{\cos^2\theta_W}T^2\right)T^2}
\end{eqnarray}
 and transverse components
 \begin{eqnarray}
     \Pi_{W_T}=\Pi_{Z_T}=\Pi_{\gamma_T} = 0.
 \end{eqnarray}

Using the Parwani scheme, the photon is now introduced into the CW potential via the finite temperature mass. This term has a similar IR divergence as the Goldstone bosons, which we regulate in the same way.

%% file: App_gravitational_waves.tex
A first order EWPT can produce detectable GWs sourced by bubble wall collisions, pressure waves in the plasma, and turbulence. We consider contributions from bubble wall collisions and sound waves. In the following, it is understood that $\Omega_{GW} = d\Omega_{GW}/d\ln(f)$.

\textbf{Bubble Collisions:}  Following~\cite{Konstandin_2018}, the GW power spectrum from bubble collisions is determined through

\begin{eqnarray}
    \Omega_{GW}^{env}h^2 &=& \tilde{\Omega}_{GW}^{env} h^2 \frac{(a + b) \tilde{f}^bf^{a}}{b \tilde{f}^{(a+b)} + a f^{(a+b)}}
\end{eqnarray}
where 
\begin{eqnarray}
    \tilde{f} = 2.62\times10^{-6}\frac{1.96}{1-0.051 v_w + 0.88 v_w^2}\left(\frac{\beta}{H_*}\right)\left(\frac{T_*}{100 \text{ GeV}}\right) \left( \frac{g_*}{100}\right)^{1/6} \text{Hz}
\end{eqnarray}
and 
\begin{eqnarray}
    \tilde{\Omega}^{env}_{GW} &=& 1.67\times10^{-5}\frac{0.44 v_w^3}{1 + 8.28 v_w^3}\left(\frac{H_*}{\beta}\right)^2\left( \frac{\kappa_c \alpha}{\alpha + 1}\right)^2 \left( \frac{100}{g_*}\right)^{1/3}
\end{eqnarray}
with $\{a,b\}=\{2.9, 0.9\}$ and  $\kappa_c$ corresponds to the efficiency factor  with numerical fits given in~\cite{Espinosa_2010}. Because we only calculate GW sourced by bubble collisions for points with $v_w=1$, we use 
\begin{eqnarray}
    \kappa_{c} = \frac{\alpha}{0.73 + 0.083\sqrt{\alpha} + \alpha}.
\end{eqnarray}

\textbf{Soundwaves:} As described in~\cite{Caprini:2019egz,Hindmarsh:2017gnf} the gravitational wave power spectrum from sound waves is calculated as
\begin{eqnarray}
\Omega_{GW}^{sw} h^2 &=& \tilde{\Omega}_{GW}^{sw} h^2 \, 7.35\times 10^{-5}\left(\frac{100}{g_*} \right)^{1/3}\left( \frac{\kappa_{sw}\alpha}{\alpha + 1}\right)^2(H_*R_*) \left( \frac{f}{f_{p,0}} \right)^3 \left( \frac{7}{4 + 3 (f/f_{p,0})^2} \right)^{7/2}\label{eq:soundwave}
\end{eqnarray}
where  $\tilde{\Omega}_{GW}^{sw} = 0.012$,
and $f_{p,0}$ is the peak frequency today, given by
\begin{eqnarray}
    f_{p,0} &=& 26\times 10^{-6}\left( \frac{1}{H_*R_*}\right)\left( \frac{z_p}{10}\right) \left( \frac{T_*}{10^2 \text{ GeV}}\right)\left( \frac{g_*}{100}\right)^{1/6}\text{Hz}
\end{eqnarray}
with $z_p\simeq 10$ and the mean bubble separation, $R_* = (8\pi)^{1/3}v_w/\beta$. Additionally, Eq.~(\ref{eq:soundwave}) assumes the source lasts near a Hubble time. For GW sourced by sound waves, the efficiency factors are taken from the fits in Appendix A of~\cite{Espinosa_2010}. To be conservative, we reduce the total GW power by a factor of 
\begin{eqnarray}
    \frac{H_*R_*}{\sqrt{\frac{\kappa\alpha}{(1+\alpha)} }}
\end{eqnarray} which is the statement that the shock front formation timescale is shorter than a Hubble time.

%% file: App_Details.tex
\begin{figure}[ht!]
\begin{center}
\includegraphics[width=0.99\textwidth]{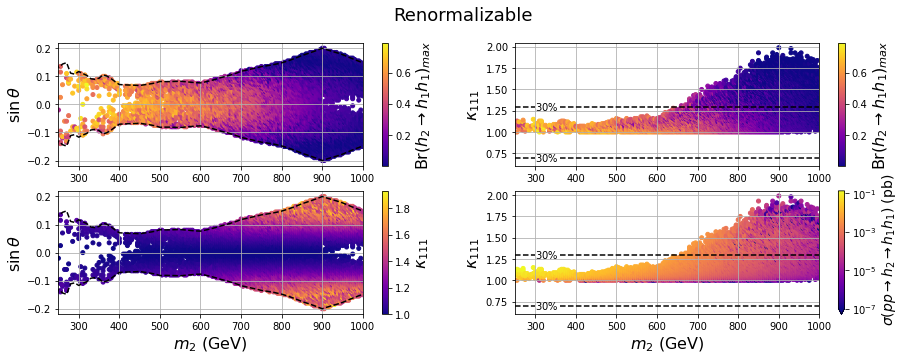}\\
\includegraphics[width=0.99\textwidth]{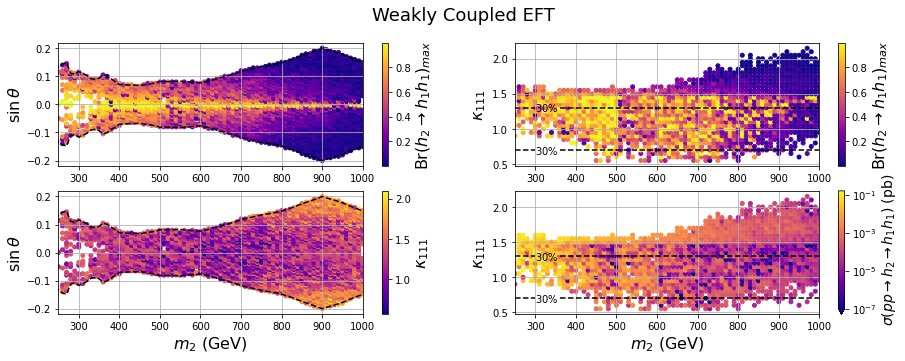}
\end{center}
\caption{\label{fig:collider_EWPT_tree}Same as Fig.~\ref{fig:collider_tree} but now requiring a strong first order EWPT. }
\end{figure}

\begin{figure}[ht!]
\begin{center}
\includegraphics[width=0.99\textwidth]{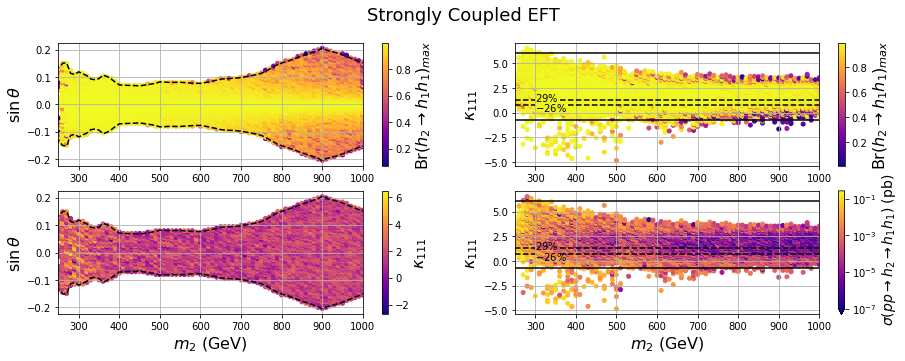}
\includegraphics[width=0.99\textwidth]{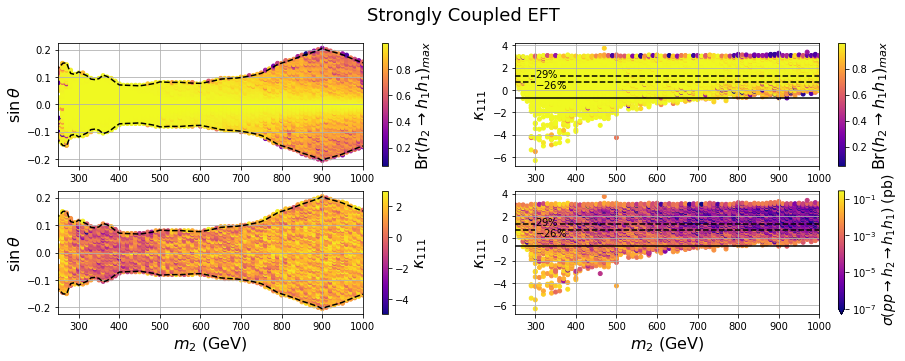}
\includegraphics[width=0.99\textwidth]{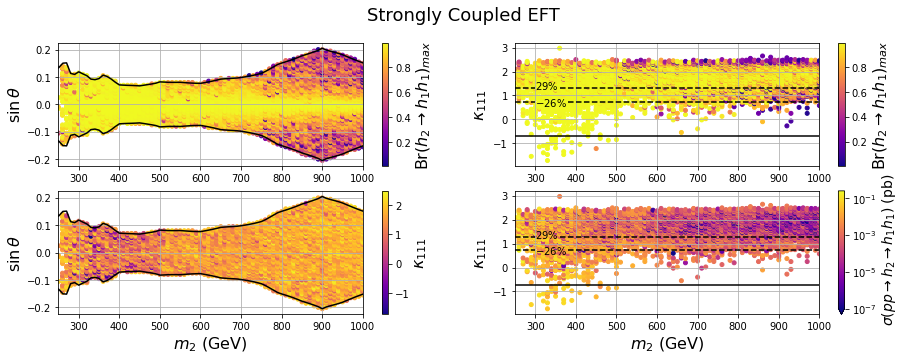}
\end{center}
\caption{\label{fig:collider_strong} Results for the strongly coupled EFT with $|C_i|\leq 4\pi$: (top) is the same as Fig.~\ref{fig:collider_tree}, (middle) is the same as Fig.~\ref{fig:collider_CW}, (bottom) is the same as Fig.~\ref{fig:collider_EWPT_oneloop}.}
\end{figure}

In this appendix we provide supplemental numerical results for the collider phenomenology. Fig.~\ref{fig:collider_EWPT_tree} is the same as Fig.~\ref{fig:collider_tree} but with the scalar trilinears evaluated at loop level. In Fig.~\ref{fig:collider_strong} we show results for the collider phenomenology in the strongly coupled EFT with $|C_i|\leq 4\pi$. The top is the same as Fig.~\ref{fig:collider_tree}, the middle is the same as Fig.~\ref{fig:collider_CW}, and the bottom is the same as Fig.~\ref{fig:collider_EWPT_oneloop}.